\documentclass[onecolumn,amsmath,amssymb,nofootinbib,12pt]{article}
\usepackage{jheppub}
\usepackage{ifpdf}

\usepackage{graphicx,subcaption}
\usepackage{amsfonts}
\usepackage{amsmath}
\usepackage{amssymb}
\usepackage{epsfig}
\usepackage{pdfpages}
\usepackage{graphicx,epstopdf}
\usepackage[makeroom]{cancel}
\usepackage{hyperref}
\hypersetup{pdftex,colorlinks=true,allcolors=blue}
\usepackage{array}
\usepackage{ulem}

\newcommand{\RN}[1]{%
  \textup{\uppercase\expandafter{\romannumeral#1}}%
}

\newcommand{\del}{\partial}
\newcommand{\vac}{\mt{vac}}

\newcommand{\BH}{\mt{BH}}

\newcommand{\mx}{\mt{max}}
\newcommand{\mn}{\mt{min}}
\newcommand{\past}{\mt{past}}

\newcommand{\be}{\begin{equation}}
\newcommand{\ee}{\end{equation}}
\newcommand{\bea}{\begin{eqnarray}}
\newcommand{\eea}{\end{eqnarray}}
\newcommand{\beq}{\begin{equation}}
\newcommand{\eeq}{\end{equation}}
\newcommand{\beqa}{\begin{eqnarray}}
\newcommand{\eeqa}{\end{eqnarray}}
\newcommand{\beqar}{\begin{eqnarray*}}
\newcommand{\eeqar}{\end{eqnarray*}}

\newcommand{\labell}[1]{\label{#1}} 

\newcommand{\eg}{{\it e.g.,}\ }
\newcommand{\ie}{{\it i.e.,}\ }
\newcommand{\reef}[1]{(\ref{#1})}

\newcommand{\mt}[1]{\textrm{\tiny #1}}

\newcommand{\eps}{\epsilon}
			
\newcommand{\tr}{{\tilde r}}

\newcommand{\rh}{{r_h}}
\renewcommand{\l}{\ell}
\newcommand{\veps}{\varepsilon}
\def\S{\Sigma}
\newcommand{\cv}{{\cal C}_\mt{V}}
\newcommand{\ca}{{\cal C}_\mt{A}}
\newcommand{\E}{P}

\newcommand{\tL}{\ell_\mt{ct}}

\usepackage{xspace}

\begin{document}

\preprint{arXiv:1804.nnnnn [hep-th]}
\title{\Large Holographic Complexity in Vaidya Spacetimes\ \ I}

\author[a]{Shira Chapman,}
\author[a, b]{Hugo Marrochio,}
\author[a]{Robert C. Myers}
\affiliation[a]{Perimeter Institute for Theoretical Physics, Waterloo, ON N2L 2Y5, Canada}
\affiliation[b]{Department of Physics $\&$ Astronomy,
University of Waterloo,\\ Waterloo, ON N2L 3G1, Canada}

\emailAdd{schapman@perimeterinstitute.ca}
\emailAdd{hmarrochio@perimeterinstitute.ca}
\emailAdd{rmyers@perimeterinstitute.ca}

\date{\today}

\abstract{We examine holographic complexity in time-dependent Vaidya spacetimes with both the complexity$=$volume (CV) and complexity$=$action (CA) proposals. We focus on the evolution of the holographic complexity for a thin shell of null fluid, which collapses into empty AdS space and forms a (one-sided) black hole. In order to apply the CA approach, we introduce an action principle for the null fluid which sources the Vaidya geometries, and we carefully examine the contribution of the null shell to the action. Further, we find that adding a particular counterterm on the null boundaries of the Wheeler-DeWitt patch is essential if the gravitational action is to properly describe the complexity of the boundary state. For both the CV proposal and the CA proposal (with the extra boundary counterterm), the late time limit of the growth rate of the holographic complexity for the one-sided black hole is precisely the same as that found for an eternal black hole. }

\maketitle


\section{Introduction}\label{sec:Intro}

Information theory and entanglement have long been seen to play a role in quantum gravity \cite{jb1,jb2,Sorkin:2014kta,Jacobson:1995ab}, however, this perspective has become central to many recent investigations of holography. In particular, it is now evident that quantum entanglement of the microscopic degrees of freedom is a key element leading to the emergence of the semi-classical spacetime geometry in the bulk \cite{VanRaamsdonk:2009ar, VanRaamsdonk:2010pw}. The novelty of the gauge/gravity duality \cite{AdSCFT,AdSCFTbook} is that the holographic dictionary between the bulk and boundary theories provides a framework where new tools and techniques from quantum information science can be precisely tested in quantum gravity. Much of this discussion has focused on the idea of holographic entanglement entropy \cite{Ryu:2006bv,Ryu:2006ef,Hubeny:2007xt, HoloEntEntropy},
which has led to a remarkably rich and varied range of new insights, \eg \cite{Swingle:2009bg,Myers:2010tj,Blanco:2013joa, Dong:2013qoa,Faulkner:2013ica, Almheiri:2014lwa, Pastawski:2015qua, Czech:2015qta,deBoer:2015kda,
Jafferis:2015del, Czech:2017zfq}.

However, it was recently observed that holographic entanglement entropy will not capture certain features of the late time behaviour of eternal black hole geometries or of the dual boundary thermal states \cite{EntNotEnough}. This motivated the suggestion that quantum circuit complexity may play a role in understanding holography. In the holographic context, we think about the quantum complexity of states, which is a measure of the resources required to prepare a particular state of interest, by applying a series of (simple) elementary gates to a (simple) reference state, \eg see \cite{johnw,AaronsonRev} for  reviews. Two parallel proposals have been developed in holography to describe the quantum complexity of states in the boundary theory, namely, the complexity=volume (CV) conjecture \cite{Susskind:2014rva,Stanford:2014jda} and  the complexity=action (CA) conjecture \cite{Brown1,Brown2}. The CV conjecture equates the complexity of the boundary state with the volume of an extremal (codimension-one) bulk surface anchored on a time slice $\S$ in the boundary where the state is defined. More precisely,
\begin{equation}
\cv(\S) =\ \mathrel{\mathop {\rm
max}_{\scriptscriptstyle{\S=\partial \mathcal{B}}} {}\!\!}\left[\frac{\mathcal{V(B)}}{G_N \, L}\right] \, ,\labell{defineCV}
\end{equation}
where $\mathcal B$ corresponds to the bulk surface of interest, while $G_N$ and $L$ denote Newton's constant and the AdS curvature scale, respectively, in the bulk theory. Instead, the CA conjecture equates the complexity with the gravitational action evaluated on a region of spacetime, known as the Wheeler-DeWitt patch (WDW), corresponding to the causal development of any of the bulk surfaces $\mathcal B$ appearing above. It reads
\beq
\ca(\S) =  \frac{I_\mt{WDW}}{\pi\, \hbar}\,. \labell{defineCA}
\eeq
Currently, both conjectures seem to provide viable candidates for holographic complexity but it still is far from clear how to construct a derivation for either of these proposals, \ie how to translate a known calculation of complexity in the boundary theory into a geometric procedure in the bulk. However, the past few years have seen extensive interest in studying these new gravitational observables, complexity in quantum field theory and the corresponding conjectures, \eg \cite{Roberts:2014isa, Cai:2016xho,RobLuis,Format,diverg,2LawComp, EuclideanComplexity1,EuclideanComplexity2,EuclideanComplexity3,Reynolds:2017lwq, koji, qft1, qft2,qft3,Growth, fish, brian, Alishahiha:2018tep, Zhao:2017isy, Fu:2018kcp,Bridges,Moosa}.

This paper presents another step in this research program, in which we investigate the full time evolution of holographic complexity for a class of time-dependent geometries. In particular, we study the time evolution of complexity in Vaidya shock wave spacetimes \cite{Vaid0,OriginalVaidya,VaidyaAdS}, with a collapsing shell of null matter in asymptotic AdS spacetimes. In fact, holographic complexity has already been studied for these geometries both for one-sided black holes, \eg \cite{Bridges,Moosa}, where the shell is injected into empty AdS space, and for two-sided black holes, \eg \cite{Stanford:2014jda,Brown1,Brown2, Roberts:2014isa}, where the shell falls into an existing eternal black hole. In the present paper, we focus on the case of black hole formation, \ie one-sided black holes, but we also consider shock waves falling into an eternal black hole in a companion paper \cite{Vad2}. First, we demonstrate that the null fluid action vanishes on-shell, and hence does not contribute to the WDW action. The standard prescription to evaluate the WDW action chooses the generators of the null boundaries to be affinely parametrized \cite{RobLuis}. However, we demonstrate that this prescription yields unsatisfactory results, \eg the complexity actually decreases in the case of a two-dimensional boundary CFT. However, this situation can be corrected
by supplementing the gravitational action with an additional counterterm on the null boundaries. This counterterm was introduced in \cite{RobLuis} to establish the invariance of $I_\mt{WDW}$ under reparametrizations of the null boundaries. For stationary spacetimes, the addition of this counterterm does not significantly change the properties of the holographic complexity, \eg see \cite{Format,Growth}. However, it appears to be an essential ingredient of the CA proposal \reef{defineCA} if the WDW action is to properly describe the holographic complexity of dynamical spacetimes, such as the Vaidya geometries. We also evaluate the holographic complexity for these spacetimes using the CV proposal \reef{defineCV} and compare the behaviour of the complexity for these two approaches. Our results are stated for general spacetime dimensions, as well as for both planar and spherical horizons.

The remainder of the paper is organized as follows: In section \ref{sec:NullDust}, we begin by constructing an action for a null fluid and we demonstrate that the on-shell fluid action vanishes. While this simplifies the evaluation of the WDW action, in section \ref{nfca}, we carefully examine the contribution of the region containing a narrow shell of null fluid and show that it vanishes as the width of the shell shrinks to zero.  Hence with an infinitely thin shell, the WDW action can be evaluated as the sum of the actions for two separate regions, the first inside the shell and the second outside the shell. In section \ref{duchess}, we consider the counterterm for null boundaries and consider its contribution in presence of a collapsing shell of null fluid. In section \ref{sec:evolve}, we study the evolution of the holographic complexity, using both the CA and CV conjectures, in the formation of a black hole modeled by the Vaidya geometry for a null shell collapsing into the AdS vacuum spacetime. In section \ref{sec:Discussion}, we briefly discuss our results and indicate some possible future directions.

\section{Null Fluid and the Vaidya Geometry}\label{sec:NullDust}

We start by introducing the background spacetime for our present studies of holographic complexity, namely the AdS-Vaidya spacetime. Vaidya geometries are a special class of metrics which among other things provide an analytic description of the formation of black holes by a gravitational collapse \cite{Vaid0,OriginalVaidya}. The collapse that can be studied here is generated by sending in a homogeneous shell composed of null fluid (or null dust), and the construction is easily extended to the case of asymptotically AdS boundary conditions, \eg \cite{VaidyaAdS}. In the latter holographic setting, the limit of sending in an infinitely thin, spherically symmetric shell of matter with finite energy has been studied extensively --- \eg see \cite{BatMin09, Das:2010yw,esp, ThermAl0,ThermaAlice, Garfinkle:2011hm, HubenyV, HubenyV2,ShenkerStanfordScrambling,multiple}.

We will be studying holographic complexity for a $d$-dimensional boundary CFT dual to an asymptotically AdS$_{d+1}$  Vaidya spacetime with a metric given by
\begin{align}\label{MetricV}
\begin{split}
&d s^{2} =  - F(r,v)\, d v^2 + 2\, dr\, dv + r^2
\,d \Sigma^{2}_{k,d-1}   \\
&\quad{\rm with}  \quad
F(r,v) = \frac{r^2}{L^2}+k -  \frac{f_{\text{p}}(v)}{r^{d-2}} \, .
\end{split}
\end{align}
If we fix the profile $f_{\text{p}}(v)=\omega^{d-2}$ to be a fixed constant,  these metrics would correspond precisely to the  black hole geometries in $d\geq 3$ for which the holographic complexity was studied in \cite{Format,Growth}.\footnote{It is straightforward to extend these metrics to the special case of $d=2$, and we treat the corresponding process of BTZ black hole formation separately in section \ref{sec:evolve}.} In particular, they are written in terms of the Eddington-Finkelstein  coordinate $v$, parameterizing ingoing null rays. Further, $L$ denotes  the  AdS  curvature  scale  while
$k$ indicates  the  curvature  of  the  horizon\footnote{As usual, $k$ takes three different values, $\lbrace +1,0,-1\rbrace$, which correspond to spherical, planar, and hyperbolic horizon geometries, respectively.  Following the notation of \cite{Format,Growth}, we will use $\Omega_{k,d-1}$ to denote the dimensionless volume of the corresponding spatial geometry in the expressions below.  For $k=+1$, this is just the volume of a ($d$--1)-dimensional unit sphere, \ie $\Omega_{1,d-1} = 2\pi^{d/2}/\Gamma(d/2)$, while for hyperbolic and planar geometries, we must introduce an infrared regulator to produce a finite volume. \label{footy22}} situated at $r=r_h$ where
\beq
\omega^{d-2}=r_h^{d-2}\left(\frac{r_h^2}{L^2}+k\right)\,.
\label{horz}
\eeq
However, the profile $f_{\text{p}}(v)$ may be taken from a large class of functions and then the metric \reef{MetricV} describes the collapse of a shell of null fluid. Generally, one would require that the profile is positive to ensure that the total mass is positive at all times,\footnote{In fact, the stress-tensor depends on the derivative of the profile function with respect to $v$ (see eq.~\reef{cronk} below), so that one should choose the profile to increase monotonically to ensure the energy density is everywhere positive. Note that for $k=-1$, the mass can take negative values in a restricted range, \eg see \cite{Format}.} and monotonically increasing so that the energy density of the shell is everywhere positive --- see below. As an example, consider the profile
\beq
f_{\text{p}}(v)=\omega_1^{d-2} \left(1- {\cal H}(v-v_s)\right)
+\omega_2^{d-2} \, {\cal H}(v-v_s)\,,
\label{heavyX}
\eeq
where ${\cal H}(v)$ is the Heaviside step function. This profile describes an infinitely thin shell collapsing along the null surface $v=v_s$, and it generates a sharp transition connecting one black hole geometry with mass proportional to $\omega_1^{d-2}$ to another black hole with mass proportional to $\omega_2^{d-2}$. In section \ref{sec:evolve}, we will choose $\omega_1=0$ in which case this profile \reef{heavyX} corresponds to a shell collapsing into the AdS vacuum and forming a (one-sided) black hole.

\subsection{Action for a Null Fluid}

To evaluate the holographic complexity using the CA conjecture, we need to take into account the action of the matter fields in the collapsing shell. Hence, we present here a construction of the action principle for a null fluid, which is inspired in part by the fluid actions given in \cite{PerfFluidAct, HartnollPerfFluid}.\footnote{There is an enormous literature on the subject of the action principle for relativistic fluids, \eg see \cite{schutz,Nicolis1} for further discussions of perfect fluids and \cite{Nicolis2, Kovtun:2014hpa, Felix1, Crossley:2015evo,Torrieri:2016dko} for recent developments in describing dissipative hydrodynamics. }  Let us also note that, a null fluid action was also constructed in \cite{kuchar1} using a complementary set of variables.\footnote{We note that the on-shell action also vanishes using this alternative approach.} Further, in a particular limit, it is also possible to use a massless scalar field as the source in the Vaidya metric \cite{BatMin09}.\footnote{However, this description breaks down near the singularity, \ie the solution is not well approximated by the Vaidya metric there. Therefore we did not adopt this approach since in general, the near-singularity region makes a finite contribution to the holographic complexity in CA calculations.}

The stress tensor of a null fluid takes the following simple form
\begin{equation}\label{stressnull}
T_{\mu \nu} = \veps(x^{\mu})\,\ell_{\mu}\, \l_{\nu} \, ,
\end{equation}
where $\ell^\mu$ is a null vector, \ie $\ell^\mu\ell_\mu=0$.
We can compare the above expression to the stress tensor for a conventional relativistic fluid: $T_{\mu \nu} = (\veps + p)u_{\mu} u_{\nu} + p \,g_{\mu \nu}$ where $\veps$ and $p$ are the local energy density and pressure, respectively. Further, $u^\mu$ is the local four-velocity of the fluid elements, with $u^{\mu} u_{\mu} = -1$. Hence eq.~\reef{stressnull} can be thought of as the limit where the fluid velocity becomes null and the pressure vanishes, \ie $u^\mu\to\l^\mu$ and $p=0$. Now one can show that the on-shell action for a conventional fluid is simply an integral of the local pressure \cite{PerfFluidAct} and hence this result suggests that the on-shell action for a null fluid should vanish. We demonstrate below this intuitive result is in fact correct. We follow in part the construction in \cite{HartnollPerfFluid}, but adapt it to describe the null fluid stress tensor \reef{stressnull}.

We take the following ansatz for the fluid action
\beq\label{ActionFluid}
I_\mt{fluid} = \int d^{d+1} x \sqrt{-g}\, \mathcal{L}_\mt{fluid}\qquad{\rm where}\ \ \
\mathcal{L}_\mt{fluid}(\lambda, \phi, s, \l^{\mu}, g_{\mu \nu}) =  \lambda \, g_{\mu \nu} \l^{\mu} \l^{\nu}+ s \,\l^{\mu} \partial_{\mu} \phi   \, .
\eeq
This action involves a number of auxiliary fields, beginning with $\lambda$ which is a Lagrange multiplier imposing the constraint that $\l^\mu$ is null on shell. With only the first term in the Lagrangian, we would obtain equations of motion which set $\l^\mu=0$ (or $\lambda=0$) everywhere, and hence the corresponding stress tensor would also vanish. Therefore, the second term, involving a contraction of $\l^\mu$ with the derivative of a new scalar $\phi$, is added in eq.~\reef{ActionFluid}. The field $s$ can in principle be reabsorbed with a redefinition of $\l^\mu$ (and in this sense it represents a redundancy in the description) but we will keep it to allow for an arbitrary rescaling of $\l^\mu$. The equations of motion for the full action \reef{ActionFluid} are:
\begin{subequations}\label{EoMFluid}
\begin{align}
&\frac{1}{\sqrt{-g}}\frac{\delta I_\mt{fluid}}{\delta \lambda} = \l_{\mu} \l^{\mu} = 0  \label{subfig:EoMFluid1} \, ,
\\
&\frac{1}{\sqrt{-g}}\frac{\delta I_\mt{fluid}}{\delta \l^{\mu}} =2 \lambda\, \l_{\mu} + s \, \partial_{\mu} \phi  =0 \label{subfig:EoMFluid4} \, ,
\\
&\frac{1}{\sqrt{-g}}\frac{\delta I_\mt{fluid}}{\delta \phi} =  -\nabla_{\mu} (s\, \l^{\mu}) = 0 \label{subfig:EoMFluid3} \, ,
\\
&\frac{1}{\sqrt{-g}}\frac{\delta I_\mt{fluid}}{\delta s} =  \l^{\mu} \partial_{\mu} \phi = 0 \label{subfig:EoMFluid2} \, .
\end{align}
\end{subequations}
These equations of motion provide us with an interpretation of the various fields. Of course,
eq.~\eqref{subfig:EoMFluid1} enforces that  $\l^{\mu}$ is null on-shell. Eq.~\eqref{subfig:EoMFluid4} indicates that the null `fluid velocity' $\l^\mu$ and the gradient of $\phi$ point in the same direction and fixes the prefactor in the proportionality relation between them in terms of the fields $s$ and $\lambda$. In this sense, $\phi$ plays a role analogous to the  velocity potential in potential flows \cite{HartnollPerfFluid}.
Eq.~\eqref{subfig:EoMFluid3} implies that $s$ has an interpretation of a conserved charge density. Since all fields are real (\ie the fluid is neutral) $s$ can be understood as the entropy density \cite{HartnollPerfFluid}.
Eq.~\eqref{subfig:EoMFluid2} follows automatically by contracting eq.~\eqref{subfig:EoMFluid4} with $\l^{\mu}$.
Varying the action with respect to the metric yields the stress tensor
\begin{equation}
T_{\mu \nu}  \equiv - \frac{2}{\sqrt{-g}} \frac{\delta I_\mt{fluid}}{\delta g^{\mu \nu}}
=  - s(\l_{\mu} \partial_{\nu} \phi + \l_{\nu} \partial_{\mu} \phi) - 2 \lambda \l_{\mu} \l_{\nu}  + g_{\mu \nu} (s \l^{\sigma} \partial_{\sigma} \phi + \lambda \l^{\sigma} \l_{\sigma})  \, .
\end{equation}
On-shell, this expression reduces to the desired form
\begin{equation}\label{StrEnerF}
T_{\mu \nu} =   2 \lambda\, \l_{\mu}\, \l_{\nu}   \,  ,
\end{equation}
and comparing to eq.~\reef{stressnull}, we see that $\lambda$ is proportional to the energy density \ie $\veps=2\lambda$.
Further, we note that imposing the equations of motion \eqref{EoMFluid} yields a vanishing action \eqref{ActionFluid}, \ie
\begin{equation}
\left[I_\mt{fluid}\right]_{\text{on-shell}} = 0 \, .\label{vanish}
\end{equation}
Therefore, in evaluating the holographic complexity using the CA conjecture, our calculations reduce to evaluating the geometrical quantities  in the gravitational action \eqref{THEEACTION} with the Vaidya metric \eqref{MetricV} and there will be no explicit contribution from the matter fields.

Upon substituting the metric \reef{MetricV} into the Einstein equations, only the $vv$ component is nontrivial with
\begin{equation}\label{cronk}
E_{vv} = \frac{(d-1) }{2 \, r^{d-1}  } \, f^{'}_{\text{p}}(v) = 8\pi G_N \, T_{vv} \, .
\end{equation}
We see from eq.~\eqref{StrEnerF} that this
forces $\l_\mu$ to point in the $v$ direction, \ie $\l_{\mu} d x^{\mu} \propto d v$. Recall that retaining the parameter $s$ in eq.~\reef{ActionFluid} meant that we could rescale $\l_\mu$ at will, and we use this freedom to pick an affine parametrization of the form $\l_{\mu} d x^{\mu} = d v$. In this case, combining eqs.~\reef{StrEnerF} and \reef{cronk} yields
\begin{equation}\label{ghosh}
\lambda =  \frac{(d-1)}{32\pi G_N  } \,
\frac{f^{'}_{\text{p}}(v)}{r^{d-1} }
\, .
\end{equation}
Since we identified $\lambda=2\veps$, we see here that the energy density is proportional to the derivative of the profile $f_{\text{p}}(v)$.
Next, eq.~\eqref{subfig:EoMFluid3} yields
\begin{equation}
\partial_{r} ( r^{d-1} s) = 0
\end{equation}
and as a result, the entropy density is given by
\begin{equation}
s = \frac{s_{0}}{r^{d-1}}\, .
\end{equation}
In eq.~\eqref{subfig:EoMFluid4}, we see that we must have $\phi=\phi(v)$
and the full equation becomes
\begin{equation}
s_{0} \, \partial_{v} \phi +  \frac{d-1}{16\pi G_N}  \,f^{'}_{\text{p}}(v) = 0 \, .
\end{equation}
Integrating this equation then produces
\begin{equation}
\phi = \phi_{0} - \frac{(d-1) }{16\pi G_N s_{0}} \, f_{\text{p}}(v) \label{PhiInt} \, .
\end{equation}
The integration constants, $s_0$ and $\phi_0$, will be fixed by the asymptotic boundary conditions for the matter.

\subsection{Null Fluids \& Complexity=Action} \label{nfca}

\begin{figure}
\centering
\includegraphics[scale=0.1]{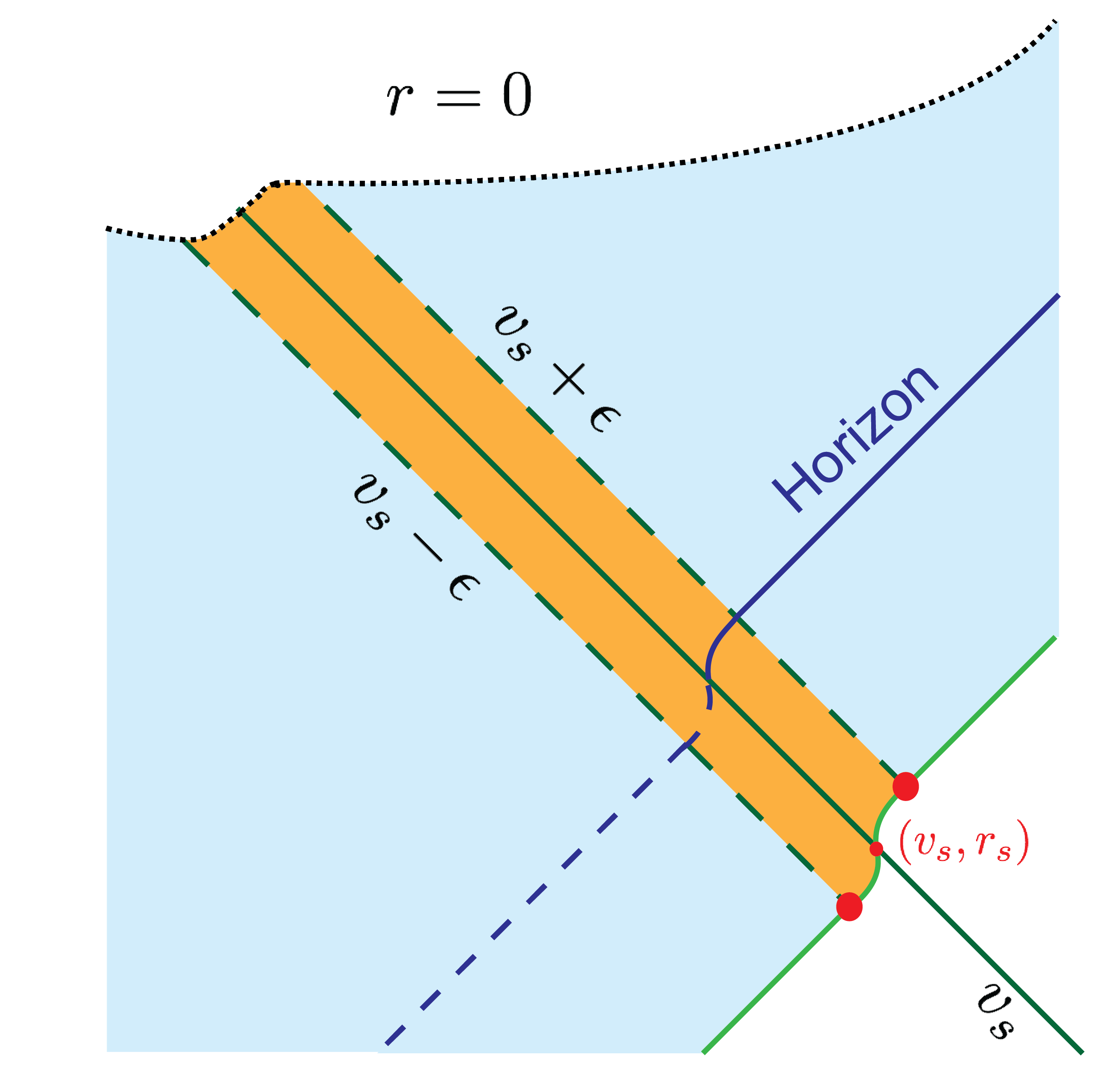}
\caption{The null shell has a finite thickness $2\veps$ around the null ray $v=v_s$. The portion enclosed by the WDW patch is shaded in orange. The contribution of the two joints indicated by red dots exactly cancels the surface term for the portion of the null boundary connecting the joints, where we have a time dependent $\kappa (v)$.}
\label{VaFocus}
\end{figure}

Having constructed a consistent null fluid action, which we showed vanishes on-shell, and found the corresponding source for the AdS-Vaidya geometry \reef{MetricV}, we can begin to study the holographic complexity in these dynamical spacetimes. In particular, to study the complexity$=$action proposal \reef{defineCA}, we showed that the null fluid action vanishes and so we need only to consider the gravitational action in the Vaidya spacetimes sourced by a collapsing shell of null fluid. Further, in section \ref{sec:evolve} and in a companion paper \cite{Vad2}, we focus on the case where the shell is very thin, \ie the profile takes the form given in eq.~\reef{heavyX}. Using the additivity of the gravitational action \cite{RobLuis}, the problem essentially then factorizes into evaluating the action for two stationary spacetime regions: one before the collapse, characterized by the mass parameter $\omega_1$, and one after, characterized by $\omega_2$. However, in this section, we wish to verify that the null shell does not contribute to the Wheeler-DeWitt (WDW) action by first considering a thin but finite-width shell --- see figure \ref{VaFocus}. That is, we split the spacetime into three regions: the stationary region before the collapse, the shell of finite width, and the stationary region after the collapse. In this section, we will only focus on the contribution of the null shell to the WDW action and we will confirm that in the limit that the width of the shell shrinks to zero this contribution vanishes, as expected. However, this analysis will also reveal a new boundary condition on the null normal to the past boundary of the WDW patch as it crosses the collapsing shell.

Recall that the CA conjecture \reef{defineCA} proposes that the complexity of the CFT state on some time slice $\S$ in the boundary is given by the bulk action evaluated on the corresponding WDW patch. In this work, we follow the conventions of \cite{diverg}.\footnote{We noticed a typo in the null surface contribution to the action, proportional to $\kappa$, in \cite{diverg, RobLuis}. Correcting for this mistake, we have flipped the sign of the $\kappa$ term above. We comment further on this issue below where this sign becomes important --- see eq.~\reef{VaKappa}.} For the Vaidya geometries with a null fluid, the bulk action becomes \cite{RobLuis}
\begin{equation}\label{THEEACTION}
\begin{split}
I = &I_\mt{grav} +  I_\mt{fluid} \\
I_\mt{grav} = & \frac{1}{16 \pi G_N} \int_\mathcal{M} d^{d+1} x \sqrt{-g} \left(\mathcal R + \frac{d(d-1)}{L^2}\right) \\
&\quad+ \frac{1}{8\pi G_N} \int_{\mathcal{B}} d^d x \sqrt{|h|} K + \frac{1}{8\pi G_N} \int_\Sigma d^{d-1}x \sqrt{\sigma} \eta
\\
&\quad  + \frac{1}{8\pi G_N} \int_{\mathcal{B}'}
d\lambda\, d^{d-1} \theta \sqrt{\gamma} \kappa
+\frac{1}{8\pi G_N} \int_{\Sigma'} d^{d-1} x \sqrt{\sigma} a \, ,
\end{split}
\end{equation}
and $I_\mt{fluid}$ was given in eq.~\reef{ActionFluid}.
Recall that, as we showed in eq.~\reef{vanish}, the fluid action $I_\mt{fluid}$ vanishes when evaluated for a solution of the fluid equations of motion \reef{subfig:EoMFluid1}-\reef{subfig:EoMFluid2}. Of course, this does not imply that there is no consequence of the shock wave, but rather that its effect only appears through the backreaction of the geometry, namely, in forming the collapsing geometry \eqref{MetricV}. For the gravitational action $I_\mt{grav}$, we have the standard geometric quantities and boundary terms, which include contributions from null boundaries and joints \cite{RobLuis}. The bulk integral contains the Einstein-Hilbert action with the Ricci scalar $\mathcal{R}$ and the cosmological constant $\Lambda=-d(d-1)/(2L^2)$. Next, we have the Gibbons-Hawking-York (GHY) surface term \cite{York,GH} for smooth timelike and spacelike segments of the boundary, which is defined in terms of the trace of the extrinsic curvature $K$. There is also an analogous boundary term for the null segments that depends on $\kappa$, which indicates by how much the coordinate $\lambda$ along the null boundaries departs from affine parametrization. Further, there are the Hayward joint terms \cite{Hay1,Hay2}, which appear at the intersection of two timelike or spacelike boundary segments and which are defined in terms of the ``boost angle" $\eta$ between the corresponding normal vectors. Finally, the last contribution in $I_\mt{grav}$ involving an analogous ``angle'' $a$ appears for the joints including at least one null segment \cite{RobLuis}.

As discussed in \cite{RobLuis}, there are inherent ambiguities in calculating the gravitational action for regions delimited by null boundaries. However, we follow the suggestion of \cite{RobLuis}, of choosing affine parametrization for the null normals (\ie setting $\kappa=0$) and fixing their overall normalization constant by normalizing with vectors at the boundary. In particular,  at infinity there is an asymptotic timelike Killing vector $\hat t = \partial_{t}$ generating time translations in the boundary and we fix  $\hat t \cdot k = \pm \alpha$, with the $+$ ($-$) for the normal to the future (past) null boundary. Both of these choices have the advantage that they do not make any reference to the background for which we are evaluating the complexity and so they allow for an unambiguous comparison of the complexities evaluated on different bulk geometries or on different boundary time slices in a given bulk geometry. Hence our evaluation of the WDW action in the following and in the next section will use both of these choices.

As discussed above, we want to consider an AdS-Vaidya spacetime \reef{MetricV} where the shell of null fluid is narrow but still has a finite width. In particular, the shell will extend from $v_\mn=v_s-\veps$ to $v_\mx=v_s+\veps$, as shown in figure \ref{VaFocus}. Further, the shell will separate two stationary\footnote{That is, $\partial_t$ satisfies the usual Killing equations in either region.} spacetime regions characterized by the mass parameter $\omega_1^{d-2}$ inside the shell and by $\omega_2^{d-2}$ outside the shell. The details of the profile $f_{\text{p}}(v)$ in the metric will not be important but we assume that it is continuous (and smoothly increasing). Of course, from integrating eq.~\reef{cronk} across the shell, the profile must also satisfy
\beq\label{across}
f_{\text{p}}(v_s+\veps)-f_{\text{p}}(v_s-\veps)=\int_\mt{shell}\!\!\!\!dv\, f'_{\text{p}}(v)
=\omega_2^{d-2}-\omega_1^{d-2}\,.
\eeq
With these choices, in the limit $\veps\to0$, the profile reduces to that given in eq.~\reef{heavyX}.\footnote{Our calculations here are general enough to accommodate both black hole formation which we examine in section \ref{sec:evolve}, and null shocks in an eternal black hole background which we will study in \cite{Vad2}.}

Now we will evaluate the contribution of the null shell to the WDW action, but we will be particularly interested in the limit where the shell becomes infinitely thin, \ie $\veps\to0$. Let us examine the various terms in eq.~\reef{THEEACTION}. First, of course, the fluid action $I_\mt{fluid}$ vanishes on-shell, as we showed in the previous section. The bulk term in $I_\mt{grav}$ is (approximately) proportional to the volume of the shell and so vanishes in the limit that $\veps\to0$. Similarly evaluating the GHY term at the $r=0$ singularity (following the prescription in \cite{Format})  yields a result which vanishes as $\veps\to0$. The Hayward joint terms are not relevant for this particular region and hence we turn to the null surface and null joint terms.

First, we must introduce (outward-directed) normals for the upper and lower null boundaries,\footnote{We have chosen the same normalization constant $\beta$ for the two normals to simplify the final result, \ie this choice ensures that the null joint terms exactly cancel with the surface term below. Of course, another choice would yield the same result for the total action of the WDW patch after summing with the relevant boundaries and joints for the portions of the WDW patch above/below the null shell, since these are all inner boundaries of the WDW patch that we have introduced. However, we note that if the two normals in eq.~\reef{shelln} were not normalized with the same constant, the null shell would make a nonvanishing contribution to the total action.}
\beqa
v=v_s+\veps\ :&&\qquad k^\mt{s+}_{\mu} d x^{\mu}  = \ \  \beta\, d v \, ,
\nonumber\\
v=v_s-\veps\ :&&\qquad k^\mt{s--}_{\mu} d x^{\mu}  =  -\beta\, d v \, .
\labell{shelln}
\eeqa
With this choice, these null normals are affinely parameterized and therefore the null surface term vanishes, \ie $\kappa = 0$, for these two boundaries. We might add that the null joint terms vanish where these boundaries meet the singularity at $r=0$ because there the transverse volume vanishes for these two joints.

The final boundary for the shell region is a portion of the past null boundary of the WDW patch.
 From the metric \eqref{MetricV}, we can see that the normal to this boundary can be written as
\begin{equation}\label{VecVad}
k^{\mu} \partial_{\mu} = H(r, v)\left(\frac{2 }{F(r, v)} \partial_{v} +  \partial_{r}\right)  \, ,
\end{equation}
where $F(r, v)$ is the usual metric function --- see eq.~\reef{MetricV}. Note that with eq.~\reef{VecVad}, we are describing the null normal for the entire past null boundary $\mathcal{B}_\past$. Hence in the regions beyond the shell, the metric function $F$ simplifies to
\beq
F(r,v)=f_i(r) =\frac{r^2}{L^2}+k -  \frac{\omega_i^{d-2}}{r^{d-2}}\,,
\label{ffii}
\eeq
with $i=1$ and 2 denoting the region inside ($v<v_\mn$) and outside ($v>v_\mx$) of the shell, respectively. Of course, across the shell, $F$ depends on both $r$ and $v$ as shown in eq.~\reef{MetricV}.
Further, we have introduced an overall factor $H(r,v)$ in eq.~\reef{VecVad} to allow for the possibility that the normalization of the null normal changes when the past boundary crosses the shell of null fluid. For $v>v_\mt{max}$, we will set $H$ to be a fixed constant, \ie $H(r,v)=\alpha$ to match the asymptotic boundary condition $k \cdot \hat{t} = - \alpha$ (see discussion above). As we will see below, this simple choice also ensures that $\kappa=0$ on this outer portion of the past boundary. Similarly for $v<v_\mt{min}$, we set $H(r,v)=\tilde\alpha$ which is again a positive constant in order for the null generators to be affinely parametrized on the inner portion of the null boundary. However, we have taken the liberty to choose an independent constant $\tilde\alpha$ since this portion of the boundary never reaches asymptotic infinity. Expressing this normal \reef{VecVad} as a form, we have
\beq
k_\mu\,dx^\mu = H(r,v)\left(- dv+\frac{2}{F(r,v)}\,dr\right)\,.
\label{formVad}
\eeq
Then in the region above the shell where $F(r,v)=f_2(r)$,  eq.~\reef{formVad} takes the expected form $k_\mu\,dx^\mu =- \alpha\,du\equiv-\alpha\left(dv-2dr/f_2(r)\right)$. That is, in the region above the null shell, the past null boundary $\mathcal{B}_\past$ is a surface where the outgoing Eddington-Finkelstein coordinate is constant (see eqs.~\reef{tort}  and \reef{time3} below) and fixing $H$ to be a constant ensures that $\kappa$ vanishes there. Similarly in the region below the null shell, we find $k_\mu\,dx^\mu =- \tilde\alpha\,du$ and again $\kappa=0$ on this portion of the past boundary.

On the other hand, because of the $r$ and $v$ dependence of $F(r,v)$ and $H(r,v)$ within the shell of null fluid, the null normal \reef{VecVad} will only be affinely parametrized on this portion of $\mathcal{B}_\past$ with a special choice of $H$. We will return to this special choice below, but for now we consider more general possibilities for which $\kappa\ne0$. In particular, using $k^\rho\nabla_{\!\rho}\, k_\mu =\kappa\, k_\mu$, we find
\beqa
\kappa &=& \left(\frac{2}F\,\partial_{v}+\partial_r\right) H(r,v)-\frac{2H}{F^2}\,\partial_v F(r,v)
\nonumber\\
 &=& k^\mu\partial_\mu \log H(r,v)-k^\mu\partial_\mu\log F(r,v) + H\,\partial_r \log F(r,v)
 \labell{pink1}\\
 &=&\frac{d\ }{d\lambda} \log\!\frac{H(r,v)}{F(r,v)} + \frac{H}{F}\,\partial_r  F(r,v)\, ,
 \nonumber
\eeqa
where in the second line, we have used eq.~\reef{VecVad} to express $\kappa$ in terms of derivatives along the null boundary. In the final line, we have introduced $\lambda$, which parametrizes the null rays in the boundary such that $k^{\mu} = {\partial x^{\mu}}/{\partial \lambda}$. Note that in the extra term in the last line, there is a partial derivative with respect to $r$, but to evaluate $\kappa$, we must calculate this quantity on the null boundary.

The resulting surface term in the gravitational action is then\footnote{We would like to point out a crucial typo in \cite{RobLuis} and \cite{diverg}. In \cite{RobLuis}, there is a typo in the conventions established in appendix C (but not in the main text). In particular, in eq.~(C1), the overall sign of the surface term for null boundaries should be flipped. Similarly, there should be an overall flip of the sign of this surface term in appendix A of \cite{diverg}, \ie the sign in front of the null boundary term in eq.~(A.1) should be a plus.}
\begin{align}
&I_{\kappa} =\frac{1}{8\pi G_N} \int_{\mathcal{B}_\past}
d\lambda\, d^{d-1} \theta\ \sqrt{\gamma}\ \kappa \nonumber \\
&= \frac{\Omega_{k,d-1}}{8 \pi G_N} \int_{\mathcal{B}_\past\cap\,
{\rm shell}}\!\!\!\!\!
d\lambda\ r^{d-1}\left[ \frac{d\ }{d\lambda} \log\!\frac{H(r,v)}{F(r,v)} + \frac{H}{F}\,\partial_r  F(r,v) \right] \, .
\label{VaKappa}
\end{align}
Now at the center of the shock (\ie $v=v_s$), the radial coordinate takes some value\footnote{Of course, this position matches that described in the main text for an infinitely thin shell to within ${\cal O}(\veps)$.} $r=r_s$ and throughout the shell $r=r_s+{\cal O}(\veps)$. Hence to leading order in $\veps/r_s$, we can fix $r=r_s$ in the above integral, in which case the first term reduces to
\beqa
I_{\kappa,1} &=&  \frac{\Omega_{k,d-1}}{8 \pi G_N}\,r_s^{d-1} \int_{\lambda_\mn}^{\lambda_\mx}\!\!
d\lambda\ \frac{d\ }{d\lambda} \log\!\frac{H(r,v)}{F(r,v)} \ +{\cal O}(\veps/r_s)  \nonumber\\
&=&   \frac{\Omega_{k,d-1}}{8 \pi G_N}\,r_s^{d-1}\  \log\frac{F(r_s,v_\mn)  \,H(r_s,v_\mx)}{F(r_s,v_\mx)\, H(r_s,v_\mn)}\ +{\cal O}(\veps/r_s) \,.
\labell{VaKappa3}
\eeqa
Now we must still evaluate the integral over the second term in eq.~\eqref{VaKappa}. Here it is convenient to convert this to an integration over $v$ along the boundary using $dv/d\lambda=2H/F$ from eq.~\reef{VecVad}. Then this contribution to the boundary term becomes
\beqa
I_{\kappa,2} &=& \frac{\Omega_{k,d-1}}{16 \pi G_N} \int_{v_\mn}^{v_\mx}dv\ \left[r^{d-1}\,\partial_r  F(r,v)\right]_{r=h(v)}\,,
\labell{VaKappa3a}
\eeqa
where we have expressed the null boundary as a constraint equation $r=h(v)$. Of course, for the present thin shell, we have $r\simeq r_s+\veps\, \tilde h(v/r_s)$ where $\tilde h(v/r_s)$ is a smooth dimensionless function. Similarly, $\partial_r  F$ remains finite across the shell,\footnote{That is, $\partial_r F$ does acquire any terms proportional to a delta-function $\delta(v)$ in the limit $\veps\to0$.} and so we have $I_{\kappa,2}=O(\veps)$ since the range of integration is $\delta v=v_\mx-v_\mn=2\veps$. Therefore, in the limit $\veps\to0$, the surface term reduces to
\beq
I_{\kappa} = \frac{\Omega_{k,d-1}}{8 \pi G_N}\,r_s^{d-1}\  \log\frac{2\beta\, \alpha}{f_2(r_s) }
-\frac{\Omega_{k,d-1}}{8 \pi G_N}\,r_s^{d-1}\  \log\frac{2\beta\, \tilde\alpha}{f_1(r_s) } \,,
\labell{VaKappa4}
\eeq
where we have written the final result as a sum of two terms, in a suggestive manner.\footnote{Recall that $\beta$ is the normalization constant for the normals to the surfaces $v=v_s\pm\veps$ (see eq.~\reef{shelln}), but note that the $\log\beta$ terms cancel in the difference between the two terms in eq.~\reef{VaKappa4}.} Note that in converting the expression in eq.~\reef{VaKappa3} to the above result, we have used the fact that at either edge of the shell, $F(r,v)$ precisely matches the metric function $f_i(r)$ in the corresponding region beyond the shell, \eg $F(r,v_\mx)=f_2(r)$. A similar matching applies for the normalization function, as we described above, \ie $H(r_s,v_\mx)=\alpha$ and $H(r_s,v_\mn) = \tilde\alpha$.

Now the final contribution to the action of the null shell comes from the null joints where the two edges (\ie $v=v_s\pm\veps$) intersect the past boundary of the WDW patch (indicated by red dots in figure \ref{VaFocus}). Given the null normals in eqs.~\reef{shelln} and \reef{VecVad}, it is straightforward to evaluate these contributions using the prescription given in \cite{diverg} with the result
\beqa
I_\mt{joint} &=& -\left[\frac{\Omega_{k,d-1}}{8 \pi G_N}\,r^{d-1}\  \log\frac{\beta\, \alpha}{f_2(r) }\right]_{r=h(v_\mx)}
+\left[\frac{\Omega_{k,d-1}}{8 \pi G_N}\,r^{d-1}\  \log\frac{\beta\, \tilde\alpha}{f_1(r) }\right]_{r=h(v_\mn)}
\nonumber\\
&\underset{\scriptscriptstyle{\veps\to0}}{=}& -\frac{\Omega_{k,d-1}}{8 \pi G_N}\,r_s^{d-1}\  \log\frac{\beta\, \alpha}{f_2(r_s) }
+\frac{\Omega_{k,d-1}}{8 \pi G_N}\,r_s^{d-1}\  \log\frac{\beta\, \tilde\alpha}{f_1(r_s) }
\labell{VaJnt1}
\eeqa
where $r=h(v)$ again denotes the position of $\mathcal{B}_\past$. In the second line, we have used that within our narrow shell, the radial position of this boundary is fixed up to order $\veps$ corrections, \ie $r=h(v)\simeq r_s+\veps\, \tilde h(v/r_s)$. Now we see that the two nonvanishing contributions to the action evaluated on the thin null shell precisely cancel! That is, combining eqs.~\reef{VaKappa4} and \reef{VaJnt1}, we have
\beq
I_\mt{shell} \underset{\scriptscriptstyle{\veps\to0}}{=} I_\kappa + I_\mt{joint}=0\,.
\label{not}
\eeq
Therefore we have shown that in the limit of an infinitely thin shell, evaluating the WDW action in the Vaidya spacetime \reef{MetricV} reduces to two separate calculations: one for evaluating the action $I_2$ of the region outside of the shell ($v>v_s$) and another for evaluating the action $I_1$ of the inside region ($v<v_s$).

Notice that our result \reef{not} for the vanishing of the shell action did not require that we specify the value of $\tilde\alpha$, the normalization constant for the null normal on the portion of the past boundary $\mathcal{B}_\past$ before the collapse. Hence we are left with an ambiguity in evaluating $I_1$, the part of the WDW action coming from the region inside the null shell. This ambiguity is, of course, related to the ambiguities discussed in \cite{RobLuis} and it arises here because our calculations left $\kappa$
unspecified on the portion of the past boundary inside the shell --- see eq.~\reef{pink1}. As discussed above, the most natural way to fix this ambiguity is to simply set $\kappa=0$. In fact, we already made this choice for all of the other null boundaries above and it is certainly possible to fix $\kappa=0$ on $\mathcal{B}_\past$ inside the shell as well. One would simply treat eq.~\reef{pink1} with $\kappa=0$ as a (first order) differential equation for $H(r,v)$, or rather $H(\lambda)$ since we are only interested in the value of $H$ on the null boundary. The integration constant in this equation is fixed by setting $H =\alpha$ at the upper edge of the shell, \ie at $v=v_\mx$. Solving the differential equation will then determine the value of $\tilde\alpha$ as the value that $H$ reaches at the lower edge of the shell, \ie $v=v_\mn$. However, we can easily determine this value (at least in the limit $\veps\to0$) by examining the result for $I_\kappa$ in eq.~\reef{VaKappa4}. If $\kappa=0$ everywhere along the boundary, this contribution must vanish and so we must have
\beq
\tilde{\alpha}=\alpha\,\frac{f_1(r_s)}{f_2(r_s)}\,.
\label{magic22}
\eeq
We might also observe that the sum of the null joint terms in eq.~\reef{VaJnt1} also vanishes with this particular choice for $\tilde\alpha$. In any event, as expected, we see that fixing $\kappa=0$ everywhere removes the ambiguity in evaluating $I_1$ by fixing the value of $\tilde\alpha$ along the corresponding portion of the past null boundary.

\subsection{Counterterm for Null Boundaries} \label{duchess}

As we discussed above, various ambiguities arise in calculating the WDW action coming from contributions associated with the null boundaries \cite{RobLuis}. We followed a standard approach suggested in \cite{RobLuis} to fix the corresponding null normals, however, an alternate approach which was also suggested there was to add to the following counterterm action
\beq
I_\mt{ct} =  \frac{1}{8 \pi G_N} \int_{\mathcal{B}'} d \lambda\, d^{d-1}\theta\, \sqrt{\gamma} \ \Theta \log\left({\tL \Theta}\right) \, ,
\label{counter}
\eeq
where $\tL$ is an arbitrary (constant) length scale and $\Theta$ is the expansion scalar of the null boundary generators, \ie
\beq
\Theta = \partial_{\lambda} \log \sqrt{\gamma} \, .
\label{count3}
\eeq
The expansion $\Theta$ only depends on the intrinsic geometry of the null boundaries and so this additional surface term \reef{counter} is not required to ensure that the gravitational action \reef{THEEACTION} produces a well-defined variational principle. However, this counterterm was constructed to eliminate the dependence of the action on the parametrization of the null generators. Including this surface term does not effect certain key results for the CA proposal, \eg the complexity of formation \cite{Format} or the late-time rate of growth for an eternal black hole \cite{Growth}. On the other hand, it was found to modify the structure of the UV divergences in an interesting way \cite{Simon2} and it also modifies the details of the transient behaviour in the time evolution for an eternal black hole \cite{Growth}.\footnote{In particular, see appendices A and E of \cite{Growth}.} We note that these previous studies involved stationary spacetimes, and we will see below and in \cite{Vad2} that the inclusion of the counterterm is essential in dynamical spacetimes, such as the Vaidya geometries \reef{MetricV}, in order to reproduce some key properties of complexity.

We will explore the effect of the counterterm \reef{counter} in detail in the next section, but here we will extend some of the previous calculations to include the contributions of this surface term. In particular, let us consider including this term on the past null boundary $\mathcal{B}_\past$. In evaluating this contribution, the essential behaviour will be determined by the normalization function $H(r,v)$ appearing in the null normal \reef{VecVad}.
Hence, considering the limit $\veps\to0$,\footnote{The following results would remain unchanged if we first evaluate the counterterm contribution with a small but finite width and only take the limit $\veps\to0$ afterwards.} we have $H(r,v)=\alpha$ above the shell (\ie for $v>v_s$) and $H(r,v)=\tilde\alpha$ below the shell (\ie for $v<v_s$). Recall that $H(r,v)$ is only defined along the null boundary, and so in the following, it will be helpful to treat $H$ as a function of the radial coordinate (along $\mathcal{B}_\past$), \ie
\begin{equation}\label{Funcg}
H (r,v) = \alpha\ \mathcal{H} (r - r_s) + \tilde\alpha \, \left(1 - \mathcal{H} (r - r_s)  \right) \, ,
\end{equation}
where $\mathcal{H}$ stands for the Heaviside function. Further, the inner normalization constant $\tilde \alpha$ is determined by eq.~\reef{magic22}. Further, from eq.~\reef{VecVad}, we have $dr/d\lambda = H(r,v)$.
Hence we evaluate the null expansion \reef{count3} as
\begin{equation}
\Theta = \frac{H(r,v)}{r^{d-1}}\, \frac{d}{d r} \left( r^{d-1} \right) = \frac{(d-1) H(r,v)}{r} \, .
\end{equation}

Now the counterterm contribution \eqref{counter} becomes
\begin{equation}\label{count1}
 I_{\mt{ct}} = \frac{\Omega_{k, d-1} (d-1)}{8 \pi G_N} \int_{r_\mn}^{r_{\text{max}}} \!\!\! d r \, r^{d-2} \, \log{\left( \frac{ (d-1) \tL H(r,v)}{r} \right)} \, ,
\end{equation}
where we replaced $d \lambda = d r / H(r,v)$. The upper limit of the radial integral will be $r_\mx = L^2/\delta$, where $\delta$ is the short-distance cutoff in the boundary CFT. The lower limit  $r_\mn$ will depend on the details of the situation for which we are evaluating the holographic complexity. Using eq.~\reef{Funcg}, we may evaluate the integral in eq.~\reef{count1} to find
\beqa
 I_{\mt{ct}} &=& \frac{\Omega_{k, d-1}}{8 \pi G_N} \, r_\mx^{d-1} \left[ \log\!{\left( \frac{(d-1) \tL\alpha}{r_\mx} \right)} +\frac1{d-1}\right] \labell{count02}\\
 &&\qquad - \frac{\Omega_{k, d-1}}{8 \pi G_N} \, r_\mn^{d-1} \left[ \log\!{\left( \frac{(d-1) \tL\tilde\alpha}{r_\mn} \right)} +\frac1{d-1}\right]
+ \frac{\Omega_{k, d-1} }{8 \pi G_N}\, r_s^{d-1} \, \log\!\left( \frac{\tilde\alpha}{\alpha} \right)
\nonumber
\eeqa
and hence upon substituting for $\tilde\alpha$ using eq.~\reef{magic22}, we find
\beq
I_{\mt{ct}} =\qquad\cdots\qquad + \frac{\Omega_{k, d-1} }{8 \pi G_N}\, r_s^{d-1} \, \log\!\left( \frac{f_1(r_s)}{f_2(r_s)} \right)\,.
\labell{count2}
\eeq
In this expression, we have focused on the contribution that appears where the past boundary crosses the null shell (\ie $r=r_s$). Note that this term appears similar to the expressions appearing in eq.~\reef{VaKappa4} or \reef{VaJnt1} if we substituted $\tilde\alpha=\alpha$ in the latter. It will turn out that this particular surface contribution will play an essential role in determining the (proper) behaviour of the holographic complexity.

\section{Complexity in Black Hole Formation}\label{sec:evolve}

In this section, we study the case of a thin shell of null fluid collapsing in empty AdS to form a black hole. In these geometries describing a one-sided black hole, we evaluate the holographic complexity, using the Complexity=Action proposal in section \ref{CollapseCA}, and the Complexity=Volume proposal in section \ref{sec:VaidyaVolume}. From the perspective of the boundary CFT, this geometry describes a quantum quench, \eg see \cite{Das:2010yw,esp, ThermAl0,ThermaAlice,quench1,quench2,quench3}.  The CFT begins in the vacuum state and then, say, at $t=0$, we act with a (homogeneous) operator which injects energy into the system creating an excited state.

The bulk geometry is described by eq.~\eqref{MetricV} with the profile
\beq
f_{\text{p}}(v)=\omega^{d-2} \ {\cal H}(v)\,,
\label{heavy}
\eeq
where ${\cal H}(v)$ is the Heaviside step function.  This is a simplified version of the profile in eq.~\reef{heavyX} where we set $\omega_1=0$ and $v_s=0$, as well as $\omega_2=\omega$. Here we focus on dimensions $d\ge3$, and the special case of  BTZ black holes (\ie $d=2$) will be treated separately below. Hence the metric function $F$ becomes
\beqa
v<0\ :&& \qquad  F(r, v) = f_\mt{vac} (r) = \frac{r^2}{L^2} + k  \,,\labell{fVac}\\
v>0\ :&& \qquad  F(r, v) = f_\mt{BH} (r) =  \frac{r^2}{L^2} + k - \frac{\omega^{d-2}}{r^{d-2}} \, .\labell{fBH}
\eeqa
We consider these collapses for planar and spherical shells (and horizons), \ie $k=0$ and $k=+1$.\footnote{The case $k=-1$ with hyperbolic spatial sections is somewhat different since a time slice only covers half of a constant time surface in the global AdS boundary, \eg see \cite{Format}. The present discussion could be extended to cover this case if  shells of null fluid were injected symmetrically from both halves of a global boundary time slice.} As noted above, these AdS-Vaidya geometries can be interpreted as the holographic dual of the quantum quenches described above for the boundary CFT in the $d$-dimensional geometry:\footnote{As is conventional, the AdS curvature scale $L$ also appears here as the curvature scale of the boundary. However, a simple Weyl scaling in the boundary theory can be used to separate these two scales, \eg see \cite{Growth}.}
\begin{equation}
d s_{bdry}^{2} = - d t^2 + L^2\, d \Sigma_{k, d-1}^{2} \, .\label{bgeometry}
\end{equation}
Here we have simply defined the boundary time $t=v$ at $r\to\infty$. In the regime $t>0$ in the boundary CFT (\ie $v>0$), the energy is determined as usual by the black hole mass from $f_{\text{BH}} (r)$ in eq.~\reef{fBH}, \ie
\beq
M = \frac{(d-1) \, \Omega_{k,d-1}}{16 \pi \, G_N}\,\omega^{d-2} \,,
\label{energy}
\eeq
where $\Omega_{k,d-1}$ denotes the (dimensionless) volume of the spatial geometry (see footnote \ref{footy22}). In this part of the geometry, we determine the horizon radius with $f_{\text{BH}} (r=r_h)=0$ which corresponds to
\beq
\omega^{d-2}=r_h^{d-2}\left(\frac{r_h^2}{L^2}+k\right)\,.
\label{horiz}
\eeq
Then, using the usual gravitational expressions, we can assign an effective temperature and entropy to the corresponding excited state:
\beq
T
=\frac{1}{4\pi }\left.\frac{\partial f}{\partial r}\right|_{r=r_h}=\frac{1}{4\pi \,r_h}\left(d\,\frac{r_h^2}{L^2} + (d-2)\,k \right) \, ,\quad\qquad  S = \frac{\Omega_{k,d-1}}{4G_N}\,r_h^{d-1}    \, .
\label{effect}
\eeq

\begin{figure}
\centering
\includegraphics[scale=0.35]{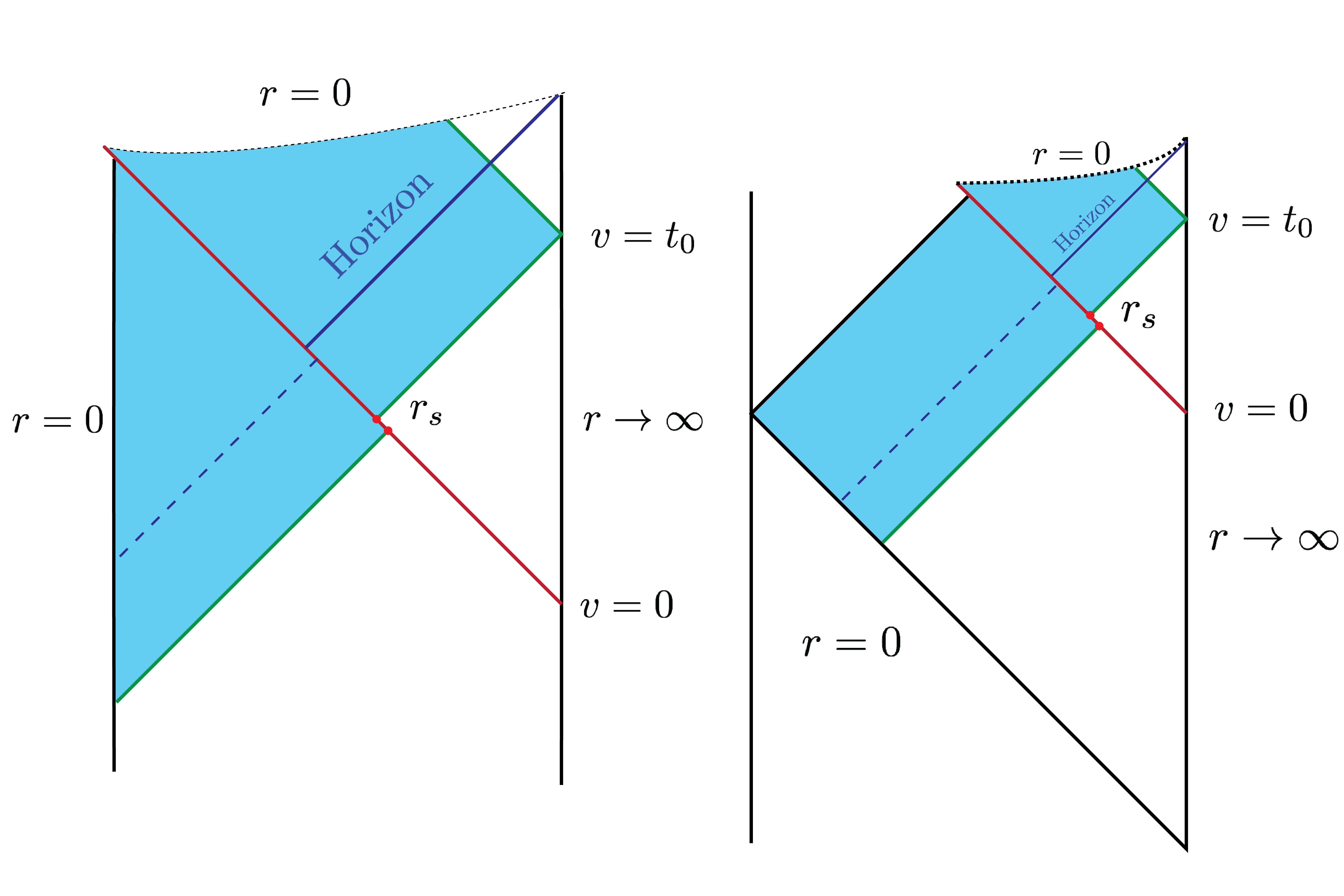}
\caption{Penrose-like diagrams for the thin shell collapsing geometries, we represent spherical horizon collapse from global AdS (left) and planar horizon from Poincar\'e patch (right). In order to not distort the diagrams, we represent the discontinuity in the outgoing coordinate $u$ by a jump while crossing the collapsing shell, \eg the dashed blue line indicates the extension of the event horizon into the region before the collapsing shell. We use $r_s$ to denote the radial position where the null boundary of the WDW patch crosses the shock wave. }
\label{CollapseOneSided}
\end{figure}

In the following, it will also be useful to construct the radial tortoise coordinates on each side of the shock wave as:
\beqa
v>0\ :&& \qquad  r^*_{\text{BH}}(r)= -\int^\infty_r \frac{d \tr}{f_{\text{BH}}(\tr)} \, ,\labell{tort}\\
v<0\ :&& \qquad  r^*_{\text{vac}}(r)= -\int^\infty_r \frac{d \tr}{f_{\text{vac}}(\tr)} =\left\lbrace
\begin{matrix}
-L^2/r\hfill&&{\rm for}&& k=0\ \\
L\left(\tan^{-1}\left(r/L\right)-\frac{\pi}2\right)&&
{\rm for}&& k=+1
\end{matrix}
\right.\,,\nonumber
\eeqa
where $f_{\text{BH}}(r)$ and $f_{\text{vac}}(r)$  are given in eqs.~\reef{fBH} and \reef{fVac}, respectively. Note that the sign is chosen in eq.~\reef{tort} to ensure that $dr^*=dr/f$  and the range of integration ensures that the tortoise coordinates vanish at infinity, \ie
\begin{equation}
\lim_{r \to \infty} r^*_{\text{vac,\,BH}}(r) \rightarrow 0  \, .
\label{bound3}
\end{equation}
Now we can define an outgoing null coordinate $u$ and an auxiliary time coordinate $t$ as
\beq
u \equiv v - 2 r^*(r) \, ,\qquad\quad
t \equiv v - r^*(r)  \, .
\label{time3}
\eeq
Notice that these coordinates are discontinuous across the shell because $f(r)$ changes from the vacuum to a black hole spacetime, as in eqs.~\eqref{fVac} and \reef{fBH}. Of course, $r$ and the ingoing coordinate $v$ are globally defined, but it is still useful to consider $t$ and $u$ if one properly matches these coordinates across the collapsing shell. In particular, we will represent the collapsing-shell geometries with Penrose diagrams, or rather `Penrose-like' diagrams, as shown in figure \ref{CollapseOneSided}.
These diagrams can be smoothly ruled with lines of constant $u$ and $v$. Since $u$ is discontinuous, this introduces a(n unphysical) jump as the outgoing null rays cross the shell. The spacetime is, of course, continuous along this surface and the outgoing null rays are smooth, as can be seen by regulating the thin shell to have a small but finite thickness, as was discussed in section \ref{sec:NullDust}. Further, these jumps in the outgoing null rays can be removed by deforming the Penrose diagrams to the future or the past of the shell, but the undeformed figures are simpler to construct and we found that they provide a useful intuitive picture of the geometry.

In order to translate the bulk results into boundary quantities, it is useful to work in terms of the dimensionless variables (following \cite{Growth}) defined as
\begin{equation}\label{DimlessCoord}
x \equiv \frac{r}{r_h} \, , \qquad \, z \equiv \frac{L}{r_h} \, .
\end{equation}
The temperature in eq.~\reef{effect} can be recast in terms of $z$ as
\begin{equation}\label{eq:DimlessTemp}
L\,T  = \frac{1}{4 \pi \, z} \, \left(d + k \, (d-2) \, z^2 \right) \, ,
\end{equation}
or alternatively, this expression can be inverted in order to express $z$ as a function of $LT$,
\begin{equation}\label{eq:zfuncT}
z = \frac{d}{\sqrt{4 \pi ^2 (L T)^2-(d-2) \, d \, k} + 2 \pi L T} \, .
\end{equation}
Note that for the planar case (\ie $k=0$), this expression simplifies
$z=d/(4\pi LT)$. Now  any result that depends on $z$ can be regarded as a boundary quantity defined in terms of $L\,T$ with eq.~\eqref{eq:zfuncT}.

Further, following the notation in \cite{Growth}, it is useful to define a dimensionless tortoise coordinate. We write
\begin{eqnarray}\labell{eq:DimlessTortoise}
x^{*}(x, z) &\equiv& \frac{r_h}{L^2} \, r^{*}(r) =-\int_x^\infty \frac{d x}{\tilde f(x, z)}   \, , \\
{\rm where}&& \quad \tilde f(x, z) = z^2 f(r, r_h)  \, , \nonumber
\end{eqnarray}
where $\tilde f(x,z)$ is the blackening factor written in terms of the dimensionless coordinates $x$ and $z$ and rescaled by $z^2$. That is, combining eqs.~\reef{fVac}, \reef{fBH}, \reef{horiz} and \reef{DimlessCoord}, we find
\beqa
v<0\ :&& \qquad  \tilde f_{\vac} (x,z) = x^2 + k \,z^2 \,,\labell{tf}\\
v>0\ :&& \qquad  \tilde f_{\BH} (x,z) =  x^2 + k \,z^2 - \frac{1+k\,z^2}{x^{d-2}} \, .\nonumber
\eeqa
We note that for the planar case (\ie $k=0$), $\tilde f(x, z)$ is actually independent of $z$.

\subsection{Complexity=Action}\label{CollapseCA}

The CA proposal \reef{defineCA} suggests that we can calculate the complexity of the CFT state on some time slice $\S$ in the boundary by evaluating the action of the dual gravitational configuration  on the corresponding WDW patch in the bulk. We have already introduced the null fluid and gravitational actions in section \ref{sec:NullDust} --- see eqs.~\eqref{vanish} and \reef{THEEACTION}. Further in section \ref{nfca}, we showed that in the limit of a thin collapsing shell of null fluid, the WDW action is given by the sum of the actions separately evaluated on the portion of the WDW patch outside of the shell and on the portion inside the shell. As we will see below, this greatly simplifies the calculation since the spacetime geometry is stationary in each of these regions.

Let us begin by examining in more detail the structure of the WDW patch, as shown in the Penrose-like diagrams of the collapsing geometries in figure \ref{CollapseOneSided}. We anchor the WDW patch to a constant time slice in the boundary, with some $t=t_0\ge 0$ --- recall that the collapsing shell starts at the asymptotic boundary at $t=0$. The future null boundary of the WDW patch is then defined by the surface $v=t_0$ --- see eqs.~\reef{bound3} and \reef{time3} --- and this boundary segment terminates at the curvature singularity at $r=0$. The past null boundary of the WDW patch is defined by $u=t_0=v-2r^*_\BH(r)$, where the outgoing null coordinate is defined in eq.~\reef{time3}. However, at this point, we must recall from eq.~\reef{tort} that the definition of the radial tortoise coordinate $r^*(r)$, and hence the null coordinate $u$, depends on whether we are to the future or the past of the collapsing shell. The null boundary of the WDW patch meets the collapsing shell at $r=r_s$ which is given by
\beq
2 r^*_\mt{BH}(r_s) + t_0 = 0\, ,
\label{meetingV1}
\eeq
where we are using the tortoise coordinate defined for $v\ge0$, and it will be useful in the following to note that
\beq
\frac{d \, r_s}{d \, t_0} = -  \frac{1}{2}\,f_\mt{BH}(r_s) \, .
\label{meetingV2}
\eeq
Now $v$ and $r$ are continuous as we cross the collapsing shell, but since the form of the tortoise coordinate changes here, there is a jump in $u$ (and in our Penrose diagrams). Hence to the past of the shell, the past boundary of the WDW patch is described by $u=-2r^*_\mt{vac}(r_s)=v-2r^*_\vac(r)$, which then reaches $r=0$ at $v=  2r^{*}_{\vac}(0)- 2 r^{*}_{\vac}(r_s) $. In this description of the WDW patch, we have overlooked various cut-off surfaces, \eg at the UV boundary or at the curvature singularity, but these details will be the same as in \cite{Format}.

In our Vaidya geometry with an infinitely thin shell, the surface $v=0$ naturally divides the WDW patch into two regions: 1) for $v<0$, the geometry is simply the AdS vacuum, and 2) for $v>0$, the geometry matches that of a static AdS black hole. In section \ref{nfca}, we smoothed out the geometry by giving the shell a (small) finite width and we were able to rigorously show that the action of the shell (within the WDW patch) vanishes when the width shrinks to zero. Hence in evaluating $I_\mt{WDW}$, we can simply calculate the action on regions 1 and 2 separately, and then simply add the results together.

\subsubsection{Action Calculation}

The evaluation of the action \eqref{THEEACTION} on the WDW patch was carefully analyzed in \cite{RobLuis}, and in calculating the various contributions below. The bulk integral can be written in the $r, v$ coordinates as
\begin{equation}
I_{\text{bulk}} = - \frac{d \, \Omega_{k,d-1}}{8 \pi L^2 \, G_{N}} \int_{\text{WDW}} r^{d-1} \, d r \  d v \ \, ,
\end{equation}
with the WDW patch as described above --- see also figure \ref{CollapseOneSided}. Integrating over $v$ first, yields
\begin{equation}
I_{\text{bulk}} = - \frac{d \, \Omega_{k, d-1}}{8 \pi L^2 \, G_{N}} \left[ \int_{0}^{r_s} r^{d-1} (2 r^*_{\text{vac}}(r_s)  -2 r^*_{\text{vac}}(r) +  t_0) dr -2 \int_{r_s}^{\infty} r^{d-1}  r^*_{\text{BH}}(r) d r \right] \, .
\end{equation}
Using eq.~\eqref{meetingV2} for $dr_s/dt_0$,  it is possible to show that the time derivative of this integral becomes
\begin{equation}\label{VaBulk2}
\frac{d I_{\text{bulk}}}{d t_0} = - \frac{\Omega_{k, d-1}}{8 \pi \,L^2 \, G_N} \, r_s^{d} \, \left( 1- \frac{f_{\text{BH}}(r_s)}{f_{\text{vac}}(r_s) } \right).
\end{equation}
We can write the above expression in terms of the black hole mass using eq.~\eqref{energy} and $d>2$, which then leads to
\begin{equation}\label{VaBulk}
\frac{d I_{\text{bulk}}}{d t_0} = - \frac{2 M}{(d-1)} \frac{x_s^2}{\left( k \, z^2 + x_s^2 \right)} \, ,
\end{equation}
where we have used the dimensionless coordinate $x_s \equiv r_{s}/r_h$.

We evaluate the GHY boundary term at the future singularity with the prescription discussed in \cite{Format}, but with total time lapse equal to $t_0$. Therefore,
\begin{eqnarray}
I_{\text{GHY}} &=&- \lim_{r \to 0} \frac{\Omega_{k, d-1} }{16 \pi G_{N}}\, r^{d-1} \left( \partial_{r} f_{\text{BH}}(r) + \frac{2 (d-1)}{r}  f_{\text{BH}}(r) \right) \,t_0  \,
\nonumber\\
\frac{d I_{\text{GHY}}}{d t_0} &=& \frac{d \,  \Omega_{k, d-1} }{16 \pi \,  G_{N}}\, \omega^{d-2} = \frac{d \, M}{(d-1)} \, . \labell{VaGHY}
\end{eqnarray}
As usual, we demand that the null boundaries are affinely parametrized, which yields $\kappa=0$. Hence the null surface terms do not contribute to the WDW action or its time derivative.

The only nonvanishing joint contributions to the time derivative of the holographic complexity arise where the past null boundary intersects with the collapsing shell, as indicated by the two big red dots in figure \ref{CollapseOneSided}. These joints are codimension-two surfaces at the intersection of two null hypersurfaces, and so we need to define the appropriate null normals. The null normal for the past boundary of the WDW patch was defined quite generally in eq.~\reef{formVad}. For the present geometry described by eqs.~\reef{fVac} and \reef{fBH}, this expression becomes
\beq
k_{\mu} d x^{\mu}  =\left\lbrace \begin{matrix}
&\alpha \big(  - d v + \frac{2}{f_{\BH}(r)} d r \big)&&
{\rm for}&& r>r_s\,,\\
&\tilde\alpha \big( - d v + \frac{2}{f_{\vac}(r)} d r \big)&&
{\rm for}&& r<r_s\,.
\end{matrix}\right.
\label{past}
\eeq
As we are dividing the WDW patch into two regions along the collapsing shell, we also need to introduce two (outward-directed) null normals which we denote
\beqa
v>0\ :&&\qquad k^2_{\mu}\, d x^{\mu}  =  - \beta d v \, ,
\nonumber\\
v<0\ :&&\qquad k^1_{\mu}\, d x^{\mu}  =  \ \  \beta d v \, ,
\labell{shellnX}
\eeqa
where $\beta$ is some arbitrary normalization constant.\footnote{We can compare these normals to those in eq.~\reef{shelln} for the edges of the finite-width shell.}
Combining the two joint contributions of interest then yields
\beqa
I_{\text{joint}} &=& \frac{\Omega_{k,d-1}}{8 \pi G_N}\,r_s^{d-1}\  \log\frac{2\beta\, \alpha}{f_{\BH}(r_s) }
-\frac{\Omega_{k,d-1}}{8 \pi G_N}\,r_s^{d-1}\  \log\frac{2\beta\, \tilde\alpha}{f_{\vac}(r_s) }
\nonumber\\
&=&
\frac{\Omega_{k, d-1} r_s^{d-1}}{8 \pi G_N} \,\log{\!\left[ \frac{\alpha\,f_{\vac}(r_s)}{\tilde\alpha\,f_{\BH}(r_s)}\right]}\, . \labell{VaJointAct}
\eeqa
However, at this point we recall that if we demand that $\kappa=0$ all along this past boundary,
then the normalization constant $\tilde\alpha$ must be fixed as in eq.~\reef{magic22}, which yields
\beq\label{locker}
\tilde\alpha=\alpha\,\frac{f_{\vac}(r_s)}{f_{\BH}(r_s)}\,,
\eeq
for the present situation. However, we easily see that substituting this result into eq.~\reef{VaJointAct} yields $I_{\text{joint}}=0$! Of course, this result might have been anticipated by realizing that the past null boundary is perfectly smooth and so without our division of the WDW patch into various regions the only way in which this boundary could contribute to $I_\mt{WDW}$ would be through the $\kappa$ surface term. However, if we demand that $\kappa=0$ everywhere along this boundary, then all of the contributions coming from this surface must vanish. Of course, since $I_{\text{joint}}$ vanishes, it will not contribute to the time derivative of the WDW action.

\subsubsection{Time Dependence of Complexity, Version 1}

Hence combining all of the terms in eq.~\eqref{THEEACTION}, we found that there are only two nonvanishing contributions to the time derivative of the WDW action. These come from the bulk integral in eq.~\eqref{VaBulk2}, and GHY surface term on the spacelike boundary at the future singularity in eq.~\eqref{VaGHY}. Combining these two expressions, we find (for $d>2$)
\begin{equation}\label{dCdt_noct}
\frac{d \mathcal{C}_{A}}{d t_0} = \frac{d-2}{d-1}\,\frac{M}{\pi} \left( 1+\frac{2}{d-2}\,\frac{k z^2 }{k z^2 + x_s^2} \right) \, .
\end{equation}
For $k=0$, this expression simplifies and the rate of growth of the complexity is simply a constant,
\begin{equation}\label{dCdt_noctk0}
\frac{d \mathcal{C}_{A}}{d t_0} \bigg{|}_{k = 0}  = \frac{d-2}{d-1} \ \frac{M}{\pi} \, .
\end{equation}
However, we observe that this growth rate is much lower than the late time limit found in an eternal black hole background \cite{Brown1,Brown2} \ie ${d \ca}/{d t_0}|_\mt{eternal}=2M/\pi$ as $t_0\to\infty$.

For $k=+1$, the rate acquires a time dependence through the coordinate $x_s$. At early times, $x_s$ is close to the boundary, \ie $x_s\to\infty$ as $t_0 \rightarrow 0$, and hence the rate of change in eq.~\reef{dCdt_noct} starts at the same value of the planar geometry \reef{dCdt_noctk0}. On the other hand, at very late times, the meeting point approaches the horizon, \ie $x_s\to1$ as $t_0 \rightarrow \infty$ and hence the growth rate approaches
\beq
 \frac{d \mathcal{C}_{A}}{d t_0} \bigg{|}_{\mt{late time}}  =
\frac{d-2}{d-1}\,\frac{M}\pi\left(1+\frac{2}{d-2}\,\frac{k z^2 }{k z^2 + 1} \right)\,.
\label{LateTimenoCT}
\eeq
Hence for spherical black holes (\ie $k=+1$), the late time limit yields a slightly larger growth rate that in the planar case. For very high temperatures, the increase is very small since in this regime the horizon radius is much larger than the AdS curvature scale and hence $z=L/\rh\ll1$. The correction is largest at the Hawking-Page transition, for which $z=1$ and we find
$d\ca/dt_0= M/\pi$ at late times. Hence the late time limit in eq.~\eqref{LateTimenoCT} is always smaller than the corresponding result \cite{Brown1,Brown2} for the eternal black hole geometry with any $d$ and for both $k=0$ and $+1$. This mismatch may seem somewhat surprising since at late times, the WDW patch in figure \ref{CollapseOneSided} is almost entirely in region 2, where the geometry matches that of a static black hole, as given in eq.~\reef{fBH}. Further, the above expressions suggest that the rate vanishes for $d=2$. Strictly speaking the previous calculations must be redone for the case of BTZ black holes, but the new calculations reproduce ${d \ca}/{d t_0}=0$ for $d=2$ --- see below.

We will see in a moment that adding the boundary counterterm \reef{counter} to the gravitational action restores the expected late time limit, however, we first examine the late and early time limits in more detail. In eq.~\eqref{dCdt_noct}, we have written the rate of complexity growth in terms of dimensionless boundary quantities. Hence, it is useful to write eq.~\eqref{meetingV1} as an equation determining $x_s$ as a function of the time (normalized by the temperature),
\begin{equation}\label{eq:DimlessxmEq}
2 \, x_\mt{BH}^{*}(x_{s}, z) + \frac{4 \pi\,T\,t_0 }{d + k \, z^2 \, (d-2)} =0 \, ,
\end{equation}
where $x_\mt{BH}^{*}$ is given by eq.~\reef{eq:DimlessTortoise} with $\tilde f_{\BH} (x,z)$ in eq.~\reef{tf}. Again, the dynamical variable in the problem is the (dimensionless) distance $x_s$, that ranges from infinity (\ie close to the asymptotic boundary) at early times, to one (\ie close to the event horizon) at late times.

\vspace{0.5em}
\noindent{\bf Early times:} We begin by examining the early time behaviour of the meeting point $x_s$, \ie immediately after the shell appears with $T\,t_0\ll1$. Again, we restrict the analysis to $d\ge3$ and consider $d=2$ separately below. From eq.~\eqref{eq:DimlessxmEq}, we can expand $x_s$ for early times to find\footnote{The corrections in eqs.~\reef{lack} and \reef{EarlyExp} are slightly different for $d =3$. In particular, we find $x_s (d=3)=\frac{3+k z^2}{2 \pi}\frac1{T t_0 }  -\frac{2 \pi  k z^2}{3 \left(3+k z^2\right)}\,Tt_0  +\frac{\pi ^2 \left(1+k z^2\right) }{\left(3+k z^2\right)^2}\left(Tt_0\right){}^2+ \mathcal{O}(t_0^3 T^3)$, and $\frac{d \mathcal{C}_{A}}{d t_0} \big|_{d=3}  =\frac{2 M}{\pi} \big(\frac{1}{2} + \frac{ 2 \pi ^3 \left(1+k z^2+1\right)}{\left(3+k z^2\right)^3} T^3t_0^3 + \mathcal{O}(t_0^6 T^6)\big)$. Note the additional ${\cal O}(T^2t_0^2)$ term in the first expression while the ${\cal O}(T^5t_0^5)$ correction vanishes in the second expression.}
\begin{equation}\label{lack}
x_s = \frac{d + (d-2) k z^2}{2 \pi} \frac{1}{T t_0} - \frac{2 \pi}{3} \frac{k z^2}{d + (d-2) k z^2} \, T t_0 + \mathcal{O} \left( T^3 t_0^3 \right) \, .
\end{equation}
Substituting the above expression into eq.~\reef{dCdt_noct} then yields
\begin{equation}
\frac{d \mathcal{C}_{A}}{d t_0} \bigg{|}_{\text{early time}}  = \frac{d-2}{d-1} \, \frac{M}{\pi}  + \frac{8\pi\,M}{d-1} \left( \frac{k z^2}{d+ (d-2)k z^2} \right) \, T^2 t_0^2 + \mathcal{O} \left( T^4 t_0^4 \right) \, .
 \label{EarlyTimenoCT}
\end{equation}
Hence to leading order, we recover the limit given by eq.~\reef{dCdt_noctk0} and above, we see that the rate begins to grow at order $(Tt_0)^2$.

\vspace{0.5em}

\noindent{\bf Late times:} To examine the late time behaviour, we follow the arguments in \cite{Growth}. Suppose that we rewrite the rescaled blackening factor by factoring out the root corresponding to the horizon. In this way, we find
\beq\label{eq:Factorf}
 \tilde f(x) = \tilde F(x) (x-1) \, , \qquad{\rm where}\ \ \ \
\tilde F(x=1) = d + k (d-2) z^2 \, .
\eeq
In the second expression, we have used eq.~\eqref{eq:DimlessTemp} to evaluate the function $\tilde F(x=1)$ at the horizon. At late times $x_s$ approaches $1$, and we can solve the meeting condition in eq.~\eqref{eq:DimlessxmEq} in this limit by using the decomposition
\begin{equation}
\frac{1}{\tilde f(x,z)} = \frac{1}{\tilde F(1) (x-1)} + \frac{\tilde F(1) - \tilde F(x) }{\tilde F(1) \tilde F(x) (x-1)} \, .
\end{equation}
Then we can write the tortoise coordinate as
\begin{equation}
x^{*}(x) = \frac{1}{\tilde F(1)} \, \log \frac{|x-1|}{\tilde \ell} + \int^{x}\!d\tilde x\, \frac{\tilde F(1) - \tilde F(\tilde x) }{\tilde F(1) \tilde F(\tilde x) (\tilde x-1)}  \, ,
\end{equation}
and $\tilde\ell$ is some integration constant. With this decomposition, we can solve eq.~\reef{eq:DimlessxmEq} for  late times
\begin{equation}\label{xsVeryLate}
x_s = 1 + c_1 \, e^{- 2 \pi T \, t_0} + \cdots  \, ,
\end{equation}
and the constant $c_1$ is given by
\begin{equation}
c_1 = \lim_ {x_{max} \rightarrow \infty} \left( x_{max} -1 \right) e^{\int_{1}^{x_{max}}  \, d \tilde x \, \frac{\tilde F(1) - \tilde F(\tilde x)}{\tilde x (\tilde x-1)}  }  \, ,
\end{equation}
which is a (finite) positive constant.

Substituting eq.~\eqref{xsVeryLate} into the growth rate \reef{dCdt_noct}, the late time limit becomes
\beq
 \frac{d \mathcal{C}_\mt{A}}{d t_0} \bigg{|}_{\mt{late time}}  = \frac{M \left(d-2+d k z^2\right)}{\pi  (d-1) \left(k z^2+1\right)} -\frac{ 4 c_1 \,M \, k z^2  }{\pi  (d-1) \left(k z^2+1\right)^2} \, e^{-2 \pi  t_0 T}  + \mathcal{O} \left(  e^{- 4 \pi T t_0 }\right) \, .
 \label{LateTimenoCTExp}
\eeq
The first term matches our previous expression \reef{LateTimenoCT} for the late time limit.
The second term shows that the limiting growth rate is approached from below, and that this behaviour corresponds to an exponential decay controlled by the thermal length scale, \ie $1/T$.

\vspace{0.5em}

In fact, given the expression in eq.~\reef{dCdt_noct}, it is not hard to show that the growth rate (for $k=+1$) begins at $t_0=0$ with value given in eq.~\reef{dCdt_noctk0} and then rises monotonically to reach the late time rate \reef{LateTimenoCT} in a time of order $t_0\sim 1/T$. Further,
it is straightforward to explicitly evaluate eq.~\reef{dCdt_noct} and plot $d\ca/dt_0$ as a function of time in various examples. Below in figures \ref{VadActBTZ} and \ref{VadActSphd34_Both}, we show the growth rates (both without and with the counterterm) for various temperatures with $d=2$,  and with $d=3 \, \text{and} \, 4$, respectively.

\subsubsection{Time Dependence of Complexity, Version 2}

Next, we wish to examine the effect of adding the counterterm \reef{counter} for null boundaries to the gravitational action. Recall that for an eternal black hole background, adding this counterterm did not affect the late-time rate of growth of the holographic complexity but it did changed the details of the transient behaviour in the time evolution  \cite{Growth}.

In principle, this term should be evaluated on both the future and past null boundaries of the WDW patch --- see figure \ref{CollapseOneSided}. However, the future boundary is entirely in region 2, where the geometry is identical to that of the eternal black hole. In particular, in this region, the time $t$ is a Killing coordinate and so the contribution of the counterterm on the future boundary is unchanged under time translations. That is, on this boundary, the counterterm does not contribute to the complexity growth rate.

Therefore we only evaluate the counterterm on the past null boundary $\mathcal{B}_\past$. This calculation was discussed in section \ref{duchess} and the required integral is given by eqs.~\reef{Funcg} and \reef{count1}. For the present case, the limits of integration are $r_\mx=L^2/\delta$ and $r_\mn=0$. Hence the result in eq.~\reef{count2} becomes
\beq
 I_{\mt{ct}} = \frac{\Omega_{k, d-1}}{8 \pi G_N} \, \frac{L^{2(d-1)}}{\delta^{d-1}} \left[ \log\!{\left( \frac{(d-1) \tL\delta\alpha}{L^2} \right)} +\frac1{d-1}\right] + \frac{\Omega_{k, d-1} }{8 \pi G_N}\, r_s^{d-1} \, \log\!\left( \frac{f_\vac(r_s)}{f_\BH(r_s)} \right)\,,
\label{count22}
\eeq
where implicitly we have assumed that $\kappa=0$ and so the normalization constant $\tilde\alpha$ is fixed by eq.~\reef{magic22}.
The first term above contributes to the UV divergences in the complexity \cite{diverg,Simon2} and is independent of $t_0$. Hence only the second term contributes to the growth rate through the variation of $r_s$, the radius where the past boundary meets the null shell. In particular, we recall from eq.~\eqref{meetingV2} that
\begin{equation}
\frac{d r_s}{d t_0} = - \frac{1}{2} f_{\text{BH}}(r_s) \, .
\label{meetingV2a}
\end{equation}
As a result, the time derivative of eq.~\reef{count22} becomes
\beqa
\frac{d  I_{\mt{ct}}}{d t_0} &=&  -  \frac{\Omega_{k, d-1} (d-1)}{16 \pi G_N} \, r_s^{d-2} \, f_{\text{BH}}(r_s)  \log\! \left(  \frac{ f_{\text{vac}} (r_s) }{f_{\text{BH}}(r_s) } \right)  \nonumber \\
&&\qquad\qquad-  \frac{\Omega_{k, d-1}}{16 \pi G_N} \, r_s^{d-1} \, f_{\text{BH}}(r_s)  \left[ \frac{ f_{\text{vac}} ' (r_s)}{ f_{\text{vac}} (r_s)} - \frac{ f_{\text{BH}} ' (r_s)}{ f_{\text{BH}} (r_s)}  \right].  \label{tDerCounterTerm}
\eeqa
Expressing this result in terms of the dimensionless quantities \reef{DimlessCoord} then yields
\beq
\frac{d  I_{\mt{ct}}}{d t_0}=\frac{d\,M}{d-1}\left(1-\frac{2 \, k \, z^2}{x_s^2+k \, z^2 }\right)  + \frac{M \,x_s^{d-2} \, \tilde f_{\BH}(x_s, z) }{\pi (1 + k \, z^2) }\, \log
\!{\left(\frac{\tilde f_{\BH}(x_s, z)}{\tilde f_{\vac}(x_s,z)}\right)}   \,,\label{VACT}
\eeq
using eq.~\reef{horiz} for the mass, and the expression for $\tilde f(x,z)$ in eq.~\reef{tf}.


Hence when the action \reef{THEEACTION} is supplemented by the counterterm \reef{counter}, the total time derivative of the holographic complexity is given by combining the expressions in eqs.~\eqref{VaBulk2}, \eqref{VaGHY} and \reef{tDerCounterTerm}. Alternatively, we can simply add eq.~\eqref{VACT} to the previous result in eq.~\reef{dCdt_noct}, which yields
\begin{equation}\label{VaidyaDerivativeMass}
\frac{d \mathcal{C}'_\mt{A}}{d t_0} = \frac{2 M}{\pi} + \frac{M\,x_s^{d-2}  \, \tilde f_{\BH}(x_s, z) }{\pi\, (1 + k \, z^2) }\, \log
\!{\left( \frac{\tilde f_{\BH}(x_s, z)}{\tilde f_{\vac}(x_s,z)}\right)}
\end{equation}
for $t_0\ge0$. The most striking feature of the new result is that at late times, the new rate approaches the expected limit, \ie ${d \ca'}/{d t_0}|_{t_0\to\infty}=2M/\pi$ \cite{Brown1,Brown2}. In particular, as $t_0\to\infty$, $r_s$ approach the horizon sending the blackening factor $f_\mt{BH}(r_s)$ to zero (\ie at late times, $x_s\to1$ and $ \tilde f_{\BH}(x_s\to1, z)\to0$) and hence the second term in the above expression vanishes.

Further we note that at $t_0=0$, $r_s$ begins at asymptotic infinity. As $t_0$ increases from zero, $r_s$ decreases monotonically --- see eq.~\reef{meetingV2a} --- and at late times, $r_s\to \rh$. Using the explicit form of the blackening factors in eqs.~\reef{fVac} and \reef{fBH}, it is also straightforward to show that the second term in eq.~\reef{VaidyaDerivativeMass} is always negative and that ${d^2 \ca'}/{d t_0^{\,2}}\ge 0$.\footnote{Recall that we are focusing on $k=0$ and +1 in this discussion.} Therefore ${d \ca'}/{d t_0}$ is monotonically increasing and approaches the late time limit from below. These features contrast with the corresponding results for the eternal black hole \cite{Growth}, and as previously noted in \cite{Moosa}, for the process of black hole formation, ${d \ca'}/{d t_0}$ respects the proposed bound on the rate of complexity growth suggested in \cite{Brown1,Brown2}, \ie ${d \ca'}/{d t_0}\le 2M/\pi$.

We observe that for $k=0$, eq.~\reef{VaidyaDerivativeMass} simplifies somewhat yielding
\begin{equation}\label{VDMass2}
\frac{d \ca'}{d t_0} = \frac{2 M}{\pi} - \frac{M  }{\pi }\left(x_s^d-1\right)\, \log\!{\left[ \frac{x_s^d}{x_s^d-1}\right]}
 \, ,
\end{equation}
where $x_s$ is given by
\beq
\left(\frac{1}{x_s^d-1}\right)^{1/d} \, _2F_1\left(\frac{1}{d},\frac{1}{d};1+\frac{1}{d}; -\frac{1}{x_s^d - 1}\right)=\frac{2\pi\,T\,t_0}{d}\,.
\label{potter}
\eeq

Next let us apply the previous analysis for early and late times  to evaluate the behaviour of the complexity evaluated with the modified action.
In both cases, we focus on $d\ge3$ and consider the special case  $d=2$ in detail afterwards.

\noindent{\bf Early times:} Here, we apply eq.~\reef{lack} to evaluate the complexity growth rate in eq.~\eqref{VaidyaDerivativeMass} for $T\,t_0\ll1$,
\begin{equation}\label{EarlyExp}
\frac{d \mathcal{C}'_\mt{A}}{d t_0} \bigg|_{\text{early time}}  =  \frac{2 M }{\pi} \left(  \frac{1}{2} + \frac{(2 \pi)^d}{4} \frac{1+ k z^2}{(d + (d-2) k z^2)} \, T^d t_0^d + \mathcal{O} \left( T^{d+2} t_0^{d+2} \right) \right).
\end{equation}
Therefore, we see that for $d \geq 3$, the early time behavior is given by $M/\pi$, for both spherical and planar black holes, \ie
\begin{equation}
\frac{d \ca'}{d t_0 } \bigg|_{t_0 \to 0^{+}} = \frac{M}{\pi} \, . \label{friend}
\end{equation}
That is, the rate of growth of the holographic complexity begins at precisely one-half the late time limit. Recall that in \cite{Growth}, it was found that for the eternal black hole, $d \ca/d t_0$ remained zero up to a critical time, at which point it became negatively divergent. The rate then quickly rose to positive values but this transient behaviour depended on the choice of the normalization constant $\alpha$. In the bulk, this transition corresponds to the moment  when the past boundary of the WDW patch lifts off from the white hole singularity and the past null boundaries begin to meet at a joint above the past singularity.

\vspace{0.5em}

\noindent{\bf Late time expansion:} Next we apply eq.~\reef{xsVeryLate} to evaluate the late time expansion of the growth rate in eq.~\reef{VaidyaDerivativeMass},
\begin{equation}
\frac{d \mathcal{C}'_\mt{A}}{d t_0} \bigg|_{\text{late time}} = \frac{2 M}{\pi}  - 2M \frac{d + (d-2) \, k  \, z^2}{(1+k z^2)}  \,  c_{1} e^{- 2 \pi \, T  t_0} \,  T  t_0+
  \cdots \, .
\label{Late}
\end{equation}
As argued above, we  see that the late time limit is approached from below. Further, this behaviour is an exponential decay controlled by the thermal length scale, \ie $1/(2 \pi T)$.
A similar exponential decay is found in the eternal black hole geometry but there the late time limit is approached from above \cite{Growth}.

\vspace{0.5em}

\noindent{\bf Examples:}

We turn our attention to numerically evaluating  eq.~\eqref{VaidyaDerivativeMass} in $d=3$ and $d=4$ with $k=+1$, as well as investigating the special case of $d=2$ where the collapse forms a BTZ black hole. We start with the latter, for which the coordinate $x_s$ can be determined analytically as a function of time.

\vspace{0.5em}
\noindent\underline{$d=2$}:
For $d=2$, the collapsing shell produces a BTZ black hole with \cite{Banados:1992wn,Banados:1992gq}
\begin{equation}\label{blackTZ}
f_{\BH}(r) = (r^2 - r_h^2)/L^2.
\end{equation}
Hence the corresponding dimensionless blackening factor \reef{tf} simplifies to
$\tilde f_{\BH}(x) = x^2-1$ for $v>0$.
The physical parameters describing the BTZ  geometry are
\begin{equation}\label{BTZquantities}
M = \frac{\Omega_{k, 1}\, r_h^2}{16 \pi G_N L^2 }\, ,  \qquad T = \frac{r_h}{2 \pi L^2} \,,\qquad S=\frac{\Omega_{k, 1} r_h}{4 G_N} = \frac{\pi}{6}\,c\, \Omega_{k, 1} L T\, ,
\end{equation}
where $c=3L/(2G_N)$ is the central charge of the boundary CFT. The choices $k=0$ and 1 correspond to the Ramond and Neveu-Schwarz vacuum, respectively, of the boundary theory \cite{couscous}. While in principle, the results for the Ramond vacuum are already described by eqs.~\reef{VDMass2} and \reef{potter} above, we consider both possibilities in the following.\footnote{Recall that the ground state energy vanishes for the Ramond vacuum, but for the Neveu-Schwarz vacuum, it is negative: $E_\mt{R,0}=-  1/(8 \pi  G_N)=-c/(12\pi L)$.} Eq.~\reef{eq:DimlessxmEq} simplifies with $d=2$, and we can solve for $x_s$ analytically,
\begin{equation}\label{xsBTZsimple}
x_s = \coth (\pi T t_0) \, .
\end{equation}

First, we analyze the rate of change of complexity for BTZ black holes without the inclusion of the counterterm. The rate of change is then given by summing eqs.~\eqref{VaBulk2} and \eqref{VaGHY},
\begin{equation}\label{notctd2}
\frac{d \mathcal{C}_{A}}{d t_0}  = - \frac{M}{\pi} \, \frac{2 k z^2 (x_s^2-1)}{x_s^2+k z^2} \, .
\end{equation}
There are differences in the rate of change of BTZ in comparison to the higher dimensional cases ($d>2$) in eq.~\eqref{dCdt_noct}. First, for a collapse of the Ramond vacuum ($k=0$), the rate of change is exactly zero!

\begin{figure}
\centering
\includegraphics[scale=0.6]{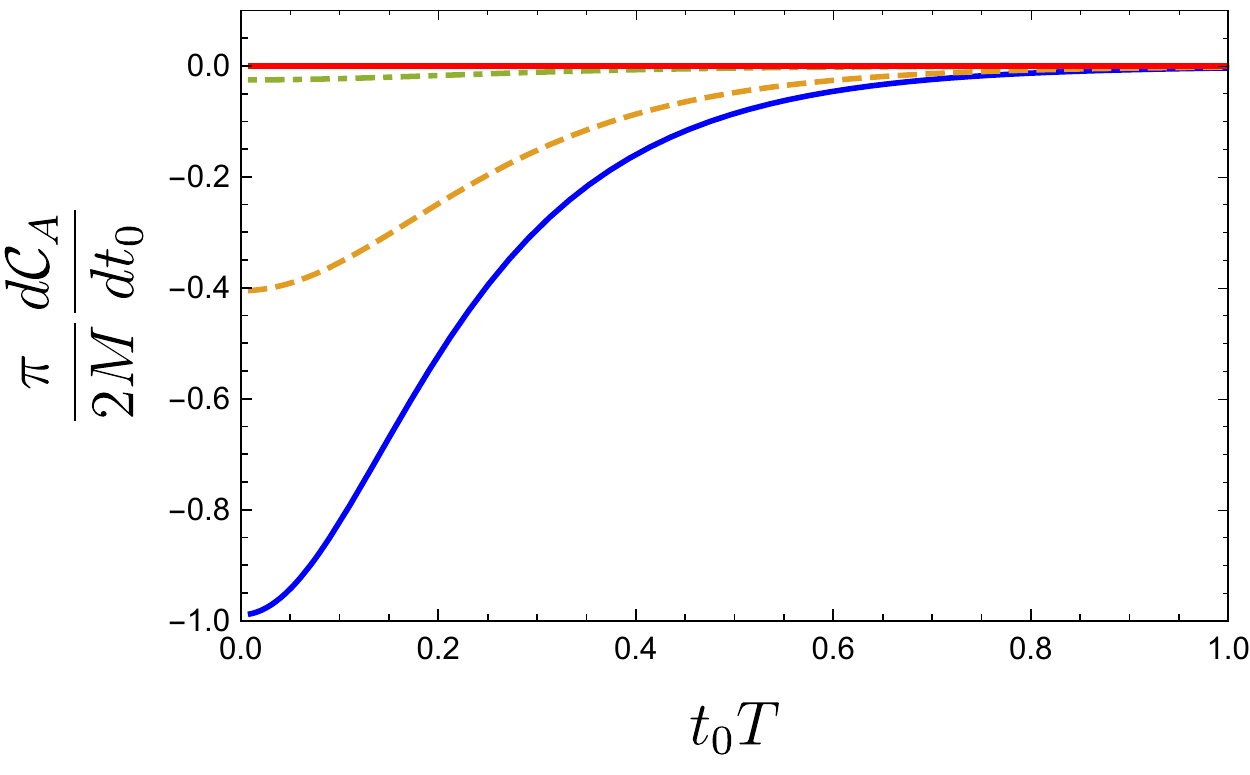}
\includegraphics[scale=0.6]{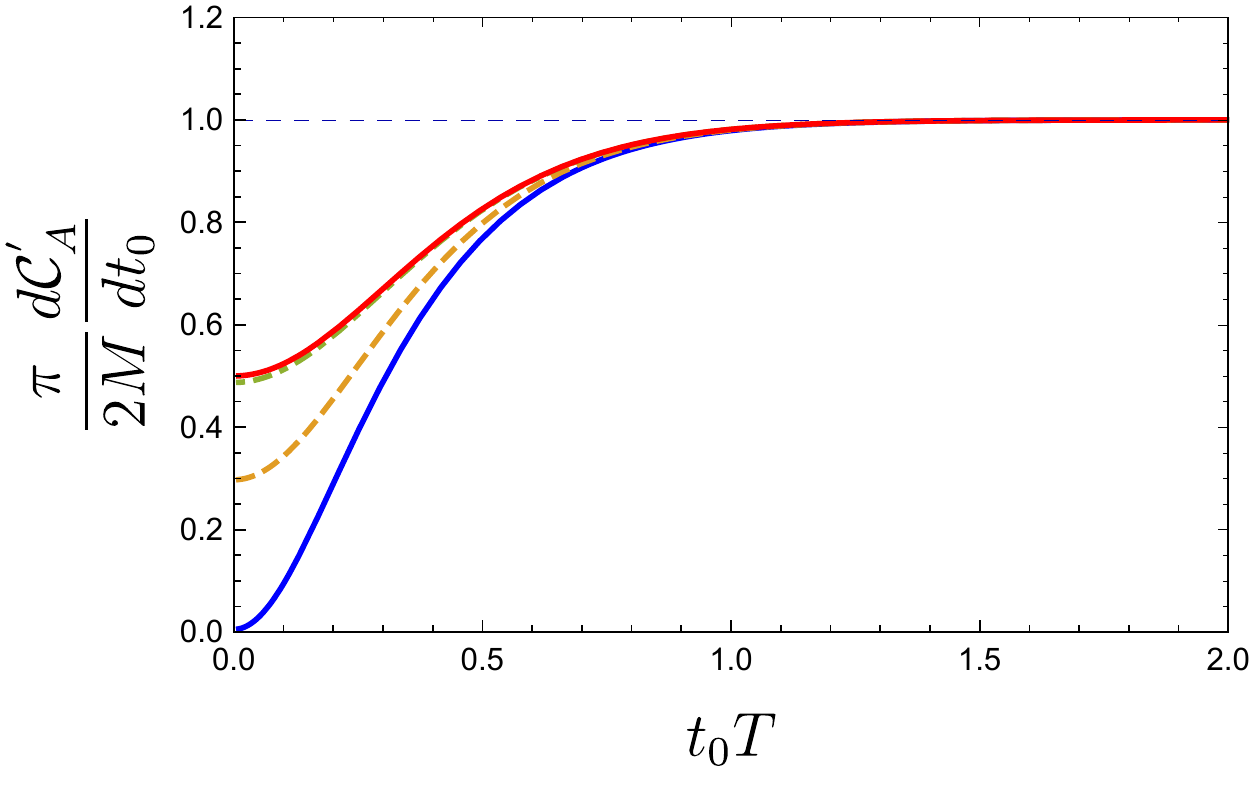}
\caption{The growth rate for the complexity, evaluated without (left) and with (right) the boundary counterterm in $d=2$. In both plots we have the collapse from Neveu-Schwarz vacuum (\ie $k=+1$) with temperatures $LT = 0.16$ (blue, solid), $LT = 0.25 $ (orange dashed) and $LT = 1.0$ (green dot-dashed). The collapse from Ramond vacuum (\ie $k=0$) is shown in red. For the NS vacuum, the growth rate always starts at different values for different temperatures, as given by eq.~\eqref{earlyRateBTZ_NoCT} (left) and eq.~\eqref{earlyRateBTZ} (right). In both cases, the high temperature limit of the NS collapse approaches the Ramond collapse. At late times, independent of the temperature, the rate of change approaches zero on the left, and $2M/\pi$ on the right. }
\label{VadActBTZ}
\end{figure}

Further, for the collapse from the Neveu-Schwarz vacuum ($k=+1$), the rate of change begins with negative values,
\begin{equation}\label{earlyRateBTZ_NoCT}
\frac{d \mathcal{C}_{A}}{d t_0} \bigg{|}_{\text{early time}} =  - \frac{2 M k z^2}{\pi} + 2 M  k z^2 \pi (1+ k z^2) T^2 t_0^2 + \mathcal{O} \left(T^4 t_0^4 \right) \, .
\end{equation}
In fact, the time derivative never becomes positive and instead approaches the late time limit (\ie $0$) from below,
\begin{equation}
\frac{d \mathcal{C}_{A}}{d t_0} \bigg{|}_{\text{late time}} = - \frac{8 k z^2 M}{\pi (1+ k z^2)} \, e^{-2 \pi T t_0} + \mathcal{O} \left(e^{-4 \pi T t_0} \right) \, .
\end{equation}
We show the full profile of the rate of change of complexity for various temperatures in the left panel of figure \ref{VadActBTZ}.

Next, we evaluate the rate of change of complexity including the contribution of the boundary counterterm. Continuing with either $k=0$ or 1, we have $ \tilde f_\vac(x) = x^2 +k z^2$ from eq.~\reef{tf}. The time derivative of complexity then reads
\begin{equation}\label{RateBTZOneSided}
\frac{d \mathcal{C}'_\mt{A}}{d t_0} = \frac{2 M}{\pi} - \frac{M }{\pi}\, (x_s^2 - 1)\,\log\! \left( \frac{x_s^2 + kz^2}{x_s^2 - 1} \right) \, .
\end{equation}
Using eq.~\reef{xsBTZsimple} for the early time limit (in which case,  $x_s\to\infty$), eq.~\eqref{RateBTZOneSided} yields
\begin{equation}\label{earlyRateBTZ}
\frac{d \mathcal{C}'_\mt{A}}{d t_0} \bigg|_{t_0 \rightarrow 0^+} = \frac{M}{\pi} \left( 1 - \frac{k}{4\pi^2\,L^2\,T^2} \right) \, ,
\end{equation}
where we substituted $z=1/(2\pi LT)$, from eq.~\reef{eq:DimlessTemp} with $d=2$. Recall that for higher dimensional black holes (\ie with $d \geq 3$), this limit was always $M/\pi$, as shown in eq.~\reef{friend}. The above result matches this previous limit for the Ramond vacuum (with $k=0$), but for
the Neveu-Schwarz vacuum (with $k=1$), the initial rate is reduced by a factor depending on the temperature. Notice that the correction factor (\ie the factor in brackets) in eq.~\reef{earlyRateBTZ} is positive above the Hawking-Page transition (\ie for $2\pi LT>1$), and it vanishes at precisely $2\pi LT=1$.

In the late time limit, combining eqs.~\eqref{xsBTZsimple} and \eqref{RateBTZOneSided} yields
\begin{equation}\label{BTZLateTimeOneSided}
\frac{d \mathcal{C}'_\mt{A}}{d t_0}  = \frac{2 M}{\pi} \left( 1 - 4 \pi T t_{0}\, e^{- 2 \pi T t_0}   +\cdots \right) \, .
\end{equation}
Hence the growth rate approaches its late time value from below in more or less the same way as in eq.~\reef{Late} for higher dimensions.

We show the full time evolution of $d\ca'/dt_0$  for a range of temperatures beginning with the Neveu-Schwarz vacuum (\ie $k=+1$) in figure \ref{VadActBTZ}. For small temperatures, it starts at a different rate from the higher dimensional examples, as shown in eq.~\eqref{earlyRateBTZ}, but the rate starts at approximately $M/\pi$ for higher temperatures. In addition, the rate of growth increases monotonically from the initial rate and the late time limit is approached from below as well. Further, the $d\ca'/dt_0$ essentially reaches $2M/\pi$ at a time $t_0\sim 1/T$

\vspace{0.5em}
\noindent\underline{$d=3$}:
Next, we turn our attention to evaluating numerically the growth rate of complexity with a spherical collapsing shell in $d=3$. The dimensionless tortoise coordinate given by eq.~\eqref{eq:DimlessTortoise} reads
\begin{equation}
x^{*}_{BH} (x,z)= \frac{\sqrt{4 k z^2+3} \left(2 \log \left(\frac{|x-1|}{\sqrt{k z^2+x^2+x+1}}\right)\right)+\left(4 k z^2+6\right) \,\tan ^{-1}\!\left(\frac{2 x+1}{\sqrt{4 k z^2+3}}\right)}{2 \left(k z^2+3\right) \sqrt{4 k z^2+3}} \, .
\end{equation}
We can then solve numerically the transcendental equation \eqref{eq:DimlessxmEq} for $x_s$, and evaluate eq.~\eqref{VaidyaDerivativeMass}.

\begin{figure}
\centering
\includegraphics[scale=0.6]{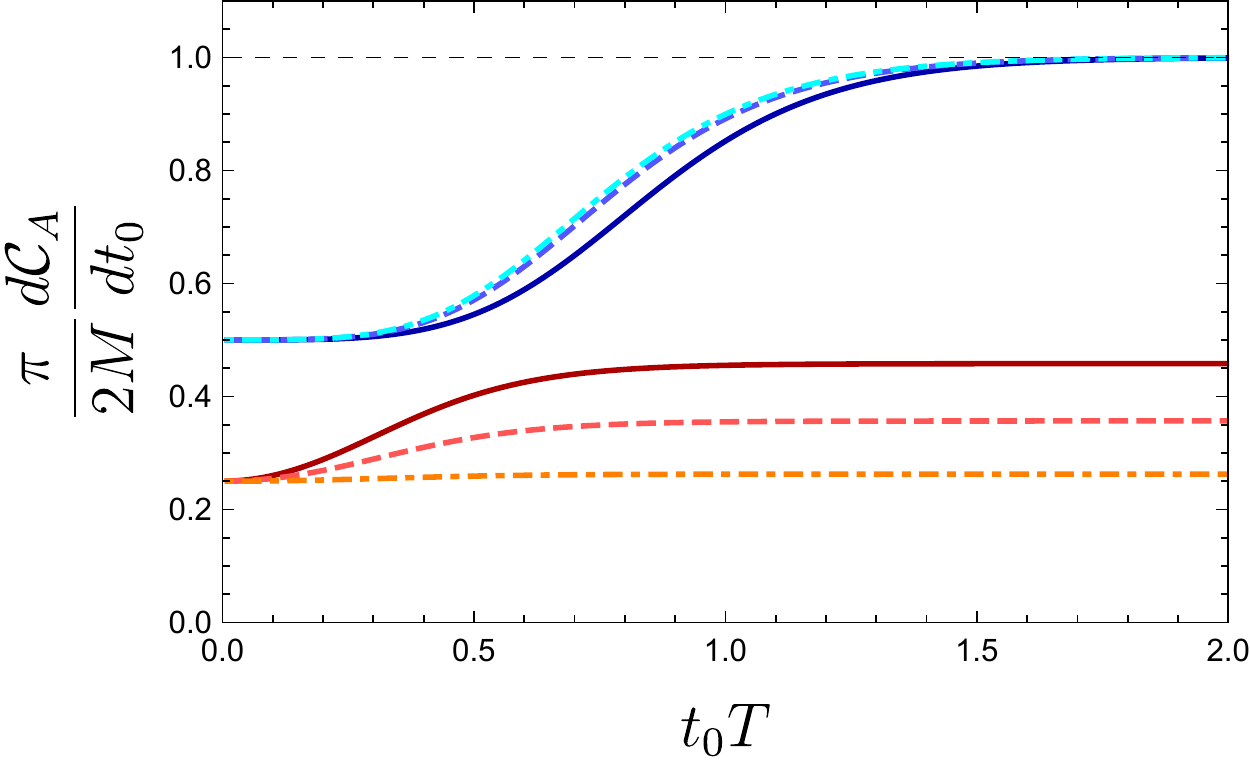}
\includegraphics[scale=0.6]{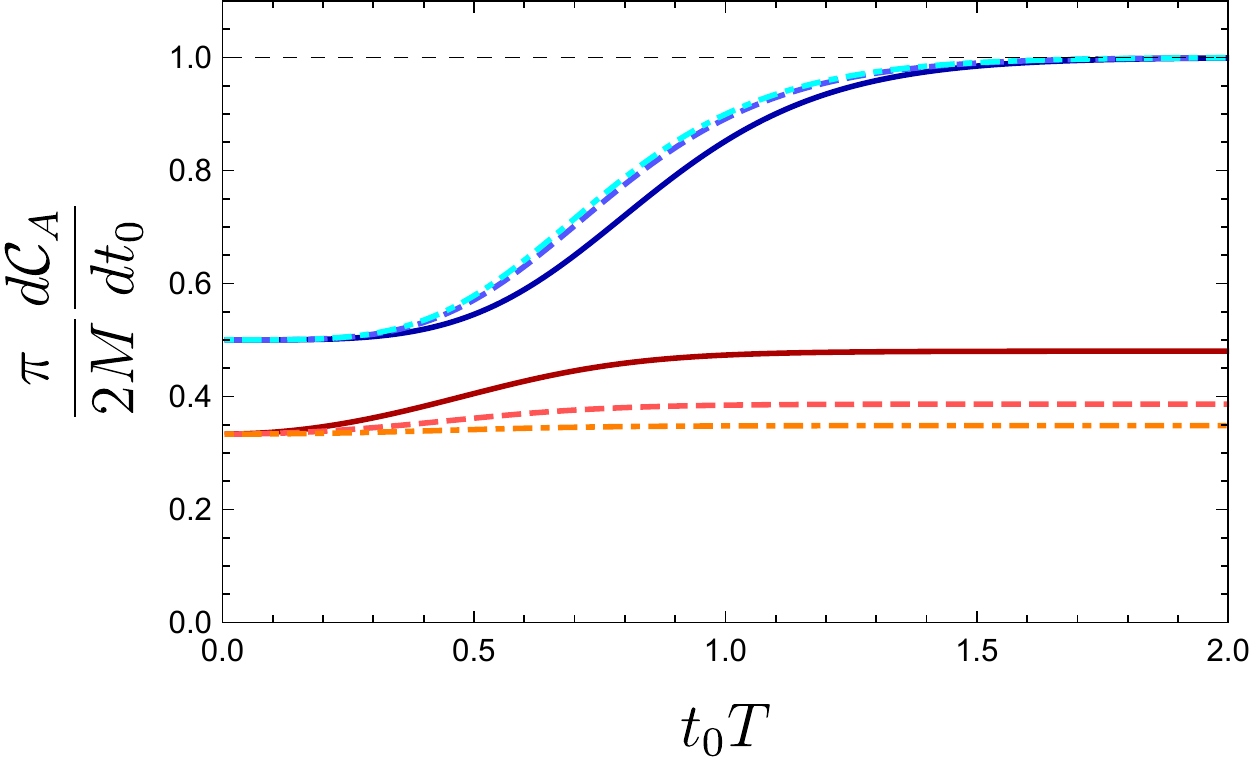}
\caption{The growth rate for the complexity in $d=3$ (left) and $d=4$ (right) and spherical geometry ($k=+1$), evaluated without (red and orange curves) and with (blue and cyan curves) the boundary counterterm \reef{counter}. In both case, we evaluate the growth rate for temperatures $TL  = 0.35$ (solid), $TL  = 0.5 $ (dashed) and $TL  = 2.0$ (dot-dashed) in the left and  $T L = 0.5$ (solid), $T L = 0.8 $ (dashed) and $T L = 1.5$ (dot-dashed) in the right figure. In both dimensions, $d\ca/dt_0$ (without the counterterm) starts at the value of the planar rate of change given by eq.~\eqref{EarlyTimenoCT} and approaches the late time limit from below in eq.~\eqref{LateTimenoCT}. The late time growth rate in this case is smaller than the one for the eternal black hole, and it depends on the temperature. With the inclusion of the counterterm, $d\ca'/dt_0$ starts at half of its late time limit, then it grows at times of the order of the thermal length, and approaches the eternal black hole bound from below.}
\label{VadActSphd34_Both}
\end{figure}

We show the time dependence of both $d\ca/dt_0$ and $d\ca'/dt_0$ for the spherical boundary geometry (\ie $k=+1$) in the left panel of figure \ref{VadActSphd34_Both} for several temperatures. Recall that $z$ is determined in terms of $L\,T$ by eq.~\reef{eq:zfuncT}.  As discussed above, $d\ca'/dt_0$ (with the counterterm) approaches ${2M}/{\pi}$ from below at late times   and starts with ${M}/{\pi}$ immediately after the shell is injected from the boundary. For $d\ca/dt_0$ (without the counterterm), the late time limit is much lower (\ie it does not match that found with eternal black holes) and depends on the value of the temperature, as in eq.~\eqref{LateTimenoCT}.

\vspace{0.5em}
\noindent\underline{$d=4$}:
For $d=4$, the relevant dimensionless tortoise coordinate in eq.~\eqref{eq:DimlessTortoise} reads
\begin{equation}
x^{*}_{BH}(x,z) = -\frac{1}{2 k z^2+4} \left[ \sqrt{k z^2+1} \left(\pi -2 \tan ^{-1}\left(\frac{x}{\sqrt{k z^2+1}}\right)\right)-\log\! \left(\frac{x-1}{x+1}\right) \right] \, .
\end{equation}
Therefore, we can solve numerically eq.~\eqref{eq:DimlessxmEq} for the meeting point $x_s$, which then allows us to evaluate the complexity growth rates with and without the inclusion of the boundary counterterm \reef{counter} in eqs.~\eqref{VaidyaDerivativeMass} and \eqref{dCdt_noct}, respectively. Recall that $z$ is determined in terms of $L\,T$ by eq.~\reef{eq:zfuncT}. We show $d\ca/dt_0$ and $d\ca'/dt_0$ for several temperatures (and the spherical geometry with $k=+1$) in the right panel of figure \ref{VadActSphd34_Both}. Again, as discussed above, we see that when the counterterm is included, $d\ca'/dt_0$ starts with ${M}/{\pi}$ at $t_0=0$ and rises monotonically to $2M/\pi$ at late times. Without the counterterm, the late time growth rate does not match the eternal black hole geometry, and it depends on the temperature, as given by eq.~\eqref{LateTimenoCT}.

\subsection{Complexity=Volume} \label{sec:VaidyaVolume}

\begin{figure}
\centering
\includegraphics[scale=1.3]{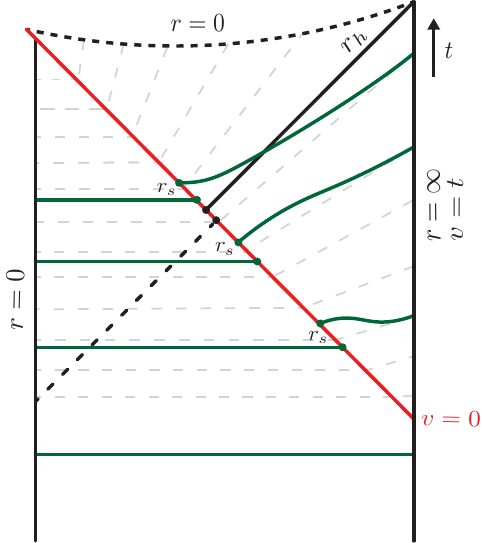}
\caption{Penrose-like diagram of maximal volume surfaces at different times embedded in the Vaidya AdS spacetime.
Constant time slices are indicated by thin dashed gray lines and the maximal volume surfaces asymptote them near the boundary. The event horizon extends past the shell, as we have indicated by a thick dashed gray line. Since the momentum \reef{ener22} of the surfaces is positive, they evolve towards decreasing time outside the horizon. Surfaces lie on constant time slices in the vacuum part of spacetime to avoid a conical singularity at $r=0$.}
\label{fig:VaidyaVolume}
\end{figure}

In this section, we evaluate the  holographic complexity following  the CV conjecture \reef{defineCV} for the same Vaidya spacetime describing the formation of a black hole with the collapse of a(n infinitely) thin shell of null fluid. Our calculations closely follow those in the CV section of \cite{Growth}. The maximal volume surfaces take the form illustrated in figure \ref{fig:VaidyaVolume}.

We are again working with the Vaidya metric in eq.~\eqref{MetricV} with $f_P(v) = \omega^{d-2}\,{\cal{H}}(v)$, as in eq.~\reef{heavy}.
To find the maximal volume slices anchored to the boundary time slice at $v=t_0$, we must extremize the following
\begin{equation}
\mathcal{V} = \Omega_{k,d-1} \int \mathcal{L} \, d\lambda = \Omega_{k,d-1} \int d \lambda\,r^{d-1} \sqrt{-F \dot{v}^2 +2 \dot{v}  \dot{r}}\, ,
\label{volu99}
\end{equation}
where we have taken advantage of the ``rotational'' symmetry to integrate out spatial boundary directions. The remaining radial direction on the (codimension-one) bulk surfaces is parameterized by $\lambda$ above and the surface is defined by its trajectory in the $rv$-plane, ($r(\lambda),v(\lambda))$.

\begin{figure}
\centering
\includegraphics[scale=0.6]{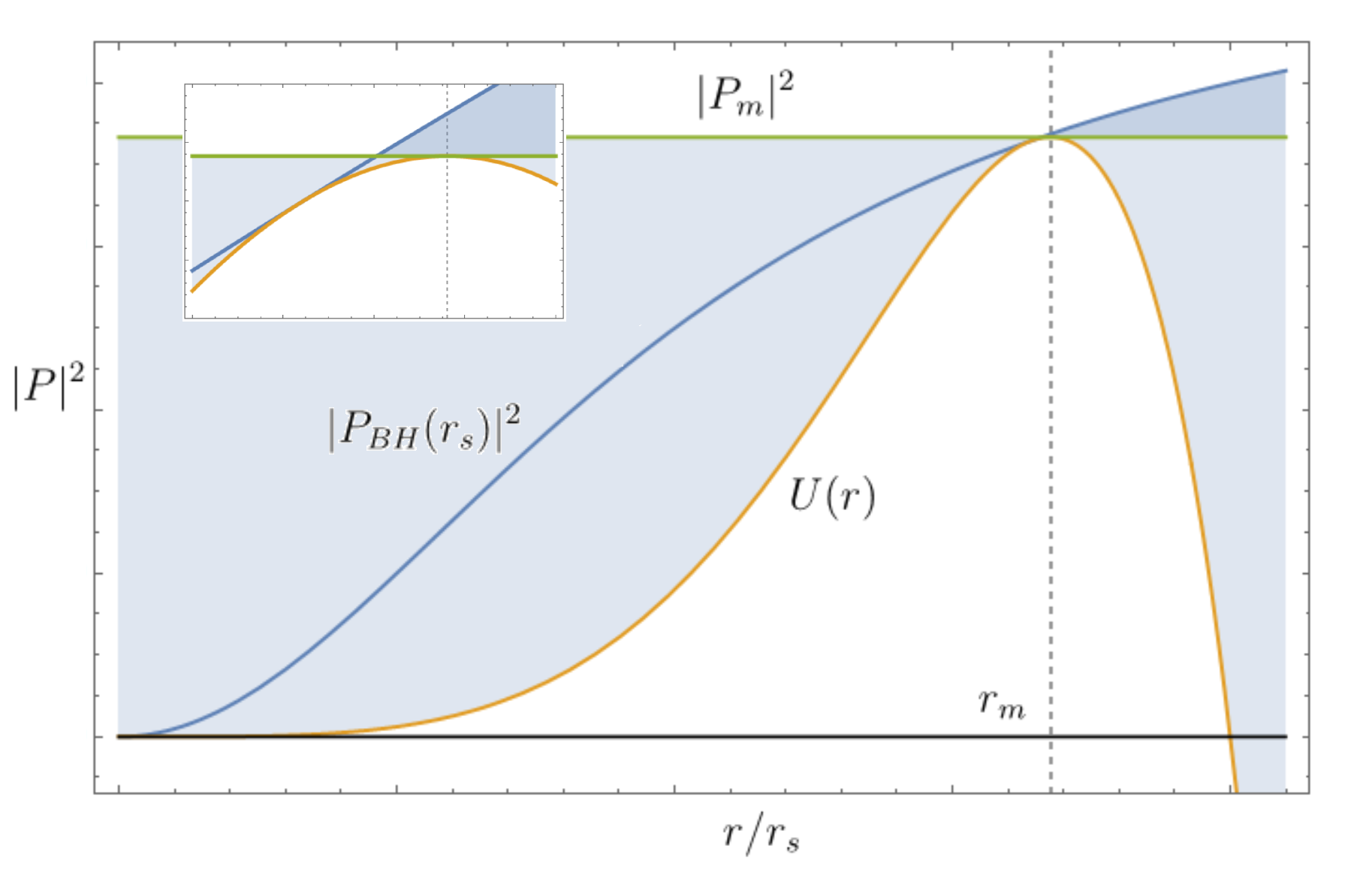}
\caption{Generic form of the potential $U(r)=-f_{\BH}(r)r^{2(d-1)}$ (yellow curve) as a function of $r$ for black holes with $k=1$ and $d\ge3$, or for BTZ black holes in $d=2$. The peak of the potential corresponds to $r_m$ (dashed gray line) and the corresponding energy $P^2_m$ (green line) is defined in eq.~\eqref{maximal_surf}. The blue curve corresponds to the energy in the black hole side as a function of $r_s$ where the shell is crossed, for $k=1$ and $d\geq 3$ or for BTZ black holes in $d=2$ with the Neveu-Schwarz vacuum. The point where the yellow and blue curves meet indicates a change in the direction of the velocity $\dot r_{BH}(r_s)$. To reach the asymptotic boundary we require $P_{BH}^2 \geq P_m^2$. That is, $r_s$ should be larger than the value at the intersection of the blue and green curves --- see inset.}
\label{fig:Potential}
\end{figure}

Our metric is independent of the coordinate $v$ in each part of the spacetime, \ie $v>0$ and $v<0$, separately. Hence, in each of these regions, we have the conserved ``momentum,''
\begin{equation}
P = \frac{\partial \mathcal{L}}{\partial{\dot v}} = \frac{r^{d-1}( \dot r-F \dot v)}{\sqrt{-F \dot{v}^2 +2 \dot{v} \dot{r}}}\,.
\label{ener22}
\end{equation}
Now the expression in eq.~\reef{volu99} is invariant under reparametrizations of $\lambda$ and we make the following convenient gauge choice:
\begin{equation}\label{eq:modiprop}
\sqrt{-F \dot{v}^2 +2 \dot{v} \dot{r}}=r^{d-1}\,.
\end{equation}
We can use this condition to simplify the $v$-momentum \reef{ener22} as follows
\begin{equation}\label{eq:consEVaid}
P = \dot r-F \dot v \,.
\end{equation}
We can then use eqs.~\eqref{eq:modiprop} and \eqref{eq:consEVaid} to express $\dot r$ and $\dot v$ in terms of $r$ and $\E$
\begin{equation}\label{eq:onedot}
\begin{split}
\dot  r  = &\, \pm {\sqrt{F(r)	 r^{2(d-1)} + \E^2}}\,,
\\
\dot v  = &\, \frac{\dot r-\E}{F(r) } =\frac{1}{F(r) } \left( -\E  \pm \sqrt{F(r)	 r^{2(d-1)} + \E^2} \right)\,,
\end{split}
\end{equation}
where in principle, either sign may play a role since $r$ may be increasing or decreasing as we move along the surface.
However, we will see that $\dot r$ (as well as  $\dot v$) will be positive in general for the solutions of interest, and $\E$ will be positive. Since $\E$ is not conserved in the full spacetime (due to the ${\cal{H}}(v)$ in the profile \eqref{heavy}), it is convenient to have the full equations of motion:
\begin{equation}
\begin{split}\label{dotsVad}
\ddot v =&\,(d-1) r^{2d-3}-\frac{\dot v ^2 }{2 }\, \del_r F\,,
\\
\ddot r =&\,
\frac{\dot v^2 }{2 }\, \del_v F
+\frac12\,\del_r\!\left(r^{2d-2} F\right)\,,
\end{split}
\end{equation}
where we simplified these expressions using eq.~\reef{eq:modiprop}. Here we see that $\del_vF$ only enters on the right-hand side of the equation for $\ddot r$. Hence integrating eq.~\reef{dotsVad} over an infinitesimal interval around the shell at $v=0$, we conclude that $\dot v$ is continuous across the shell while $\dot r$ jumps discontinuously with
\begin{equation}\label{eq:rdotJump}
\dot r_{\BH}(r_s) = \dot r_{\vac}(r_s) + \frac{\dot v(r_s)}{2} \left(f_{\BH}(r_s)-f_\vac(r_s)\right)  \, ,
\end{equation}
where $r_s$ denotes the value of the radial coordinate at which our extremal volume surface meets the collapsing shell.
It is also useful to recast the $\dot r$ equation as follows:
\beq
\dot  r^2 - F(r)\,	 r^{2(d-1)}   =  \E^2\,.
\label{yarn}
\eeq
This equation takes the form of a classical Hamiltonian constraint for a particle of mass $m=2$ and with energy $E=\E^2$ moving in a potential $U(r)=- F(r)	 r^{2(d-1)}$. This gives us an intuitive picture to understand the evolution of the surface on either side of the collapsing shell.\footnote{Note that this picture also agrees with eq.~\reef{dotsVad}, which {\it away} from the shell can be cast in the form: $m\,\ddot r = \del_r U(r)$. Of course, we must keep in mind that both $U(r)$ and $E$ will jump discontinuously at $r=r_s$ where the extremal surface crosses the shell.} The effective potential is depicted in figure \ref{fig:Potential} for the black hole geometry. We see that depending on the value of $\E_\BH^2$ certain values of $r$ may not be accessed. It will be useful in what follows to keep in mind the maximal value of the black hole potential $U_{\BH,\mt{max}}\equiv\E_m^2$ and the value $r=r_m$ for which it is obtained. They are obtained by solving the following equations:
\begin{equation}\label{maximal_surf}
\begin{split}
\del_r \!\left[f_{\BH}(r_m)	\,  r_m^{2(d-1)} \right]=0\,,
\qquad \E_m^2=- f_{\BH}(r_m)	\, r_m^{2(d-1)}  \,.
\end{split}
\end{equation}

The boundary conditions for our surface are determined as follows: In order for the extremal surfaces to avoid a conical singularity at $r=0$, we require that $\dot t=\dot v-\dot r/f=0$ there.\footnote{We observe that this boundary condition yields  $\dot t=0$ throughout the vacuum region. Further, we note that while this boundary condition is obvious for $k=+1$, it is more subtle in the planar geometry with $k=0$. In the latter case, we need to introduce a timelike regulator surface at some $r = \eps_0$ and consider the limit $\eps_0\to 0$, as in \cite{Format}.}  Eq.~\eqref{eq:consEVaid} then fixes $\E_\vac=0$, the conserved momentum in the vacuum part of the spacetime ($v<0$). When the surface crosses the collapsing shell at $r=r_s$, eq.~\eqref{eq:onedot} then determines
\begin{equation}\label{eq:vacrels}
\dot r_\vac(r_s) = r_s^{d-1} \sqrt{ f_\vac(r_s)}, \qquad \dot v(r_s) = \frac{r_s^{d-1}}{\sqrt{f_\vac(r_s)}}.
\end{equation}
Hence the value of the $v$-momentum and $\dot r$ on the black hole side of the shell can be read from eqs.~\eqref{eq:consEVaid} and \eqref{eq:rdotJump},
\begin{equation}\label{eq:BHrels}
\E_{\BH} = r_s^{d-1}\frac{f_\vac(r_s)-f_{\BH}(r_s)}{2\sqrt{f_\vac(r_s)}}=\frac{r_s\,\omega^{d-2}}{2\sqrt{f_\vac(r_s)}}\,, \qquad
\dot r_{\BH}(r_s) = r_s^{d-1}\frac{f_{\BH}(r_s)+f_\vac(r_s)}{2\sqrt{f_\vac(r_s)}}\,.
\end{equation}
The last boundary condition is that we are anchoring the extremal surface to the boundary time slice at  $v=t_0>0$. Hence using eq.~\reef{eq:onedot}, we integrate from the shell to the asymptotic boundary
\begin{equation}\label{timeeq}
t_0=\int_{0}^{t_0} dv = 	\int_{r_s}^{\infty}\frac{\dot v}{\dot r} dr = \int_{r_s}^{\infty}
 \left(1- \frac{\E_{\BH}}{ \sqrt{f_{\BH}(r)	 r^{2(d-1)} + \E_{\BH}^2}}  \right) \frac{dr}{f_{\BH}(r)}\,.
\end{equation}
Now eqs.~\eqref{eq:BHrels} and \eqref{timeeq} relate the boundary time $t_0$, the momentum $\E_{\BH}$ in the black hole part of spacetime ($v>0$), and the radius $r_s$ at which our extremal surface crosses the shell.

We can use these equations to prove that the momentum $\E_\BH$ on the black hole side is always positive. As a consequence the surfaces outside the black hole cross decreasing time slices. It is also easy to show that $\E_{\BH}^2-U(r)$ is in general positive, so that the Hamiltonian constraint \eqref{yarn} is consistent with $\dot r_\BH^2>0$ and so we have shown that the extremal surface is always able to cross the shell.
Figure \ref{fig:Potential} depicts the effective potential $U(r)=-f_\BH(r)\,r^{2(d-1)}$ (yellow line) and also the effective energy $\E_{\BH}^2$ as a function of the crossing radius $r_s$ (blue line), using eq.~\reef{eq:BHrels}. We note that if the latter energy is below the peak of the potential, \ie $\E_{\BH}^2<P_m^2$ from eq.~\reef{maximal_surf}, then the trajectory cannot escape the potential barrier and terminates on the singularity at $r=0$.
A special point in the figure is where the yellow and blue curves meet --- see inset. At that point $\dot r_\BH(r_s)$ vanishes and in fact, this is the point where the direction of $\dot r$ is flipped. That is, $\dot r_\BH(r_s)$ is positive for larger values of $r_s$, while it is negative for smaller values of $r_s$ and the extremal surface is headed towards the singularity at $r=0$ right after the crossing.
In any event, we are only interested in extremal surfaces which reach the asymptotic boundary and so we require $\E_{\BH}^2\ge P_m^2$.

We can see from eq.~\reef{timeeq} that as the latter inequality is saturated the boundary time diverges, \ie $t_0\to\infty$ when $\E_{\BH}^2\to P_m^2$.  This does not happen exactly at $r_s=r_m$ but rather at a slightly lower value of $r_s$ --- see the inset in figure \ref{fig:Potential} where the energy $\E_{\BH}(r_s)^2$ (blue line) crosses $\E_m^2$ (green line). To prove that the point for which the momentum is equal to $\E_{m}$ occurs with $r_s<r_m$ we can use the following general argument: First we note from general consideration that $\dot r_\BH$ is a monotonic function of $r_s$. In addition, we can check that $\dot r_{\BH}(r_s=r_m)$ is positive. To do that we use eq.~\eqref{maximal_surf} for $r_m$
\begin{equation}
\frac{2r_m^2}{L^2}+2 k - \frac{\omega^{d-2}}{r_m^{d-2}}=\frac{2k}{d}\,,
\end{equation}
as well as eq.~\eqref{eq:BHrels} for the velocity $\dot r_\BH$ after the crossing
\begin{equation}
\dot r_{BH} (r_s= r_m) = \frac{r_m^{d-2}}{2\sqrt{r_m^2/L^2+1}} \left(\frac{2 r_m^2}{L^2}+ 2k -\frac{\omega^{d-2}}{r_m^{d-2}}
\right)
=\frac{r_m^{d-2} k }{d \sqrt{r_m^2/L^2+1} }\,.
\end{equation}
The latter is strictly positive when $k=1$ (and is exactly zero for $k=0$). In fact, the blue curve for $k=0$ becomes a line of constant energy $\E^2=\E_m^2$.

With the gauge choice in eq.~\reef{eq:modiprop}, the maximal volume \reef{volu99} becomes
$\mathcal{V} =  \Omega_{k,d-1} \int d \lambda\,r^{2d-2}$. We evaluate the latter as
\begin{equation}
\begin{split}
\mathcal{V} =&\, \Omega_{k,d-1} \left[
\int_0^{r_s} \frac{dr}{\dot r} \,r^{2d-2}
+
\int_{r_s}^\infty \frac{dr}{\dot r}\,r^{2d-2}
\right]
\\
=&\,\Omega_{k,d-1} \int_0^{r_s} \frac{dr\,r^{d-1}}{\sqrt {f_\vac(r)}}
+ \Omega_{k,d-1} \int_{r_s}^{r_\mx}\frac{dr\,r^{2(d-1)}}{\sqrt {f_\BH(r)r^{2(d-1)}+\E_{\BH}^2}}
\end{split}
\end{equation}
and we have introduced the UV cutoff $r_{\mx}$ to produce a finite volume. It is convenient to use eq.~\eqref{timeeq} to re-express the second integral as follows
\begin{equation}\label{volume1}
\mathcal{V} =\Omega_{k,d-1} \left[ \int_0^{r_s} \frac{dr\,r^{d-1}}{\sqrt {f_\vac(r)}}
+ \int_{r_s}^{r_\mx}\!dr
\left[\frac{\sqrt {f_\BH(r)r^{2(d-1)}+\E_{\BH}^2}}{f_\BH(r)} -\frac{\E_\BH}{f_{\BH}(r)} \right]+ \E_\BH t_0\right]\,.
\end{equation}
We note that our expressions for the time and volume match those found in appendix A of \cite{Bridges} for the case of $d=2$ and $r_h=L$.

With all this technology in hand, we are ready to compute the time derivative of the holographic complexity using eq.~\reef{defineCV}. It is straightforward to check that the continuity of $\dot v$ across the shell implies that the contributions from differentiating the limits of integration vanish. Using again eq.~\eqref{timeeq}, a second cancellation arises from the derivative of the momentum inside the second integral and in the last term of \eqref{volume1}. We are finally left with
\begin{equation}\label{VolVadFinal}
\frac{d\cv}{dt_0}=\frac1{G_N L}\,\frac{d\mathcal{V}}{dt_0} = \frac{\Omega_{k,d-1}}{G_N L} \,\E_{\BH}\,.
\end{equation}
This surprisingly simple result bears some similarity to the expression for the rate of change of the volume complexity in the eternal black hole \cite{Growth}. However, we note that the expression \reef{eq:BHrels} relating the $\E_\BH$ and $r_s$ is different here than that relating the momentum and $r_{min}$ there.

The above result \reef{VolVadFinal} is implicit because in general it still requires solving eqs.~\reef{timeeq} and \reef{eq:BHrels} for $\E_\BH$ (or equivalently $r_s$) given the boundary time $t_0$. However, these equations are simply solved for the planar geometry with $k=0$ and one obtains
\begin{equation}
\E_{\BH}=\frac{r_h^d}{2 L}\,, \qquad \qquad \frac{d \cv}{dt_0} =\frac{8 \pi\,M }{d-1}\,.
\label{kaboom}
\end{equation}
Hence for the planar geometry, holographic complexity begins growing as soon as $t_0>0$ and the rate of growth is a fixed constant for all times. Further, this constant rate matches the late time rate of growth found for the eternal black hole in \cite{Stanford:2014jda,Growth}.
Our results for the spherical geometry confirm that the interpretation presented in appendix A of \cite{Bridges} for BTZ black holes also holds in higher dimensions. Namely that the main contribution in the late time limit comes from the extremal surface wrapping around a surface of constant $r=r_m$ while the contributions coming from the smaller value of $r$ reached by our surface, as well as the portions reaching to the boundary, are approximately constant and do not influence the time derivative of the holographic complexity.

\subsubsection*{Early Time Behaviour}
We can evaluate analytically the early time limit $t_0\to0$. For early times we know that $r_s\rightarrow\infty$ and using eq.~\eqref{eq:BHrels},  we see that for black holes in $d>2$:
\begin{equation}
\lim_{t_0\rightarrow 0}\E_{\BH}=\frac{L \, \omega^{d-2}}{2}\,,
\label{kapop}
\end{equation}
where as given in eq.~\reef{horiz}, $\omega^{d-2}=r_h^{d-2}(r_h^2/L^2+k)$. Now using eq. \eqref{VolVadFinal}, this leads to
\begin{equation}
\lim_{t_0\rightarrow 0} \frac{d\cv}{dt_0} = \frac{8\pi\, M}{d-1}\,.
\label{kapop2}
\end{equation}
That is, as noted above for the planar geometry, the rate of growth of the holographic complexity immediately jumps to a nonvanishing (positive) value for $t>t_0$. We also observe that the early time rate in eq.~\reef{kapop2}, which holds for both $k=0$ and +1, matches the $k=0$ result in eq.~\reef{kaboom}, which holds for all times.

\subsubsection*{Late Time Behaviour}

Another limit that we consider is the late time limit $t_0\to\infty$:
In the late time limit, we have already explained that the value of the momentum reaches $\E_m$ defined in eq.~\eqref{maximal_surf}.
In this case, our surface wraps around the surface of constant $r=r_m$, but the volume required to reach the minimal value of $r_s$ below $r_m$ and to reach the boundary above $r_m$ remains  (approximately) constant.
The contribution to the increasing growth of complexity at late times comes from the part of the surface which wraps around the $r=r_m$ surface. This will give us the value of $P_{\BH}$ in the late time limit for our numerical solutions below. One then finds that the rate of growth of
the holographic complexity (for $d>2$) at late times satisfies
\begin{equation}\label{eq:cv:late_rate}
\lim_{t_0\rightarrow\infty}\frac{(d-1)}{8\pi M} \frac{d \mathcal{C}_V}{dt_0}=\frac{2\,\E_m}{\omega^{d-2}L}\,.
\end{equation}
In general, eq.~\eqref{maximal_surf} cannot be solved analytically. However we can solve it in a large temperature expansion (or equivalently for small $z=L/r_h$ --- see eq.~\eqref{DimlessCoord})
\beqa
r_m &= &\frac{r_h}{2^{\frac{1}{d}}} \left[1 - \frac{ \left(2^{2/d} (d-1)-d\right)}{d^2} k z^2
\right.\labell{rsrs}\\
&&~~~~~ \left.
+\frac{ (d-1) \left(-d^2+2^{\frac{2}{d}+1} d+2^{4/d} (d-3) (d-1)\right) }{2\,d^4}\,k^2z^4+ \mathcal{O}(z^4)
\right]\,.
\nonumber
\eeqa
It is then possible to extract $\E_m$ using  eq.~\eqref{maximal_surf} and to use eq.~\eqref{eq:cv:late_rate} to determine the late time rate of change of the holographic complexity. Finally relating
 $z$  to the temperature with eq.~\eqref{eq:DimlessTemp}, we conclude
\begin{equation}\label{highT}
\begin{split}
&\lim_{t_0\rightarrow\infty}\frac{(d-1)}{8\pi M} \frac{d\cv }{dt_0}=
1-\frac{2^{\frac{2}{d}-1} d^2 k}{(4\pi) ^2 (L T)^2}+\frac{2^{\frac{2}{d}} \left(\gamma- d(d-3)\right) d^2 k^2}{(4\pi) ^4
(LT)^4} + \mathcal{O}\left(\frac{1}{L^6T^6}\right),
\end{split}
\end{equation}
where we have introduced the parameter $\gamma\equiv 2^{\frac2d-3}(3 d-2)(d-2)$. Hence with a spherical spatial geometry (\ie $k=+1$), there are curvature corrections which reduce the late time growth rate of the holographic complexity. That is, ${d\cv }/{dt_0}$ begins with the value $8\pi M/(d-1)$ shown in eq.~\reef{kapop2}, but then it decreases to a smaller growth rate at late times --- see also figure
\ref{fig:VaidyaVolAdS5Spherical}.  However, for the planar geometry (\ie $k=0$), the growth rate remains a fixed constant, as shown in eq.~\reef{kaboom}.

\subsubsection*{Two Boundary Dimensions ($d=2$)}

The collapse with $d=2$ forms a BTZ black hole with
$f_{\BH}(r) = (r^2 - r_h^2)/L^2$ \cite{Banados:1992wn,Banados:1992gq}. Recall that the mass, temperature and entropy are given in eq.~\reef{BTZquantities} and the choices $k=0$ or 1 in $f_\vac(r)$ correspond to the Ramond and Neveu-Schwarz vacuum, respectively, in the boundary theory.

\begin{figure}
\centering
\includegraphics[scale=0.5]{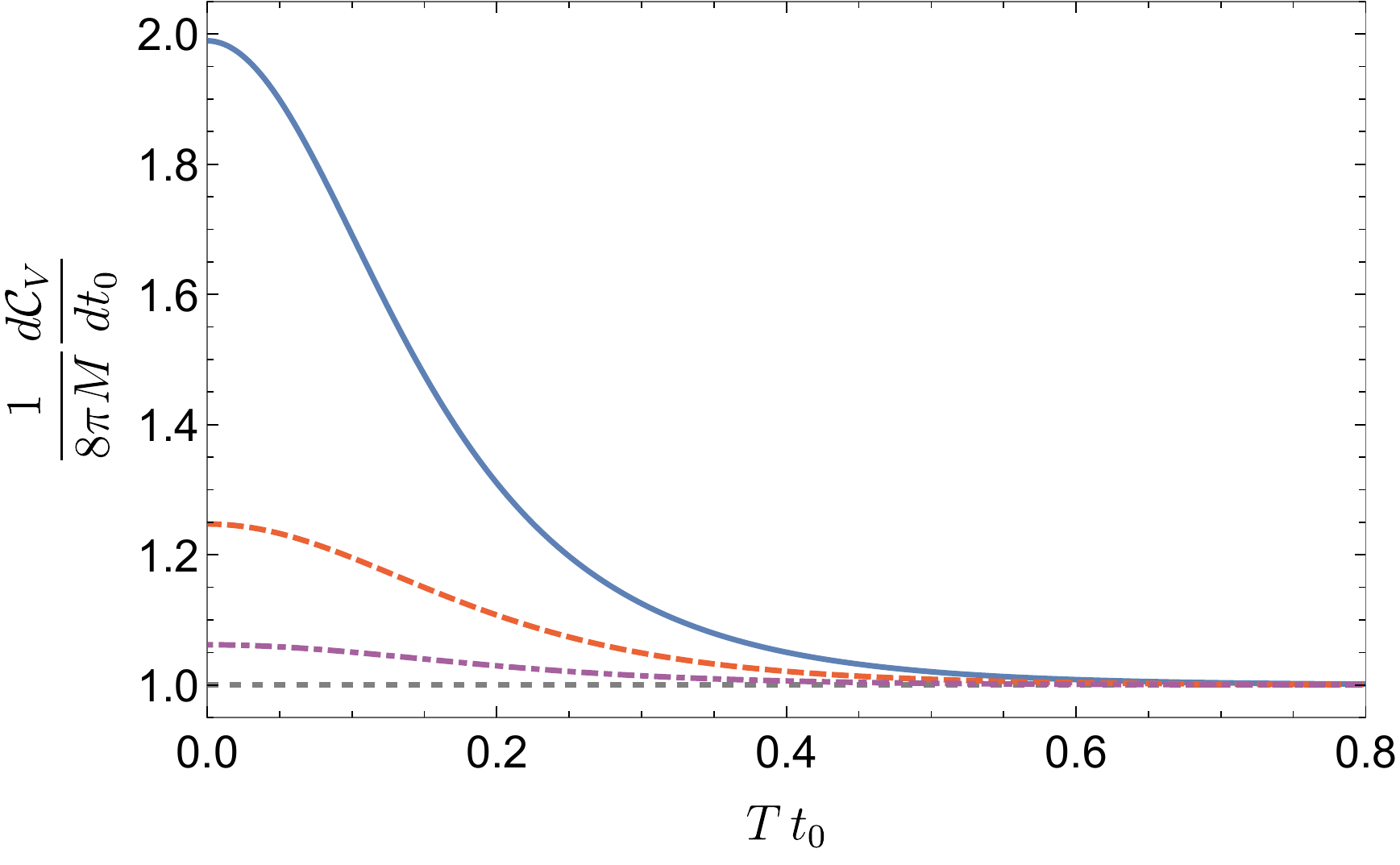}
\caption{Rate of change of complexity evaluated from the complexity=volume conjecture in the Vaidya-AdS spacetime for the BTZ black hole with the Neveu-Schwarz vacuum for several values of the temperature, \ie for $TL=0.16$ (blue), $TL=0.32$ (red, dashed) and $TL=0.64$ (purple, dot-dashed).}
\label{fig:VaidyaVolBTZ}
\end{figure}

The analysis follows identically to the previous case, with the obvious replacement of the blackening factors.
For the early time limit we may use eq.~\eqref{eq:BHrels} with $r_s\rightarrow \infty$ to obtain
\begin{equation}
\lim_{t_0\rightarrow 0}\E_{\BH} = \frac{L}{2}\left(\frac{r_h^2}{L^2}+k\right)\,.
\end{equation}
Next, using eq.~\eqref{VolVadFinal} for the rate of change of the holographic  complexity, we obtain:
\begin{equation}
\lim_{t_0\rightarrow 0} \frac{d\cv}{dt_0} = 8\pi M \left(1+k \frac{L^2}{r_h^2}\right) = 8\pi M \left(1+\frac{k}{(2 \pi L\, T)^2}\right).\label{kapop4}
\end{equation}
For the late time limit, we can solve eq. \eqref{maximal_surf} analytically and obtain
\begin{equation}
r_m=\frac{r_h}{\sqrt{2}}, \qquad \qquad  \E_m = \frac{r_h^2}{2 L}\label{kapop3}
\end{equation}
and hence we find
\begin{equation}\label{kapop99}
\lim_{t_0\rightarrow\infty} \frac{d \cv}{d t_0} = 8\pi M \,,
\end{equation}
which is independent of the value of $k$.  For the Ramond vacuum (\ie $k=0$),  $d \cv/d t_0$ is a fixed constant for all times, as expected from eq.~\reef{kaboom} for the planar geometry. However, we see that for the Neveu-Schwarz vacuum (\ie $k=1$), the early rate of growth is higher that the late time rate of growth.\footnote{Our results for the BTZ black hole are the same as those in appendix A of \cite{Bridges}. There the authors have set $r_h=L$ and the early and late time limits in their eqs.~(A.65) and (A.69) match with ours above.} In figure \ref{fig:VaidyaVolBTZ}, we numerically evaluate the rate of growth of the holographic complexity for intermediate times.

\subsubsection*{Numerical Results}
\begin{figure}
\centering
\includegraphics[scale=0.433]{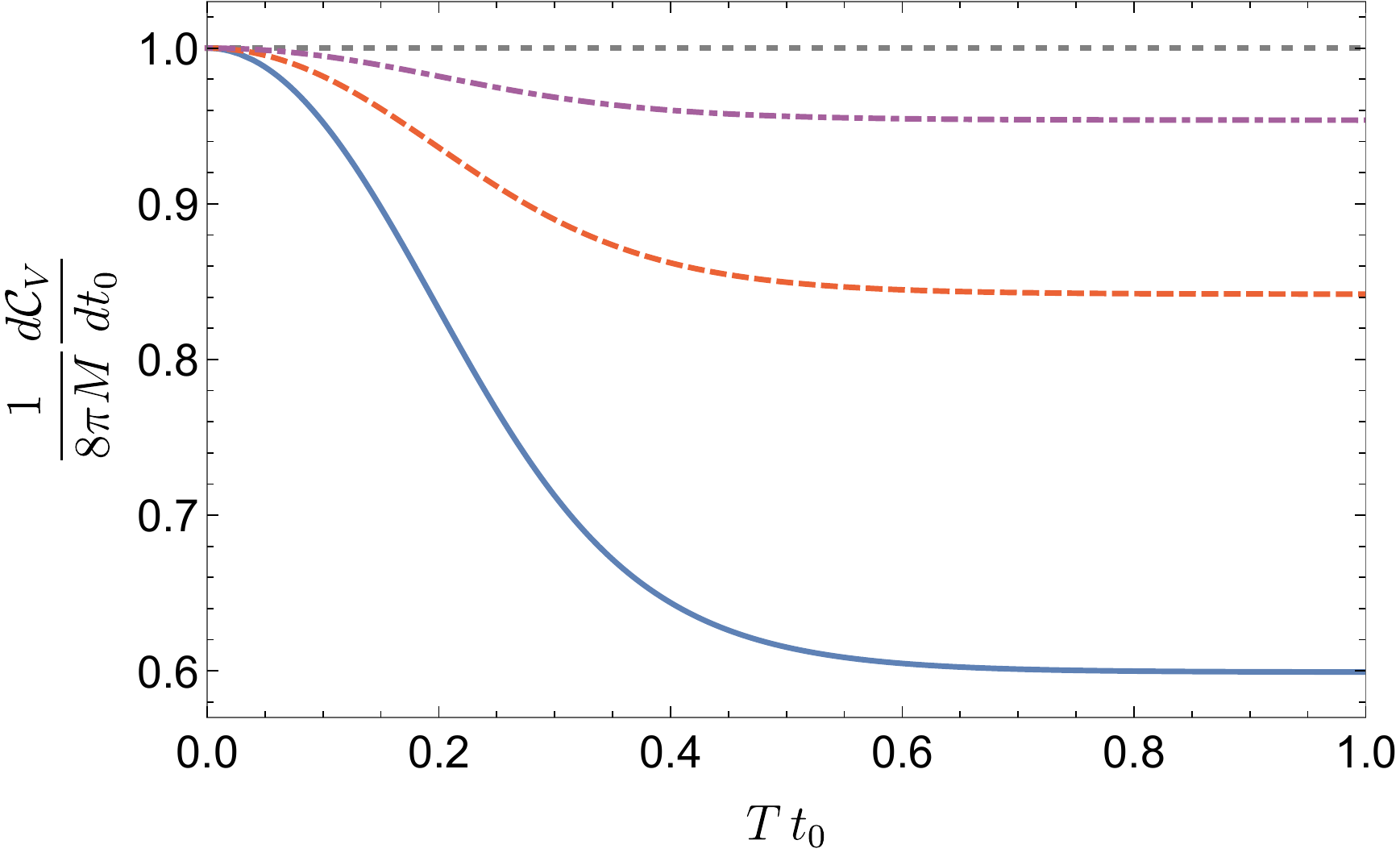}
\includegraphics[scale=0.41]{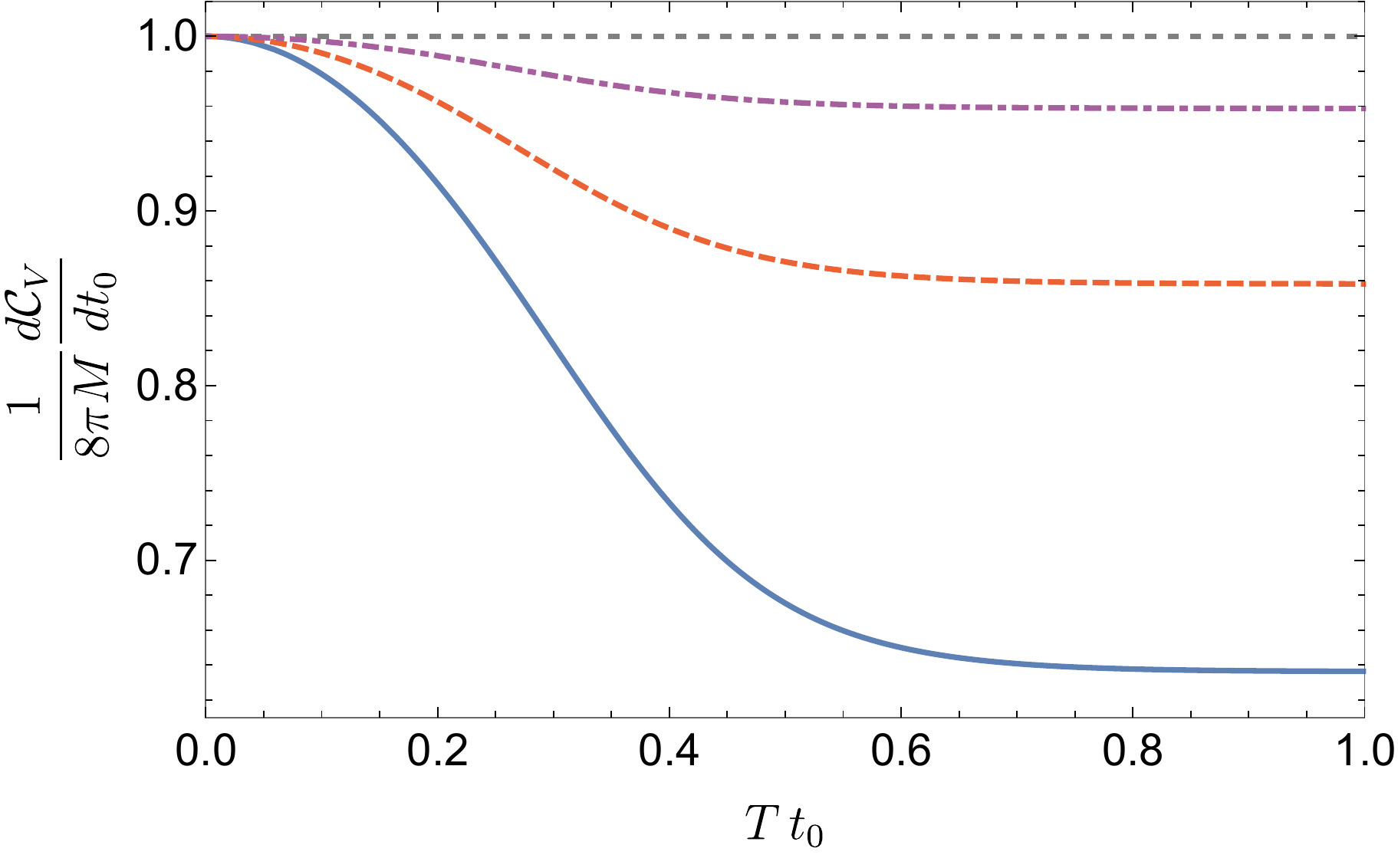}
\caption{Rate of change of complexity evaluated from the complexity=volume conjecture in the spherical ($k=1$) Vaidya-AdS spacetime in $d=3$ (left) for $TL=0.32$ (blue), $TL=0.52$ (red, dashed) and $TL=0.98$ (purple, dot-dashed) and $d=4$ (right) for $TL=0.48$ (blue), $TL=0.72$ (red, dashed) and $TL=1.32$ (purple, dot-dashed). }
\label{fig:VaidyaVolAdS5Spherical}
\end{figure}

We evaluated numerically the rate of growth of the holographic complexity for the spherical geometries with $d=3$ and $4$, shown in  figure \ref{fig:VaidyaVolAdS5Spherical}. We note a number of interesting features: First, of course, the early and late time rates match those discussed above. Second, in all of the cases shown, the rate of growth decreases at early times and the late time limit is approached from above. Recall from eq.~\reef{kapop4}, that the rate of growth is highest at early times for the Neveu-Schwarz vacuum in $d=2$ --- see also figure \ref{fig:VaidyaVolBTZ}.

\section{Discussion}\label{sec:Discussion}

In section \ref{sec:evolve}, we examined holographic complexity in the Vaidya geometry \reef{MetricV} for the case where a shell of null fluid is injected into empty AdS and collapses to form a black hole. Hence these geometries describe one-sided black holes, a situation which was previously considered in, \eg \cite{Bridges,Moosa} in the context of holographic complexity. Of course, using either the CA or CV approaches, we found that holographic complexity remains constant until the moment when the thin shell is injected. After that the complexity immediately begins to grow and the rate of growth monotonically approaches the corresponding late time limit.

In fact, using the CV conjecture, we found that for planar geometries in $d\ge 3$, the complexity grows at a constant rate which is equal to the late time rate of planar eternal black holes \cite{Stanford:2014jda,Growth}, \ie
\beq
\frac{d\cv}{dt} = \frac{8\pi\, M}{d-1}\,.
\label{nice9}
\eeq
For spherical geometries, the growth rate at early times is the same as the above expression, but the rate then decreases monotonically, as shown in figure \ref{fig:VaidyaVolAdS5Spherical}. Hence the (positive) curvature of the boundary geometry reduces the late time growth rate below that in eq.~\reef{nice9}, but this reduction is smaller for high temperature black holes. In eq.~\reef{highT} for large temperatures, we expressed the final rate in terms of an expansion in $1/(LT)^2$, \ie the curvature of the boundary geometry divided by the temperature, and we can see that the late time rate approaches eq.~\reef{nice9} for very high temperatures. Comparing eq.~\reef{highT} to eq.~(3.26) in \cite{Growth}, we can see that the curvature corrections for the present one-sided black holes precisely match those found for the analogous eternal black hole backgrounds. This agreement becomes obvious when we realize that eq.~\reef{maximal_surf} which determines late time limit \reef{eq:cv:late_rate} is identical to the corresponding equation for the eternal black hole \cite{Growth}.

Similar results were found for the BTZ black hole. In particular, beginning with the Ramond vacuum (with $k=0$), the growth rate is a fixed constant for all $t_0>0$ and matches the expression in eq.~\reef{nice9} with $d=2$. Starting with the Neveu-Schwarz vacuum (with $k=+1$), the growth rate decreases, similar to what was observed above for $d\ge3$. However, in this case, the initial rate is increased, as shown in eq.~\reef{kapop4}, and the final rate \reef{kapop99} matches eq.~\reef{nice9}, corresponding to the final rate for the eternal black hole background --- see also figure  \ref{fig:VaidyaVolBTZ}. The rate of change in complexity relaxes to its late time limit at times of the order of $t\sim 1/T$. All of these results are in accord with the expectations and calculations presented in \cite{Bridges}. In particular, the geometry of a one-sided black hole naturally includes regions behind the event horizon where time slices are growing to infinite volume (or as we discuss below, where the gravitational action grows without bound).

Above, we highlighted ways in which the CV results were the same for the one-sided and eternal (two-sided) black holes. However, we must also point out how the complexity for the Vaidya geometry differs from that for the eternal black holes \cite{Growth,run}. First, for planar black holes, the rate of growth of complexity in the eternal case had a transient period in which the rate of change in complexity gradually rose to its final value. As noted above for the collapsing shell, the growth rate  jumps discontinuously at $t_0=0$ to a value in eq.~\reef{nice9} and remains constant. For spherical eternal black holes, the growth rate increased towards the final late time rate \cite{Growth}, while here we observed a decreasing rate which approaches the late time limit from above.

Turning to the CA proposal \reef{defineCA}, we began in section \ref{sec:NullDust} by constructing an action \reef{ActionFluid} for the null fluid, which sources the Vaidya metric \reef{MetricV}.\footnote{We remind the reader that a similar action was derived in \cite{kuchar1} using complementary variables.} With this construction, we confirmed that when evaluated on a solution of the equations of motion, the null fluid action vanishes. As a result, in applying the CA conjecture \reef{defineCA} to evaluate the holographic complexity of the Vaidya metric in section \ref{sec:evolve} (as well as in \cite{Vad2}), the only nonvanishing contributions come from the gravitational action \eqref{THEEACTION}.  While this simplifies the task of evaluating  the WDW action, we also carefully examined the contribution of the spacetime region containing a narrow shell of null fluid and we found that it vanishes as the width of the shell shrinks to zero. We note that this vanishing result required a precise cancellation of the $\kappa$ surface term and joint terms on the past null boundary, as indicated in eq.~\reef{not}.  Hence with an infinitely thin shell, the WDW action can be evaluated as the sum of the actions for two separate regions, the first inside the shell and the second outside the shell.
We might observe that a similar statement holds for the calculations with the CV proposal \reef{defineCV}, where the extremal volume was found by evaluating separately the corresponding equations of motion inside and outside of the shell. Further in passing, we note that the vanishing of  the gravitational action for the spacetime region containing the null fluid shell, was an implicit assumption in various previous studies of holographic complexity, \eg \cite{Moosa,brian, Alishahiha:2018tep}.

In evaluating the holographic complexity on the collapsing null shell geometry in section \ref{CollapseCA}, one of our most striking results was that the late time growth rate did not match that found in an eternal black hole background. To be precise, the result in eq.~\reef{LateTimenoCT} for $d\ge3$ was evaluated using the gravitational action \reef{THEEACTION} and the standard prescription that the generators of the null boundaries are affinely parameterized (\ie $\kappa=0$).\footnote{Recall that this prescription also fixes the relative normalization of the null normals on the two sides of the shell, see eq.~\eqref{magic22}.} Perhaps even more striking is the result for $d=2$. Combining eqs.~\reef{xsBTZsimple}  and \reef{notctd2}, we have
\begin{equation}\label{revamp}
\frac{d \mathcal{C}_{A}}{d t_0}  = - \frac{2M}{\pi} \, \frac{k }{4\pi^2 L^2T^2 +(4\pi^2L^2T^2+k) \sinh^2(\pi Tt_0)} \, ,
\end{equation}
where we have also substituted $z=1/(2\pi LT)$ from eq.~\reef{eq:zfuncT}. Hence with $k=0$, the growth rate simply vanishes in $d=2$, while with $k=+1$, it is actually negative and only approaches zero at late times. That is, for $k=+1$ and $d=2$, the standard CA prescription yields a holographic complexity that decreases in time!

Clearly, this is an unsatisfactory result, however, we also found that the situation was corrected by adding the boundary counterterm \reef{counter} on the null surfaces. In particular, with this slightly modified prescription, the late time rate of growth was identical to that found for eternal black holes. That is, eq.~\reef{VaidyaDerivativeMass} yields ${d \ca'}/{d t_0}|_{t_0\to\infty}=2M/\pi$ for both $k=0,+1$ and any $d\ge3$. Hence, both the CV and CA approaches yield a late time growth rate which matches the rate found for the analogous eternal black hole backgrounds, as long as the gravitational action includes the extra counterterm. Further, from eq.~\reef{RateBTZOneSided}, we can see that for $d=2$,
\begin{equation}\label{revamp2}
\frac{d \mathcal{C}'_\mt{A}}{d t_0} = \frac{2 M}{\pi} - \frac{M }{\pi\,\sinh^2(\pi Tt_0)}\, \log\! \left[1+\left(1+ \frac{ k}{4\pi^2L^2T^2}\right)\sinh^2(\pi Tt_0) \right] \, .
\end{equation}
Hence  for both $k=0$ and $+1$, there is a transient behaviour at early times but the growth rate reaches the expected late time limit by $t_0\sim 1/T$. We also note that  eq.~\reef{revamp2} yields $d\ca'/d t_0$ which is positive for all times. Hence adding the boundary counterterm repairs the previous problematic result \reef{revamp} for $d=2$. Therefore, we conclude that it is essential in defining the CA proposal \reef{defineCA} to supplement the gravitational action \reef{THEEACTION} with the boundary counterterm \reef{counter}. This conclusion will be reinforced by our analysis of shock waves in an eternal black hole background in \cite{Vad2} --- see further comments below.

We also discuss this conclusion further below, but first let us examine the behaviour of the holographic complexity, using the modified prescription, for the collapsing null shell in more detail. Of course, as observed above, the modified prescription yields the same universal  late time limit as found for the eternal black holes \cite{Brown1,Brown2}. But closer examination of eq.~\reef{VaidyaDerivativeMass} shows that $d\ca'/d t_0$ begins at precisely half this rate at $t_0=0$ for $d\ge3$ (as well as $d=2$ with $k=0$) and that the growth rate increases monotonically towards the late time limit. As shown in eq.~\reef{Late}, $d\ca'/dt_0$ relaxes to this limit with an exponential decay controlled by the thermal time scale $1/(2\pi T)$. Generally, as shown in figure \ref{VadActSphd34_Both}, the growth rate has essentially reached this late time limit at $t\sim 1/T$.

We might recall that when the CA approach was applied to study the time evolution of the holographic complexity in the eternal black hole geometry \cite{Growth,run}, a number of unusual features arose. First, the holographic complexity does not change at all until some $t_c\sim 1/T$ for $d\ge3$. Second, at $t_c$, there is a sudden spike in $d\ca/dt$ where it  actually becomes (infinitely) negative. After this spike, $d\ca/dt$ grows rapidly and overshoots the late time rate. Then the growth rate approaches the late time limit with an exponential decay from above. Further, we note that the details of this transient behaviour depend on $\alpha$, the parameter appearing in the normalization of the null normals on the boundaries of the WDW patch. Of course, these calculations were found using
the standard prescription which did not include the null boundary counterterm. However, including the counterterm contributions does not modify the above description in any essential way, \eg see appendices A and E of \cite{Growth}, except that the undetermined normalization constant $\alpha$ is replaced the undetermined scale $\tL$, appearing in eq.~\reef{counter}.\footnote{The counterterm \reef{counter} was introduced to ensure the reparameterization invariance of the action and hence when this term is included, the action is completely independent of $\alpha$.} Hence it is interesting to observe that these unusual features are absent in $d\ca'/dt_0$ calculated for the formation of a black hole, rather than an eternal black hole. In particular, we emphasize the counterterm scale $\tL$ does not come into play in the time evolution of the complexity for the collapsing shell.

Our conclusion above was that without the null surface counterterm \reef{counter}, evaluating the gravitational action on the WDW patch did not yield an observable that could be associated with complexity in the boundary theory. For example, eq.~\reef{LateTimenoCT} shows that the late time growth rate after the formation of a black hole does not match that found for an eternal black hole, \ie $2M/\pi$. This discrepancy is somewhat surprising since at late times, the largest portion of the WDW patch is above the shell, where the geometry is precisely that of a static black hole --- see figure \ref{CollapseOneSided}. However, recall that in evaluating $d\ca/dt$ in the eternal black hole geometry, an essential contribution comes from the joint where the two past null boundaries meet behind the past event horizon \cite{RobLuis,Growth}. Of course, there is no counterpart of this joint contribution in the Vaidya geometry describing black hole formation.\footnote{Geometrically, the closest analog of this joint would be where the past null boundary reaches $r=0$ in the AdS vacuum region. However, this point does not contribution to the gravitational action, as discussed in \cite{Format}.}
However, upon adding the counterterm, an extra boundary contribution appears where the past boundary of the WDW patch crosses the null shell (see eqs.~\reef{count2} and \reef{count22}), and then its time derivative provides precisely the extra contribution needed to restore the late time growth rate for the holographic complexity (see eq.~\reef{VACT}).

Holographic complexity using the CA proposal has recently been applied in a number of situations involving null shells, \eg \cite{Moosa,brian,Alishahiha:2018tep,Ageev:2018nye,Agon:2018zso}. In particular, we should also compare our results with those of \cite{Moosa}, which evaluates the holographic complexity using the CA proposal for precisely the same Vaidya geometries (with $k=0$) that were studied here. In fact, we can see that the results for the growth rate are precisely the same by comparing eq.~\reef{VDMass2} with eq.~(56) of \cite{Moosa}. The primary way in which the two calculations differ is that in \cite{Moosa}, the author sets $\tilde\alpha=\alpha$ and so implicitly the null generators of the past boundary are not affinely parameterized as they cross the null shell. However, this choice does not affect the final answer. Imagine that we allow $\alpha$ and $\tilde\alpha$ to be arbitrary constants. Then the counterterm contribution at $r=r_s$ appears in eq.~\reef{count02} while the corresponding joint contributions appear in eq.~\reef{VaJointAct}. It is straightforward to see that combining these two contributions yields
\beq
I_{\text{joint}} + I_\mt{ct}=
\frac{\Omega_{k, d-1} r_s^{d-1}}{8 \pi G_N} \,\log{\!\left[ \frac{f_{\vac}(r_s)}{f_{\BH}(r_s)}\right]}\, , \labell{extra}
\eeq
which is completely independent of both $\alpha$ and $\tilde\alpha$. One might note that in fact the counterterm contribution at $r_s$ vanishes with the choice $\tilde\alpha=\alpha$, as in \cite{Moosa}. Hence although some of the intermediate steps may differ, the final results for the holographic complexity here and in \cite{Moosa} agree.

In some of the other recent studies of the CA proposal with null shells, the counterterm \reef{counter} was included \cite{Alishahiha:2018tep,Ageev:2018nye} but in other, it was not
\cite{brian,Agon:2018zso}. In all of these cases, it was assumed that the contribution of the (infinitely thin) null shell was zero, as we explicitly demonstrated in section \ref{nfca},  and so the WDW action was determined by adding together the action evaluated separately on the regions above and below the shell, as in our calculation. It is particularly interesting to compare \cite{brian} and \cite{Alishahiha:2018tep}, which both studied holographic complexity in hyperscaling violating geometries, but the first did not include the counterterm while the second did. The same simple but ad hoc prescription for the normalization of the null boundary normals was chosen in \cite{brian} as in \cite{Moosa}, \ie $\tilde\alpha=\alpha$. The observation above was that with this choice, the counterterm contribution generated at $r_s$ vanishes and so it is not surprising the main results for the growth rate in \cite{brian} and \cite{Alishahiha:2018tep} agree. However,  we note that differences appear in the transient behaviour if this ad hoc prescription is applied for a null shell falling into an eternal black hole \cite{Vad2}. Further, there is no obvious covariant principle which produces the choice $\tilde\alpha=\alpha$, \ie this parameterization appears to be an arbitrary coordinate-dependent choice. For example then, it is  not clear what the corresponding prescription for a null shell of finite width should be.

\subsection*{Future Directions}

One of our key results was that if the gravitational action evaluated on the WDW patch is to properly describe the complexity of the boundary state, then one must include the counterterm \reef{counter} on the null boundaries. This counterterm was originally constructed in \cite{RobLuis} to ensure that the action did not depend on the parametrization of the null boundaries. In particular, this term does not play a role in producing a well-defined variational principle for the gravitational action. Previous studies of holographic complexity using the CA proposal focused on stationary spacetimes, \eg eternal black hole backgrounds, and it was found that this extra surface term does not modify the essential properties of the holographic complexity, \eg the complexity of formation \cite{Format} or the late-time rate of growth \cite{Growth}. This points out the importance of testing various proposal for holographic complexity in dynamical spacetimes, such as the Vaidya geometries \reef{MetricV}. We extend the present work with a study of holographic complexity for shock waves falling into an eternal black hole in a companion paper \cite{Vad2}. In this case, we find that including the counterterm on the null boundaries of the WDW patch is an essential ingredient in order to reproduce the ``switchback effect''.

Additional topics to explore would include extending our results to collapsing charged shells, to shells of finite width, to shells of other kinds of matter, including higher curvature corrections as in \cite{Cano:2018aqi} or to localized shocks as in \cite{Roberts:2014isa}.

As emphasized in \cite{Moosa}, the growth rate for the collapsing null shell calculated using the (modified) CA proposal always obeys the bound $d\ca'/dt\le 2M/\pi$. It was proposed in \cite{Brown1,Brown2} that this bound may be related to Lloyd's bound for the maximum rate of computation for a system with a fixed energy \cite{Lloyd}. However, as noted above, transient violations of the proposed bound were already identified in studying the time evolution of complexity in an eternal black hole background \cite{Growth}. Further, even stronger violations were found in the dual of a noncommutative gauge theory \cite{fish} and in hyperscaling violating geometries \cite{brian,Alishahiha:2018tep}.\footnote{Violations of the analogous bound proposed for systems with a chemical potential were also found in certain instances \cite{Brown2, Couch:2016exn, Cai:2017sjv}.} Therefore, while the proposed bound can not be universal, it remains an interesting question to understand the situations when it does apply and when not, and the underlying reasons for this.

Another interesting direction would be to study the evolution of complexity for quantum quenches in a field theory context.  Some initial studies of this question appear in \cite{Alves:2018qfv,prep2}, which examine the evolution of the complexity through a mass quench in a free scalar field theory (analogous to those studied in \cite{quench1,quench2,quench3}). A remarkable feature of these quenches is that the scalar field remains in a Gaussian state throughout the entire process, and so methods developed in \cite{qft1,qft2,qft3} can still be applied to evaluate the complexity. The comparison of our holographic results with those in \cite{Alves:2018qfv} is not straightforward since, \eg the initial and final masses are nonvanishing (\ie neither the initial nor final scalar theories are CFTs). However, we might note that the QFT calculations suggest that the complexity  growth rate at early times increases as the energy injected by the quench increases. Hence this behaviour would be in rough agreement with our holographic results where the initial growth rate is proportional to the energy carried by the null shell, \ie see
eqs.~\reef{EarlyExp} and \reef{earlyRateBTZ} for the CA proposal, and eq.~\reef{kapop2} for the CV proposal. On the other hand, in \cite{Alves:2018qfv}, the authors found that in most instances, the complexity quickly saturated (at least approximately) while, of course, the holographic complexity continues to grow linearly at late times. Further, the complexity in the QFT quench showed a strong dependence on the mass scale $M$ associated with the unentangled reference state. In \cite{qft1,qft2}, it was suggested the dependence on $M$ could be associated with dependence of the holographic complexity $\alpha$, which seems to be traded for the dependence on scale $\tL$ with the addition of the boundary counterterm \cite{Vad2}. However, the holographic growth rate, \eg in eq.~\reef{VaidyaDerivativeMass} shows no dependence on $\tL$ at all, and so this points to another mismatch between the holographic and QFT results. One possible way to improve the comparison of the holographic and QFT quenches  would be to consider CFT-to-CFT quenches for a free scalar (in which the initial and final masses both vanish) using the protocol described in section 3.2 of \cite{quench3}. Another simple extension of this work would be to study the complexity for a mass quench of a free fermion, using the techniques of \cite{qft3}.

\section*{Acknowledgments}
We would like to thank Alice Bernamonti, Adam Brown, Dean Carmi, Lorenzo Di Pietro, Federico Galli, Minyong Guo, Robie Hennigar, Juan Pablo Hernandez, Nick Hunter-Jones, Shan-Ming Ruan, Sotaro Sugishita, Brian Swingle, Beni Yoshida and Ying Zhao for useful comments and discussions. Research at Perimeter Institute is supported by the Government of Canada through the Department of Innovation, Science and Economic Development and by the Province of Ontario through the Ministry of Research, Innovation and Science. HM and RCM thank the Kavli Institute for Theoretical Physics for its hospitality at one stage of this project. At the KITP, this research was supported in part by the National Science Foundation under Grant No. NSF PHY17-48958. SC acknowledges support from an Israeli Women in Science  Fellowship from the Israeli Council of Higher Education. RCM is supported by funding from the Natural Sciences and Engineering Research Council of Canada and from the Simons Foundation through the ``It from Qubit'' collaboration.

\appendix


\begin{thebibliography}{99}

\bibitem{jb1}
J.~D.~Bekenstein,
  ``Black holes and the second law,''
  Lett.\ Nuovo Cim.\  {\bf 4} (1972) 737,
  \href{https://link.springer.com/article/10.1007%2FBF02757029}{LettNuovoCim.4.737}.

\bibitem{jb2}
J.~D.~Bekenstein,
  ``Black holes and entropy,''
  Phys.\ Rev.\ D {\bf 7} (1973) 2333,
  \href{http://journals.aps.org/prd/abstract/10.1103/PhysRevD.7.2333}{PhysRevD.7.2333}.


\bibitem{Sorkin:2014kta}
  R.~D.~Sorkin,
  ``1983 paper on entanglement entropy: `On the Entropy of the Vacuum outside a Horizon',''
  Tenth International Conference on General Relativity and
  Gravitation (held Padova, 4-9 July, 1983), Contributed Papers, vol. II, pp. 734-736,
  \href{https://arxiv.org/abs/1402.3589}{gr-qc/1402.3589}.

\bibitem{Jacobson:1995ab}
  T.~Jacobson,
  ``Thermodynamics of space-time: The Einstein equation of state,''
  Phys.\ Rev.\ Lett.\  {\bf 75} (1995)  1260,
  \href{https://arxiv.org/abs/gr-qc/9504004}{gr-qc/9504004}.

\bibitem{VanRaamsdonk:2009ar}
  M.~Van Raamsdonk,
  ``Comments on quantum gravity and entanglement,''
      \href{https://arxiv.org/abs/0907.2939}{hep-th/0907.2939}.


\bibitem{VanRaamsdonk:2010pw}
  M.~Van Raamsdonk,
  ``Building up spacetime with quantum entanglement,''
  Gen.\ Rel.\ Grav.\  {\bf 42} (2010) 2323,
  Int.\ J.\ Mod.\ Phys.\ D {\bf 19} (2010) 2429,
      \href{https://arxiv.org/abs/1005.3035}{hep-th/1005.3035}.


\bibitem{AdSCFT}
O.~Aharony, S.~S.~Gubser, J.~M.~Maldacena, H.~Ooguri and Y.~Oz, ``Large N field theories, string theory and gravity,''
  Phys.\ Rept.\  {\bf 323} (2000) 183,
  \href{https://arxiv.org/abs/hep-th/9905111}{hep-th/9905111}.

\bibitem{AdSCFTbook}
  M.~Ammon and J.~Erdmenger,
  {\sl Gauge/gravity duality: Foundations and applications},
  Cambridge University Press, 2015.

\bibitem{Ryu:2006bv}
  S.~Ryu and T.~Takayanagi,
  ``Holographic derivation of entanglement entropy from AdS/CFT,''
  Phys.\ Rev.\ Lett.\  {\bf 96} (2006) 181602,
  \href{https://arxiv.org/abs/hep-th/0603001}{hep-th/0603001}.

\bibitem{Ryu:2006ef}
  S.~Ryu and T.~Takayanagi,
  ``Aspects of Holographic Entanglement Entropy,''
  JHEP {\bf 0608} (2006) 045,
    \href{https://arxiv.org/abs/hep-th/0605073}{hep-th/0605073}.

  \bibitem{Hubeny:2007xt}
  V.~E.~Hubeny, M.~Rangamani and T.~Takayanagi,
  ``A Covariant holographic entanglement entropy proposal,''
  JHEP {\bf 0707} (2007) 062,
    \href{https://arxiv.org/abs/0705.0016}{hep-th/0705.0016}.


\bibitem{HoloEntEntropy}
  M.~Rangamani and T.~Takayanagi,
  ``Holographic Entanglement Entropy,''
  Lect.\ Notes Phys.\  {\bf 931} (2017)  pp.1-246,
  \href{https://arxiv.org/abs/1609.01287}{hep-th/1609.01287}.

\bibitem{Swingle:2009bg}
  B.~Swingle,
  ``Entanglement Renormalization and Holography,''
  Phys.\ Rev.\ D {\bf 86} (2012)  065007,
  \href{https://arxiv.org/abs/0905.1317}{cond-mat.str-el/0905.1317}.



\bibitem{Myers:2010tj}
  R.~C.~Myers and A.~Sinha,
  ``Holographic c-theorems in arbitrary dimensions,''
  JHEP {\bf 1101} (2011) 125,
  \href{https://arxiv.org/abs/1011.5819}{hep-th/1011.5819}.


\bibitem{Blanco:2013joa}
  D.~D.~Blanco, H.~Casini, L.~Y.~Hung and R.~C.~Myers,
  ``Relative Entropy and Holography,''
  JHEP {\bf 1308} (2013) 060,
  \href{https://arxiv.org/abs/1305.3182}{hep-th/1305.3182}.

\bibitem{Dong:2013qoa}
  X.~Dong,
  ``Holographic Entanglement Entropy for General Higher Derivative Gravity,''
  JHEP {\bf 1401} (2014) 044,
  \href{https://arxiv.org/abs/1310.5713}{hep-th/1310.5713}.

\bibitem{Faulkner:2013ica}
  T.~Faulkner, M.~Guica, T.~Hartman, R.~C.~Myers and M.~Van Raamsdonk,
  ``Gravitation from Entanglement in Holographic CFTs,''
  JHEP {\bf 1403} (2014) 051,
  \href{https://arxiv.org/abs/1312.7856}{hep-th/1312.7856}.

\bibitem{Almheiri:2014lwa}
  A.~Almheiri, X.~Dong and D.~Harlow,
  ``Bulk Locality and Quantum Error Correction in AdS/CFT,''
  JHEP {\bf 1504} (2015) 163,
  \href{https://arxiv.org/abs/1411.7041}{hep-th/1411.7041}.

\bibitem{Pastawski:2015qua}
  F.~Pastawski, B.~Yoshida, D.~Harlow and J.~Preskill,
  ``Holographic quantum error-correcting codes: Toy models for the bulk/boundary correspondence,''
  JHEP {\bf 1506}, 149 (2015),
  \href{https://arxiv.org/abs/1503.06237}{hep-th/1503.06237}.

  \bibitem{Czech:2015qta}
  B.~Czech, L.~Lamprou, S.~McCandlish and J.~Sully,
  ``Integral Geometry and Holography,''
  JHEP {\bf 1510}, 175 (2015),
  \href{https://arxiv.org/abs/1505.05515}{hep-th/1505.05515}.


\bibitem{deBoer:2015kda}
  J.~de Boer, M.~P.~Heller, R.~C.~Myers and Y.~Neiman,
  ``Holographic de Sitter Geometry from Entanglement in Conformal Field Theory,''
  Phys.\ Rev.\ Lett.\  {\bf 116}, no. 6, 061602 (2016),
  \href{https://arxiv.org/abs/1509.00113}{hep-th/1509.00113}.

\bibitem{Jafferis:2015del}
  D.~L.~Jafferis, A.~Lewkowycz, J.~Maldacena and S.~J.~Suh,
  ``Relative entropy equals bulk relative entropy,''
  JHEP {\bf 1606} (2016) 004,
   \href{https://arxiv.org/abs/1512.06431}{hep-th/1512.06431}.

\bibitem{Czech:2017zfq}
  B.~Czech, L.~Lamprou, S.~McCandlish and J.~Sully,
  ``Modular Berry Connection,''
     \href{https://arxiv.org/abs/1712.07123}{hep-th/1712.07123}.

\bibitem{EntNotEnough}
  L.~Susskind,
  ``Entanglement is not enough,''
  Fortsch.\ Phys.\  {\bf 64}, 49 (2016),
  \href{https://arxiv.org/abs/1411.0690}{hep-th/1411.0690}.

   \bibitem 
 {johnw} J.~Watrous, ``Quantum Computational Complexity,'' pp 7174-7201 in {\it Encyclopedia of Complexity and Systems Science}, ed., R.~A.~Meyers  (Springer, 2009),  \href{https://arxiv.org/abs/0804.3401}{quant-ph/0804.3401}.

\bibitem 
 {AaronsonRev}   S.~Aaronson,
  ``The Complexity of Quantum States and Transformations: From Quantum Money to Black Holes,''
  \href{https://arxiv.org/abs/1607.05256}{quant-ph/1607.05256}.

 \bibitem{Susskind:2014rva}
  L.~Susskind,
  ``Computational Complexity and Black Hole Horizons,''
  Fortsch.\ Phys.\  {\bf 64}, 24 (2016),
  Addendum: Fortsch.\ Phys.\  {\bf 64}, 44 (2016),
    \href{https://arxiv.org/abs/1402.5674}{hep-th/1402.5674},
  \href{https://arxiv.org/abs/1403.5695}{hep-th/1403.5695}.

\bibitem{Stanford:2014jda}
  D.~Stanford and L.~Susskind,
  ``Complexity and Shock Wave Geometries,''
  Phys.\ Rev.\ D {\bf 90}, no. 12, 126007 (2014),
    \href{https://arxiv.org/abs/1406.2678}{hep-th/1406.2678}.


\bibitem 
 {Brown1}  A.~R.~Brown, D.~A.~Roberts, L.~Susskind, B.~Swingle and Y.~Zhao,
  ``Holographic Complexity Equals Bulk Action?,''
  Phys.\ Rev.\ Lett.\  {\bf 116} (2016) no.19,  191301,
  \href{https://arxiv.org/abs/1509.07876}{hep-th/1509.07876}.


\bibitem 
 {Brown2}  A.~R.~Brown, D.~A.~Roberts, L.~Susskind, B.~Swingle and Y.~Zhao,
  ``Complexity, action, and black holes,''
  Phys.\ Rev.\ D {\bf 93} (2016) no.8,  086006,
  \href{https://arxiv.org/abs/1512.04993}{hep-th/1512.04993}.

\bibitem{Roberts:2014isa}
  D.~A.~Roberts, D.~Stanford and L.~Susskind,
  ``Localized shocks,''
  JHEP {\bf 1503} (2015) 051,
   \href{https://arxiv.org/abs/1409.8180}{hep-th/1409.8180}.



\bibitem{Cai:2016xho}
  R.~G.~Cai, S.~M.~Ruan, S.~J.~Wang, R.~Q.~Yang and R.~H.~Peng,
  ``Action growth for AdS black holes,''  JHEP {\bf 1609} (2016) 161,
     \href{https://arxiv.org/abs/1606.08307}{gr-qc/1606.08307}.

\bibitem 
{RobLuis}   L.~Lehner, R.~C.~Myers, E.~Poisson and R.~D.~Sorkin,
  ``Gravitational action with null boundaries,''
  Phys.\ Rev.\ D {\bf 94}, no. 8, 084046 (2016),
\href{https://arxiv.org/abs/1609.00207}{hep-th/1609.00207}.

\bibitem 
 {Format} S.~Chapman, H.~Marrochio and R.~C.~Myers,
  ``Complexity of Formation in Holography,''
  JHEP {\bf 1701} (2017) 062,
  \href{https://arxiv.org/abs/1610.08063}{hep-th/1610.08063}.

  \bibitem
 {diverg} D.~Carmi, R.~C.~Myers and P.~Rath,
  ``Comments on Holographic Complexity,''
  JHEP {\bf 1703} (2017) 118,
  \href{https://arxiv.org/abs/1612.00433}{hep-th/1612.00433}.


\bibitem 
 {2LawComp}  A.~R.~Brown and L.~Susskind,
  ``The Second Law of Quantum Complexity,''
  \href{https://arxiv.org/abs/1701.01107}{hep-th/1701.01107}.

 \bibitem{EuclideanComplexity1}
  P.~Caputa, N.~Kundu, M.~Miyaji, T.~Takayanagi and K.~Watanabe,
  ``Anti-de Sitter Space from Optimization of Path Integrals in Conformal Field Theories,''
  Phys.\ Rev.\ Lett.\  {\bf 119}, no. 7, 071602 (2017),
  \href{https://arxiv.org/abs/1703.00456}{hep-th/1703.00456}.

  \bibitem{EuclideanComplexity2}
  P.~Caputa, N.~Kundu, M.~Miyaji, T.~Takayanagi and K.~Watanabe,
  ``Liouville Action as Path-Integral Complexity: From Continuous Tensor Networks to AdS/CFT,''
   \href{https://arxiv.org/abs/1706.07056}{hep-th/1706.07056}.

  \bibitem{EuclideanComplexity3}
  B.~Czech,
  ``Einstein Equations from Varying Complexity,''
  Phys.\ Rev.\ Lett.\  {\bf 120} (2018) no.3,  031601,
  \href{https://arxiv.org/abs/1706.00965}{hep-th/1706.00965}.


  \bibitem{Reynolds:2017lwq}
  A.~Reynolds and S.~F.~Ross,
  ``Complexity in de Sitter Space,''
  Class.\ Quant.\ Grav.\  {\bf 34}, no. 17, 175013 (2017),
 \href{https://arxiv.org/abs/1706.03788}{hep-th/1706.03788}.


   \bibitem 
 {koji} K.~Hashimoto, N.~Iizuka and S.~Sugishita,
  ``Time evolution of complexity in Abelian gauge theories,''
  Phys.\ Rev.\ D {\bf 96} (2017) no.12,  126001,
  \href{https://arxiv.org/abs/1707.03840}{hep-th/1707.03840}.


   \bibitem 
 {qft1}   R.~A.~Jefferson and R.~C.~Myers,
  ``Circuit complexity in quantum field theory,''
  JHEP {\bf 1710} (2017) 107,
  \href{https://arxiv.org/abs/1707.08570}{hep-th/1707.08570}.


\bibitem 
 {qft2} S.~Chapman, M.~P.~Heller, H.~Marrochio and F.~Pastawski,
  ``Towards Complexity for Quantum Field Theory States,''
  Phys.\ Rev.\ Lett.\  {\bf 120} (2018) no.12,  121602,
  \href{https://arxiv.org/abs/1707.08582}{hep-th/1707.08582}.

\bibitem{qft3}
  L.~Hackl and R.~C.~Myers,
  ``Circuit complexity for free fermions,''
  \href{https://arxiv.org/abs/1803.10638}{hep-th/1803.10638}.

\bibitem{fish}
J.~Couch, S.~Eccles, W.~Fischler and M.~L.~Xiao,
  ``Holographic complexity and noncommutative gauge theory,''
  JHEP {\bf 1803} (2018) 108,
  \href{https://arxiv.org/abs/1710.07833}{hep-th/1710.07833}.

\bibitem{Growth}
  D.~Carmi, S.~Chapman, H.~Marrochio, R.~C.~Myers and S.~Sugishita,
  ``On the Time Dependence of Holographic Complexity,''
  JHEP {\bf 1711} (2017) 188,
\href{https://arxiv.org/abs/1709.10184}{hep-th/1709.10184}.

\bibitem{Moosa}
  M.~Moosa,
  ``Evolution of Complexity Following a Global Quench,''
  JHEP {\bf 1803} (2018) 031,
  \href{https://arxiv.org/abs/1711.02668}{hep-th/1711.02668} .


\bibitem{brian}
B.~Swingle and Y.~Wang,
  ``Holographic Complexity of Einstein-Maxwell-Dilaton Gravity,''
  \href{https://arxiv.org/abs/1712.09826}{hep-th/1712.09826}.

\bibitem{Alishahiha:2018tep}
  M.~Alishahiha, A.~Faraji Astaneh, M.~R.~Mohammadi Mozaffar and A.~Mollabashi,
  ``Complexity Growth with Lifshitz Scaling and Hyperscaling Violation,''
  \href{https://arxiv.org/abs/1802.06740}{hep-th/1802.06740}.


\bibitem{Zhao:2017isy}
  Y.~Zhao,
  ``Uncomplexity and Black Hole Geometry,''
  \href{https://arxiv.org/abs/1711.03125}{hep-th/1711.03125}.

\bibitem{Fu:2018kcp}
  Z.~Fu, A.~Maloney, D.~Marolf, H.~Maxfield and Z.~Wang,
  ``Holographic complexity is nonlocal,''
  JHEP02(2018)072,
  \href{https://arxiv.org/abs/1801.01137}{hep-th/1801.01137}.

\bibitem{Bridges}
  L.~Susskind and Y.~Zhao,
  ``Switchbacks and the Bridge to Nowhere,''
  \href{https://arxiv.org/abs/1408.2823}{hep-th/1408.2823}.



\bibitem{Vaid0}
 P.~C.~Vaidya, ``The External Field of a Radiating Star in General Relativity," Curr. Sci. {\bf 12}
(1943) 183,
  \href{http://www.currentscience.ac.in/php/toc.php?vol=012&issue=06}{CurrSci.12.183}.

\bibitem{OriginalVaidya}
  P.~C.~Vaidya,
  ``The gravitational field of a radiating star,''
 Indian Acad. Sci. (Math. Sci.) (1951) {\bf 33}: 264,
   \href{https://link.springer.com/article/10.1007/BF03173260}{IndianAcadSci.33.264}.


\bibitem{VaidyaAdS}
  A.~Wang and Y.~Wu,
  ``Generalized Vaidya solutions,''
  Gen.\ Rel.\ Grav.\  {\bf 31} (1999) 107,
    \href{https://arxiv.org/abs/gr-qc/9803038}{gr-qc/9803038}.

\bibitem{Vad2} S.~Chapman, H.~Marrochio and R.~C.~Myers,
  ``Holographic Complexity in Vaidya Spacetimes II,''
in preparation.

\bibitem{BatMin09}
  S.~Bhattacharyya and S.~Minwalla,
  ``Weak Field Black Hole Formation in Asymptotically AdS Spacetimes,''
  JHEP {\bf 0909} (2009)  034,
  \href{https://arxiv.org/abs/0904.0464}{hep-th/0904.0464}.

\bibitem{Das:2010yw}
  S.~R.~Das, T.~Nishioka and T.~Takayanagi,
  ``Probe Branes, Time-dependent Couplings and Thermalization in AdS/CFT,''
  JHEP {\bf 1007} (2010)  071,
  \href{https://arxiv.org/abs/1005.3348}{hep-th/1005.3348}.

\bibitem{esp} J.~Abajo-Arrastia, J.~Aparicio and E.~Lopez,
  ``Holographic Evolution of Entanglement Entropy,''
  JHEP {\bf 1011} (2010) 149,
  \href{https://arxiv.org/abs/1006.4090}{hep-th/1006.4090}.

\bibitem{ThermAl0}
V.~Balasubramanian, A.~Bernamonti, J.~de~Boer, N.~Copland, B.~Craps, E.~Keski-Vakkuri, B.~Muller, A.~Schafer, M.~Shigemori and W.~Staessens,  ``Thermalization of Strongly Coupled Field Theories,''
  Phys.\ Rev.\ Lett.\  {\bf 106} (2011) 191601,
\href{https://arxiv.org/abs/1012.4753}{hep-th/1012.4753}.


\bibitem{ThermaAlice}
  V.~Balasubramanian, A.~Bernamonti, J.~de~Boer, N.~Copland, B.~Craps, E.~Keski-Vakkuri, B.~Muller, A.~Schafer, M.~Shigemori and W.~Staessens,
  ``Holographic Thermalization,''
  Phys.\ Rev.\ D {\bf 84} (2011)  026010,
   \href{https://arxiv.org/abs/1103.2683}{hep-th/1103.2683}.

\bibitem{Garfinkle:2011hm}
  D.~Garfinkle and L.~A.~Pando Zayas,
  ``Rapid Thermalization in Field Theory from Gravitational Collapse,''
  Phys.\ Rev.\ D {\bf 84} (2011) 066006,
\href{https://arxiv.org/abs/1106.2339}{hep-th/1106.2339}.


\bibitem{HubenyV}
  V.~E.~Hubeny and H.~Maxfield,
  ``Holographic probes of collapsing black holes,''
  JHEP {\bf 1403} (2014) 097,
    \href{https://arxiv.org/abs/1312.6887}{hep-th/1312.6887}.


\bibitem{HubenyV2}
  V.~E.~Hubeny, M.~Rangamani and E.~Tonni,
  ``Thermalization of Causal Holographic Information,''
  JHEP {\bf 1305} (2013) 136,
      \href{https://arxiv.org/abs/1312.6887}{hep-th/1312.6887}.


\bibitem{ShenkerStanfordScrambling}
  S.~H.~Shenker and D.~Stanford,
  ``Black holes and the butterfly effect,''
  JHEP {\bf 1403} (2014) 067,
  \href{https://arxiv.org/abs/1306.0622}{hep-th/1306.0622}.

\bibitem{multiple}
S.~H.~Shenker and D.~Stanford,
  ``Multiple Shocks,''
  JHEP {\bf 1412} (2014) 046,
  \href{https://arxiv.org/abs/1312.3296}{hep-th/1312.3296}.


\bibitem{PerfFluidAct}
  J.~D.~Brown,
  ``Action functionals for relativistic perfect fluids,''
  Class.\ Quant.\ Grav.\  {\bf 10} (1993) 1579,
  \href{https://arxiv.org/abs/gr-qc/9304026}{gr-qc/9304026}.



    \bibitem{HartnollPerfFluid}
  S.~A.~Hartnoll and A.~Tavanfar,
  ``Electron stars for holographic metallic criticality,''
  Phys.\ Rev.\ D {\bf 83} (2011) 046003,
      \href{https://arxiv.org/abs/1008.2828}{hep-th/1008.2828}.




\bibitem{schutz}
B.~F.~Schutz, ``Perfect fluids in general relativity: velocity potentials and a variational principle,''
Phys.\ Rev.\ D {\b 2} (1970) 2762,
  \href{https://journals.aps.org/prd/abstract/10.1103/PhysRevD.2.2762}{PhysRevD.2.2762}.


\bibitem{Nicolis1}
  S.~Dubovsky, T.~Gregoire, A.~Nicolis and R.~Rattazzi,
  ``Null energy condition and superluminal propagation,''
  JHEP {\bf 0603} (2006) 025,
  \href{https://arxiv.org/abs/gr-qc/0512260}{gr-qc/0512260}.


\bibitem{Nicolis2}
  S.~Dubovsky, L.~Hui, A.~Nicolis and D.~T.~Son,
  ``Effective field theory for hydrodynamics: thermodynamics, and the derivative expansion,''
  Phys.\ Rev.\ D {\bf 85} (2012)  085029,
      \href{https://arxiv.org/abs/1107.0731}{hep-th/1107.0731}.

\bibitem{Kovtun:2014hpa}
  P.~Kovtun, G.~D.~Moore and P.~Romatschke,
  ``Towards an effective action for relativistic dissipative hydrodynamics,''
  JHEP {\bf 1407}, 123 (2014),
      \href{https://arxiv.org/abs/1405.3967}{hep-th/1405.3967}.

\bibitem{Felix1}
  F.~M.~Haehl, R.~Loganayagam and M.~Rangamani,
  ``The Fluid Manifesto: Emergent symmetries, hydrodynamics, and black holes,''
  JHEP {\bf 1601}, 184 (2016),
   \href{https://arxiv.org/abs/1510.02494}{hep-th/1510.02494}.

\bibitem{Crossley:2015evo}
  M.~Crossley, P.~Glorioso and H.~Liu,
  ``Effective field theory of dissipative fluids,''
  JHEP {\bf 1709}, 095 (2017),
   \href{https://arxiv.org/abs/1511.03646}{hep-th/1511.03646}.

\bibitem{Torrieri:2016dko}
  D.~Montenegro and G.~Torrieri,
  ``Lagrangian formulation of relativistic Israel-Stewart hydrodynamics,''
  Phys.\ Rev.\ D {\bf 94}, no. 6, 065042 (2016),
   \href{https://arxiv.org/abs/1604.05291}{hep-th/1604.05291}.


\bibitem{kuchar1}
J.~Bicak and K.~V.~Kuchar, ``Null dust in canonical gravity," Phys.~Rev.~{\bf D56} (1997) 4878,
  \href{https://journals.aps.org/prd/abstract/10.1103/PhysRevD.56.4878}{PhysRevD.56.4878}.



\bibitem 
 {York} J.~W.~York, Jr.,
  ``Role of conformal three geometry in the dynamics of gravitation,''
  Phys.\ Rev.\ Lett.\  {\bf 28} (1972) 1082,
  \href{http://journals.aps.org/prl/abstract/10.1103/PhysRevLett.28.1082}{PhysRevLett.28.1082}.



\bibitem 
 {GH} G.~W.~Gibbons and S.~W.~Hawking,
  ``Action Integrals and Partition Functions in Quantum Gravity,''
  Phys.\ Rev.\ D {\bf 15} (1977) 2752,
  \href{http://journals.aps.org/prd/abstract/10.1103/PhysRevD.15.2752}{PhysRevD.15.2752}.


\bibitem 
 {Hay1} G.~Hayward,
  ``Gravitational action for space-times with nonsmooth boundaries,''
  Phys.\ Rev.\ D {\bf 47} (1993) 3275,
  \href{http://journals.aps.org/prd/abstract/10.1103/PhysRevD.47.3275}{PhysRevD.47.3275}.

\bibitem 
 {Hay2} D.~Brill and G.~Hayward,
  ``Is the gravitational action additive?,''
  Phys.\ Rev.\ D {\bf 50} (1994) 4914,
\href{https://arxiv.org/abs/gr-qc/9403018}{gr-qc/9403018}.

\bibitem{Simon2}
A.~Reynolds and S.~F.~Ross,
  ``Divergences in Holographic Complexity,''
  Class.\ Quant.\ Grav.\  {\bf 34} (2017) no.10,  105004,
     \href{https://arxiv.org/abs/1612.05439}{hep-th/1612.05439}.



\bibitem{quench1}
  S.~R.~Das, D.~A.~Galante and R.~C.~Myers,
  ``Universal scaling in fast quantum quenches in conformal field theories,''
  Phys.\ Rev.\ Lett.\  {\bf 112} (2014) 171601,
 \href{https://arxiv.org/abs/1401.0560}{hep-th/1401.0560}.

\bibitem{quench2}
  S.~R.~Das, D.~A.~Galante and R.~C.~Myers,
  ``Universality in fast quantum quenches,''
  JHEP {\bf 1502} (2015) 167,
 \href{https://arxiv.org/abs/1411.7710}{hep-th/1411.7710}.


\bibitem{quench3}
  S.~R.~Das, D.~A.~Galante and R.~C.~Myers,
  ``Quantum Quenches in Free Field Theory: Universal Scaling at Any Rate,''
  JHEP {\bf 1605} (2016) 164
 \href{https://arxiv.org/abs/1602.08547}{hep-th/1602.08547}.



\bibitem{Banados:1992wn}
  M.~Banados, C.~Teitelboim and J.~Zanelli,
  ``The Black hole in three-dimensional space-time,''
  Phys.\ Rev.\ Lett.\  {\bf 69} (1992) 1849,
  \href{https://arxiv.org/abs/hep-th/9204099}{hep-th/9204099}.

\bibitem{Banados:1992gq}
  M.~Banados, M.~Henneaux, C.~Teitelboim and J.~Zanelli,
  ``Geometry of the (2+1) black hole,''
  Phys.\ Rev.\ D {\bf 48}, 1506 (1993),
  Erratum: Phys.\ Rev.\ D {\bf 88}, 069902 (2013),
    \href{https://arxiv.org/abs/gr-qc/9302012}{gr-qc/9302012}.


\bibitem 
{couscous} O.~Coussaert and M.~Henneaux,
  ``Supersymmetry of the (2+1) black holes,''
  Phys.\ Rev.\ Lett.\  {\bf 72} (1994) 183,
\href{https://arxiv.org/abs/hep-th/9310194}{hep-th/9310194}.


\bibitem{run}
K.~Y.~Kim, C.~Niu, R.~Q.~Yang and C.~Y.~Zhang,
  ``Comparison of holographic and field theoretic complexities by time dependent thermofield double states,''
  JHEP {\bf 1802}, 082 (2018),
  \href{https://arxiv.org/abs/1710.00600}{hep-th/1710.00600}. 

\bibitem{Ageev:2018nye}
  D.~S.~Ageev, I.~Y.~Aref'eva, A.~A.~Bagrov and M.~I.~Katsnelson,
  ``Holographic local quench and effective complexity,''
  \href{https://arxiv.org/abs/1803.11162}{hep-th/1803.11162}.


\bibitem{Agon:2018zso}
  C.~A.~Ag\'on, M.~Headrick and B.~Swingle,
  ``Subsystem Complexity and Holography,''
  \href{https://arxiv.org/abs/1804.01561}{hep-th/1804.01561}.


\bibitem{Cano:2018aqi}
  P.~A.~Cano, R.~A.~Hennigar and H.~Marrochio,
  ``Complexity Growth Rate in Lovelock Gravity,''
  \href{https://arxiv.org/abs/1803.02795}{hep-th/1803.02795}.

\bibitem{Lloyd} S.~Lloyd, ``Ultimate physical limits to computation,'' Nature 406 (2000), no. 6799, 1047-€"1054,
 \href{https://arxiv.org/abs/quant-ph/9908043v3}{quant-ph/9908043}.

  \bibitem{Couch:2016exn}
  J.~Couch, W.~Fischler and P.~H.~Nguyen,
  ``Noether charge, black hole volume, and complexity,''
  JHEP {\bf 1703}, 119 (2017),
 \href{https://arxiv.org/abs/1610.02038}{hep-th/1610.02038}.

\bibitem{Cai:2017sjv}
  R.~G.~Cai, M.~Sasaki and S.~J.~Wang,
  ``Action growth of charged black holes with a single horizon,''
  Phys.\ Rev.\ D {\bf 95}, no. 12, 124002 (2017),
 \href{https://arxiv.org/abs/1702.06766}{gr-qc/1610.02038}.

\bibitem{Alves:2018qfv}
  D.~W.~F.~Alves and G.~Camilo,
  ``Evolution of Complexity following a quantum quench in free field theory,''
  \href{https://arxiv.org/abs/1804.00107}{hep-th/1804.00107}.

\bibitem{prep2}
 H.~Camargo, P.~Caputa, D.~Das, M.~P.~Heller and R.~Jefferson,
 ``Entanglement is not enough: complexity as a novel probe of quantum quenches,'' in preparation.



\end{thebibliography}
\end{document}